\documentclass[12pt,a4paper,twoside]{book}

\usepackage{graphicx,subfigure,epstopdf}
\usepackage{float}
\usepackage{amssymb,amsmath}
\usepackage{multirow}
\usepackage{cite}
\usepackage{appendix}
\usepackage{fancyhdr,braket}
\usepackage{time}
\usepackage{xspace}
\usepackage{slashed}

\usepackage{bookmark}
\hypersetup{
    colorlinks=true,
    linkcolor=blue,
    filecolor=magenta,      
    urlcolor=cyan,
    pdftitle={Overleaf Example},
    pdfpagemode=FullScreen,
    citecolor=red,
    }

\usepackage{xcolor}
\usepackage{rotating}
\usepackage{setspace}
\usepackage{type1cm}
\usepackage{enumitem}
\usepackage{chngcntr}
\usepackage{afterpage}
\usepackage{threeparttable}
\usepackage{color} 
\usepackage{enumitem}
\usepackage{amsfonts}
\graphicspath{{./figures/}}
\setlist[itemize,1]{label=$\times$}
\setlist[itemize,2]{label=$\checkmark$}
\setlist[itemize,3]{label=$\diamond$}
\setlist[itemize,4]{label=$\bullet$}
\definecolor{darkgreen}{rgb}{0.0,0.5,0.0}
\definecolor{greenblue}{rgb}{0.0,.1,.4}
\definecolor{brickred}{rgb}{0.8, 0.25, 0.33}
\definecolor{brass}{rgb}{0.71, 0.65, 0.26}
\definecolor{darkorchid}{rgb}{0.6, 0.2, 0.8}

\usepackage{amsmath}

\usepackage{soul}
\usepackage[normalem]{ulem}

\usepackage{amsmath}
\usepackage{makeidx}
\usepackage{amssymb}
\usepackage{graphicx}
\usepackage{bmpsize}
\usepackage{epstopdf}

\newcommand{\chushi}[1]{ }
\newcommand{\Tr}{ \mathrm{Tr} }

\newcommand{\angleLR}[1]{ \left \langle #1 \right \rangle }
\newcommand{\angleN}[1]{ \langle #1 \rangle }

\newcommand{\roundLR}[1]{ \left( #1 \right) }
\newcommand{\roundN}[1]{ ( #1 ) }

\newcommand{\roundB}[1]{ \biggl( #1 \biggr) }

\newcommand{\squareLR}[1]{ \left[ #1 \right] }
\newcommand{\squareN}[1]{ [ #1 ] }

\newcommand{\squareB}[1]{ \biggl[ #1 \biggr] }

\newcommand{\absLR}[1]{ \left| #1 \right| }
\newcommand{\absN}[1]{ | #1 | }

\newcommand{\braN}[1]{ \langle #1 | }
\newcommand{\ketN}[1]{ | #1 \rangle }

\let\calccommentout\iffalse 
\let\calcshow\iftrue 

\newcommand{\eq}[1]{\begin{equation}\begin{split} #1 \end{split}\end{equation}}

\newcommand {\mathsym}[1]{{}}
\newcommand {\unicode}[1]{{}}

\labelformat{chapter}{Chapter #1} 
\labelformat{section}{Section #1} 
\labelformat{subsection}{Section #1} 
\labelformat{subsubsection}{Section #1}
\labelformat{subsubsubsection}{Section #1}
\labelformat{equation}{Eq.~(#1)} 
\labelformat{figure}{Fig.~#1} 
\labelformat{subfigure}{Fig.~\thefigure#1} 
\labelformat{table}{Tab.~#1} 
\labelformat{appendix}{Appendix #1}

\DeclareUnicodeCharacter{2212}{-}
\begin{document}

\pagestyle{empty}  
\begin{titlepage}

\begin{center}
\vspace*{0.25in}

   \textbf{\Large Aspects of entanglement with background electric and magnetic fields in quantum field theoretic systems}\\


\vspace*{0.5in}



\textbf{By}\\ 
\textbf{\large \bf Shagun Kaushal}\\

\vspace*{0.25in}

 {\it Submitted \\ in fulfillment of the requirements for the degree \\ of}\\
 
{\large \bf Doctor of Philosophy}

\vfill

\begin{figure}[h!]
\begin{center}
        \includegraphics[width=4cm]{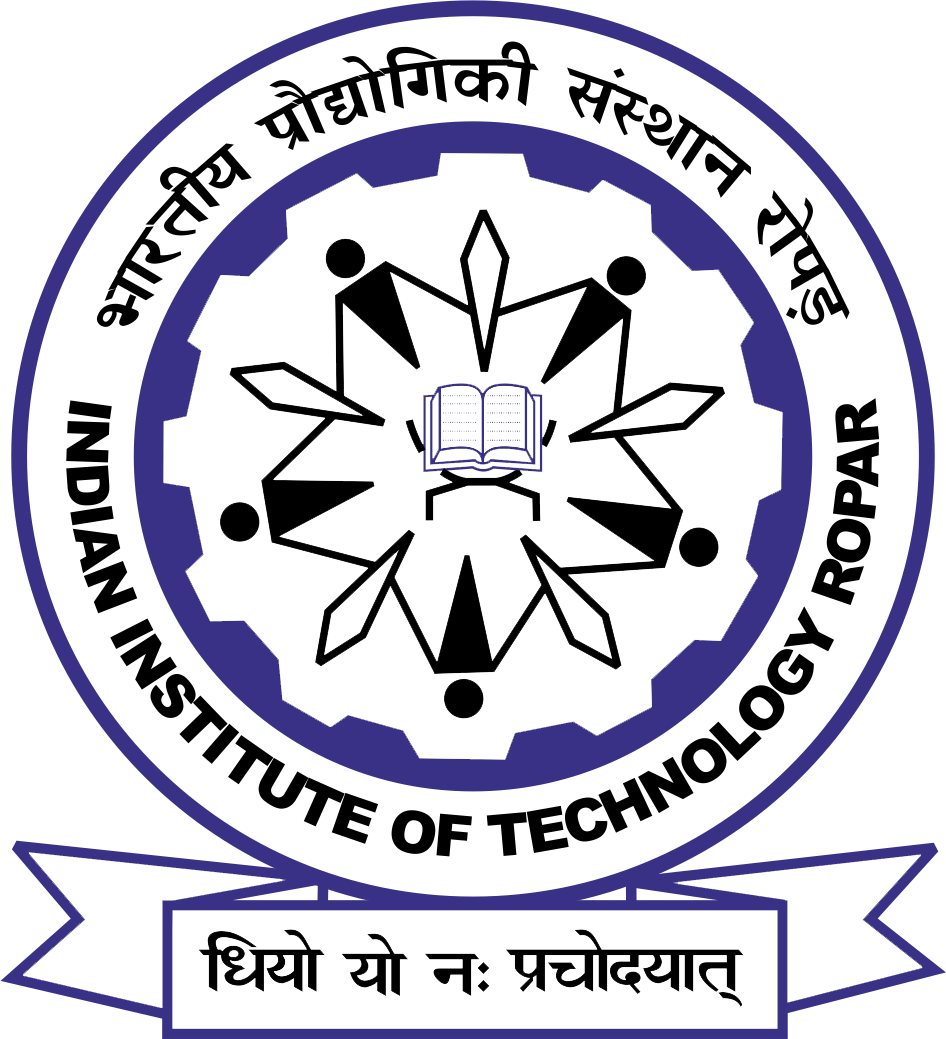}
    \end{center}
\end{figure}

\normalsize{\textbf{Department of Physics}}\\
\normalsize{\textbf{Indian Institute of Technology Ropar}}\\
\normalsize{\textbf{June 2023}}

\end{center}
\end{titlepage}

\thispagestyle{empty}
\vspace*{\fill}
\begin{center}
\end{center}
\vspace*{\fill}

\newpage

\thispagestyle{empty}
\vspace*{\fill}
\begin{center}
\large{\textcopyright Indian Institute of Technology Ropar}\\
\large All rights reserved.
\end{center}
\vspace*{\fill}

\cleardoublepage
\thispagestyle{empty}

\chapter*{}
\thispagestyle{empty}

\vspace*{\fill}
\begin{center}
 \large{This thesis is dedicated to}\\
 \textit{ My Grandfather, My Parents \\ and \\ My Supervisors}
\end{center}
\vspace*{\fill}
\thispagestyle{empty}

\chapter*{Certificate}
\thispagestyle{empty}
It is certified that the work contained in this thesis entitled ``{\bf Aspects of entanglement with background electric and magnetic fields in quantum field theoretic systems}" by {\bf Ms. Shagun Kaushal}, a research scholar in the Department of Physics, to the Indian Institute of Technology Ropar for the award of the degree of {\textbf{Doctor of Philosophy}} has been carried out under our supervision and has not been submitted elsewhere for a degree.
\vspace{1.5in}

\begin{minipage}[t]{0.4\textwidth}
\begin{flushleft}
Dr. Sourav Bhattacharya\\
Associate Professor \\
Department of Physics\\
Jadavpur University Kolkata\\
\vspace{2cm}
\textbf{June 2023}
\end{flushleft}
\end{minipage}
\begin{minipage}[t]{0.5\textwidth}
\begin{flushright}

Dr. Shankhadeep Chakrabortty\\
Assistant Professor \\
Department of Physics\\
Indian Institute of Technology Ropar \\
\end{flushright}
\end{minipage}


\chapter*{Declaration}
\thispagestyle{empty}
I hereby declare that the work presented in the thesis entitled ``{\bf Aspects of entanglement with background electric and magnetic fields in quantum field theoretic systems}" submitted for the degree of \textbf{Doctor of Philosophy} in Physics by me to Indian Institute of Technology Ropar has been carried out under the supervision of \textbf{Dr. Sourav Bhattacharya} and \textbf{Dr. Shankhadeep Chakrabortty}. This work is original and has not been submitted in part or full by me elsewhere for a degree.
\vspace{1in}

\begin{minipage}[t]{0.5\textwidth}
\begin{flushleft}
\vspace{0.6in}
\textbf{June 2023}
\end{flushleft}
\end{minipage}
\begin{minipage}[t]{0.45\textwidth}
\begin{flushright}
\textbf{} \\
Shagun Kaushal\\
PhD Research Scholar \\
Department of Physics\\
Indian Institute of Technology Ropar \\
\end{flushright}
\end{minipage}

\cleardoublepage



\newpage
%

\begin{center}
{
    \fontencoding{OT1}
    \fontfamily{ppl}
    \fontseries{b}
    \fontshape{n}
    \fontsize{17}{40}
    \selectfont
    Acknowledgements\\
  \vspace*{0.2in}
    \small{``Honesty and transparency\\
make you vulnerable\\
and confident''}
}

\vspace*{0.1in}

{
    \selectfont
    }

\end{center}

\vspace*{0.1in}

\begin{spacing}{1.2}

{These are the few words I learned from my supervisor and are now a part of my present take towards life.  This journey has been one of the most exciting, and the learning portion of my life has taught me a great deal that has been tremendously influenced by the various people I have encountered. I apologise to those who have not yet been mentioned, but I will do my best to do so. The absences and the selection made here were not intended to be valued.

First and foremost, I want to express my gratitude to my supervisors, Dr. Sourav Bhattacharya and Dr. Shankhadeep Chakrabortty, for their continuous encouragement and motivation throughout this journey. I am incredibly grateful to them for their guidance, for our many fascinating scientific and non-scientific conversations, and for believing in me.  The true gentlemen, humble and full of life, Dr. Sourav and Dr. Shankhadeep have taught me to handle things independently with a positive outlook and have always encouraged me to venture into the unknown.
This has indeed helped a lot in boosting my confidence and shaped me to be a better person. In many respects, the circle of relationships we developed and shared throughout the years has made things a lot simpler for me. I feel honoured to receive nourishment from them.

I am grateful to my Doctoral Committee members, Dr. Shubhrangshu Dasgupta, Dr. Sandeep Gautam and Dr. Tapas Chatterjee, for their valuable suggestions and discussions.

I thank Mr. Anshu Vaid for his technical and administrative help in need. My sincere thanks go out to Parkash. Without his refreshing cups of tea, this Phd journey would have been fatigue.

Among the students, I should first thank Dr. Hironori Hoshino and Dr. Md. Sabir Ali for their help in familiarising me with the techniques one needs to cope with a problem. I am thankful to Dr. Karunava Sil for his caring and motivating attitude. I am extremely grateful to my group members Dr. Rajesh, Nitin, Sanjay, Arpit, Meenu, Sudesh, Pronoy and Siddant, for maintaining a joyful environment in our group. Their support, motivation, care and surprises will remain invaluable in this journey.

I am grateful to my girls’ gang: Arzoo, Bipasha, Manju, Monika and Param, for their constant support, which eventually led to the time of submitting our thesis. I thank my friend Vasu Dev for all the laughter, late-night movies, gossip and fights that made this journey pass by so easily. I wish to thank Arpit  and Sahil Dani for their constant help whenever required. I also thank Anita, Love, Maneesha, Nancy, Prerna, Sahil Sahoo and Seema for their constant support.
The journey would have remained incomplete without Daman, Jaspreet, Katyayni, Rakhi, Raghav, Soni and Faizan bhai, members of the ``IIT Ropar family''. 

Finally, I would like to express my heartfelt gratitude towards my family. I am indebted to my mother and father, who have not only laid the foundation of my existence but also believed in me and allowed me to venture into whatever I wanted to do in my life. They shared my happiness in times of success and taught me the magic of patience and holding on to my aspirations in moments of disappointment. I want to express my gratitude towards my brother Madhav. He supported and motivated me in life's ups and downs and suggested improvement. Thanks to my caring and loving sister, Shelja, for her indomitable support and guidance. She has been a patient listener in my times of need. I am also thankful to my Dadaji, Buaji, Navwin jiju, Litika and cousin Akshit for their love and blessings.

I thank everyone who came into my life and contributed directly or indirectly to achieving this accomplishment.


    \begin{flushright}
Shagun Kaushal\\Indian Institute of Technology Ropar, India-140001\\ June 2023
    \end{flushright}
}

\end{spacing}





%
%


\frontmatter
\pagestyle{plain}
\centerline{\bf \Large List of Publications}
\vskip .4in
\thispagestyle{empty}
{\bf Peer-reviewed Journals :}
\vskip .3in
\makeatletter
\renewcommand{\theenumi}{\arabic{enumi}}
\makeatother
{\bf \underline{Included in this Thesis}}
\begin{enumerate}
\item S. Bhattacharya, S. Chakrabortty, H. Hoshino and \textbf{S. Kaushal}, \lq\lq {\it Background magnetic field and quantum correlations in the Schwinger effect} \rq\rq, Phys. Lett. B \textbf{811}, 135875 (2020) doi:10.1016/j.physletb.2020.135875 [arXiv:2005.12866 [hep-th]].

\item M.~S.~Ali, S.~Bhattacharya, S.~Chakrabortty and \textbf{S. Kaushal},
{\it``Fermionic Bell violation in the presence of background electromagnetic fields in the cosmological de Sitter spacetime''},
Phys. Rev. D \textbf{104}, no.12, 125012 (2021) doi:10.1103/PhysRevD.104.125\\012
[arXiv:2102.11745 [hep-th]].

\item \textbf{S.~Kaushal},
``Schwinger effect and a uniformly accelerated observer,''
Eur. Phys. J. C \textbf{82}, no.10, 872 (2022) doi:10.1140/epjc/s10052-022-10836-6
[arXiv:2201.03906 [hep-th]].

\end{enumerate}

{\bf \underline{Not included in this Thesis}}
\begin{enumerate}

\item  M.~S.~Ali and \textbf{S. Kaushal},
``Gravitational lensing for stationary axisymmetric black holes in Eddington-inspired Born-Infeld gravity,''
Phys. Rev. D \textbf{105}, no.2, 024062 (2022) doi:10.1103/PhysRevD.105.024062 [arXiv:2106.08464 [gr-qc]].

\item S.~Bhattacharya, N.~Joshi and \textbf{S.~Kaushal},
``Decoherence and entropy generation in an open quantum scalar-fermion system with Yukawa interaction,'' Eur. Phys. J. C \textbf{83}, 208 (2023) doi:10.1140/epjc/s10052-023-11357-6
[arXiv:2206.15045 [hep-th]].

\item M.~S.~Ali, S.~Bhattacharya and \textbf{S.~Kaushal},
``Stationary black holes and stars in the Brans-Dicke theory with \ensuremath{\Lambda}\ensuremath{>}0 revisited,''
Phys. Rev. D \textbf{106}, no.12, L121502 (2022) doi:10.1103/PhysRevD.106.L121502
[arXiv:2209.11011 [gr-qc]].

\end{enumerate}


\pagestyle{plain}

\begin{center}
{
    \fontencoding{OT1}
    \fontfamily{ppl}
    \fontseries{b}
    \fontshape{n}
    \fontsize{20}{40}
    \selectfont
    Abstract
}

\vspace*{0.1in}

{
    \selectfont
    }

\end{center}

\vspace*{0.1in}

\begin{spacing}{1.2}

{\noindent 

This thesis investigates the impact of the background magnetic field on correlations or entanglement between pairs created by the background electric field in quantum field theoretic systems in the Minkowski, the primordial inflationary de Sitter and the Rindler spacetimes. These analyses might provide insight into the relativistic entanglement in the early inflationary universe scenario, where such background fields might exist due to primordial fluctuations, and in the near-horizon of non-extremal black holes, which are often endowed with background electromagnetic fields due to the accretion of plasma onto them. 

In the beginning, with a brief introduction to relativistic quantum field theory in curved spacetimes, a review of relevant spacetimes and quantum field phenomenon in these spacetimes is provided. After that, we reviewed the Schwinger effect and different measures to quantify correlations or entanglement. In the first objective of this thesis, we begin with the simplest scenario in which we consider two complex scalar fields, not necessarily of the same rest masses, coupled to the constant background electric and magnetic fields in the $(3+1)$-dimensional Minkowski spacetime and subsequently investigate a few measures quantifying the correlations between the created Schwinger pairs. Since the background magnetic field itself cannot cause the decay of the Minkowski vacuum, our chief motivation here is to investigate the interplay between the effects due to the electric and magnetic fields. We first compute the entanglement entropy for the vacuum state of a single scalar field. Next, we consider some maximally entangled states for the two-scalar field system and compute the logarithmic negativity and the mutual information corresponding to different particle-antiparticle excitations. 

In the second objective, we consider Dirac fermions in the cosmological de Sitter spacetime in the presence of background electric and magnetic fields of constant strength. We investigate the violation of the Bell inequality and the mutual information. This scenario has two sources of particle creation: the background electric field and the time-dependent gravitational field. The orthonormal Dirac mode functions are obtained, and the relevant in-out squeezed state expansions in terms of the Bogoliubov coefficients are found. We focus on two scenarios here: $1.$ a strong electric field limit, $2.$ a heavy mass limit (with respect to the Hubble constant). Using the squeezed state expansion, we demonstrate the Bell violations for the vacuum and some maximally entangled initial states.  Our chief aim thus far is to investigate the role of the background magnetic field strength in the Bell violation. Qualitative differences in this regard for different maximally entangled initial states are shown. Further extension of these results to the one parameter family of de Sitter $\alpha$-vacua is also discussed. 

Our next natural aim is to investigate the correlations between the fermionic pairs created in the Rindler spacetime in the presence of background electric and magnetic fields of constant strength. By solving the Dirac equation in closed form in each wedge, the orthonormal local in and out modes are obtained. Next, we construct the global modes from the local ones, which are analytic in both wedges. We further quantize the field, and by comparing the local and global modes, field quantizations and the Bogoliubov transformations are obtained. Using them, the squeezed state expansion of the global vacuum in terms of local states is acquired, and accordingly, the spectra of created particles are found. This scenario also has two
sources of particle creation: the Schwinger and the Unruh effects. Our chief aim is to investigate the role of the strength of the background electric and magnetic fields on the spectra of created particles. Next, we discuss some possible implications of this
result in the context of quantum entanglement by computing the Bell inequality and the logarithmic negativity for the global vacuum.

Finally, we summarise the results discussed in the aforementioned main studies of the thesis. We also mention some future directions that might help in gaining a deeper understanding of correlations between pairs created in the presence of background fields.

}

\end{spacing}

\pagestyle{plain}
\include{abr}
\pagestyle{plain}
\centerline{\bf \Large Notations}
\vskip .5in
\thispagestyle{empty}
\makeatletter
\renewcommand{\theenumi}{\arabic{enumi}}
\makeatother
$S$ Action.

$\Gamma^{\lambda}_{\mu \nu}$  Christoffel connections.

$\Lambda$ Cosmological constant.

$\eta$ Cosmological time.

$g_{\mu \nu}$ Curved spacetime metric.

$e^-$ Electron.

$e$ Electric charge of positron. Electron carries charge $-e$.

$E$ Electric field strength.

$\psi$ Fermionic field.

$\eta_{\mu \nu}$ Flat spacetime metric.

$D_\mu$ Gauge covariant derivative.

$H$ Hubble constant.

$\mathcal{L}$ Lagrangian density.

$B$ Magnetic field strength.

$G$ Newton constant.

$e^+$ Positron.

$\phi$ Scalar field.

$\Gamma_\mu$ Spin connection.

$n_L$ the Landau level

\pagestyle{plain}
\tableofcontents
\listoffigures
\pagestyle{plain}
\pagestyle{empty}



\pagestyle{fancy} 
\renewcommand{\chaptermark}[1]{\markboth{\textbf{\thechapter}\ \emph{#1}}{}}
\renewcommand{\sectionmark}[1]{\markright{\thesection.\ #1}}

\fancyhf{} 
\fancyhead[LE,RO]{\textbf{\thepage}} 
\fancyhead[RE]{\nouppercase{\rightmark}} 
\fancyfoot[RO]{\bfseries{\leftmark}}
\renewcommand{\headrulewidth}{0.5pt} 
\renewcommand{\footrulewidth}{0.5pt} 




\mainmatter
\chapter{Motivation and Overview}
\label{Motivation and Overview}
In quantum field theory, the spacetime symmetries play a very crucial role. For example, in the Minkowski spacetime, the Poincare symmetry allows us to define a unique and stable vacuum, whereas being deprived of this symmetry one cannot do so in curved spacetimes \cite{ Parker:2009uva, Peskin:2018, Weinberg, Sean, QFTCS, QFTCS1, QFTCS2, QFTCS3, Hollands:2014eia}. The vacuum in quantum field theory has fluctuations owing to the Heisenberg uncertainty principle \cite{Reynaud:2001kc, Sakharov:1967pk, Streeruwitz:1975wzf, Zeldovich:1971mw, Ford:1975su, Chemisana:2020etd, Mainland:2018yqz, Mainland:2018aao}. As a consequence of these vacuum fluctuations, virtual particle-antiparticle pairs are created inside loops \cite{Bjorken:1965zz, Roman, Gribov:2000nhz}. If these pairs are electrically charged, such as positrons and electrons, and in addition a `strong enough' background or classical electric field is applied, these virtual pairs can be ripped apart and seen as real particle-antiparticle pairs \cite{Avramidi:1989ik}. However, the threshold electric field required for this pair creation to occur is estimated as $10^{18}V m^{-1}$, which is far too high to be achieved with nowadays technology. This phenomenon is known as the Schwinger effect \cite{Reynaud:2001kc, Sakharov:1967pk, Schwinger1}, initially predicted by Sauter \cite{Sauter} and later formalised by Schwinger \cite{Schwinger}. We note that the vacuum here, owing to the directionality of the background electric field, is no longer the Poincare invariant Minkowski vacuum, even though the spacetime is flat. Also, the creation of pairs suggests that this vacuum is unstable. Such pair creation can occur due to the presence of a gravitational field as well and is vastly studied in example \cite{Reynaud:2001kc, Sakharov:1967pk, Streeruwitz:1975wzf, Zeldovich:1971mw, Ford:1975su, Chemisana:2020etd}.

This indicates that in the presence of a background electric field, the `initial' vacuum of the quantum field theory in the Minkowski spacetime turns into a squeezed state of particle-antiparticle pairs at late times, corresponding to some suitable `out' vacuum \cite{Kim:2016xvg, Srinivasan:1998fk, Brout:1993be}. Similarly, the `initial' vacuum state in an expanding Friedmann-Lemaitre-Robertson-Walker cosmological universe is a squeezed state of particle-antiparticle pairs at some late times, corresponding to some suitable `out' vacuum. In this case, the pair creation occurs due to the time-dependent gravitational field, as was first pointed out by  Parker \cite{Parker:1968mv, Parker:1969au, Parker:1971pt, Parker:2012at}. We further refer our reader to \cite{QFTCS} for an exhaustive review of pair creation in a non-trivial background. Likewise, the Hawking radiation from a collapsing black hole is another example of pair creation and vacuum instability in a non-trivial background \cite{Hawking, Hawking1}. 

These instances deal with pair creation due to a background electric or time-dependent gravitational field. For the Schwinger effect, however, as per the covariance of the Maxwell electrodynamics, it seems natural to consider a magnetic field as well, along with the electric field. It is well known that a magnetic field alone cannot create pairs from vacuum instability \cite{Parker:2012at}. It would be interesting to see what happens if we take the background magnetic field into account alongside the background electric field. In the presence of both fields, we may speculate that the magnetic field will oppose the effect of the electric field, which can be intuitively guessed as follows. Let us imagine an $e^{+}e^{-}$ pair is created due to the application of the electric field. The electric Lorentz force will act in opposite directions to separate them eventually. However, the magnetic Lorentz
force, $e(\Vec{v}\times\Vec{B})$, will act in the same direction for both of them. Hence, one can guess that the magnetic field might stabilise the vacuum disturbed by the electric field. Will the scenario remain the same if we have the gravitational field instead of the electric field or both? How will the picture change from the perspective of a uniformly accelerated Rindler observer in the Minkowski spacetime?

We also note that one can think about realistic natural scenarios where electric and magnetic fields might coexist. These include the presence of primordial electromagnetic fields present in the early inflationary universe. The astrophysical black holes can also be endowed with electric and magnetic fields due to the accretion of plasma onto them. 

We have chiefly reviewed above the pair creation in the presence of a background electric or time-dependent gravitational field. As we have mentioned earlier, when pair creation occurs, the initial vacuum transforms into a squeezed state with respect to a suitable `out' vacuum. A remarkable aspect of these `out' particle-antiparticle states is that they are entangled. The entanglement properties between particle-antiparticle created by a background electric field have been investigated for the vacuum state in flat spacetime, example \cite{Li:2016zyv, Wu:2020dlg, Ebadi:2014ufa}. It was shown that the entanglement increases with the increase in the electric field strength for the vacuum state. Such entanglement by the time-dependent gravitational field has been studied in, example \cite{Fuentes:2010dt, Arias:2019pzy, EE, bell:2017, vaccum EE for fermions, QC in deSitter}.
The entanglement properties between created particles due to background electric and gravitational fields have been studied earlier by \cite{Ebadi2015}. It would be interesting to see how a magnetic field's strength will affect the entanglement created by electric and gravitational fields. According to the intuitive Lorentz force picture we considered earlier, we may speculate that the magnetic field will degrade the entanglement created by the electric field. What happens to the entanglement created by a gravitational field? How would this scenario change if we have mixed states instead of the vacuum, which is pure? We note that classical correlations are also present for mixed states, for example, \cite{Datta_2007, Modi_2011}. Keeping this in mind, we shall construct some initially entangled states from two complex scalar or fermionic fields and will obtain mixed states corresponding to different sectors of these particles and antiparticles. Studying how these correlations change for some non-trivial spacetime backgrounds would also be interesting. 

 The entanglement or correlations between the quantum states are the fundamental characteristics of quantum mechanics, for example, \cite{bell_4, Bell, CHSH, Werner:1989zz, seprability, bell1:2003, bell_1, bell_3}. It has gained tremendous attention in the last few years due to its significant role in quantum information and quantum computation. Not only in the non-relativistic sector, but various methodologies also exist nowadays to study these correlations in relativistic quantum field theory. These chiefly include, firstly, characterising quantum entanglement using modes identified by coordinates or momenta and, second, by using unequal-time correlation functions. For details, we refer our reader to this \cite{Anastopoulos:2022owu, Casini:2022rlv, Preskill, RevModPhys, Jordan:2011ci, Giddings:2012bm, Calabrese:2004eu, Wu:2022glj} and references therein. In this thesis, we aim to work with the first methodology. For quantifying entanglement or correlations, several measures are studied in a wide range by researchers, for example, \cite{bell_1, measure, Monogamy, Vidal:1998re, Horodecki:2009zz}. Various studies try to understand the significance of entanglement in black hole thermodynamics and the information loss issues, for example, \cite{Unruh:2017uaw, Penington:2019npb, Bekenstein:1973ur, Fiola:1994ir}. Much effort have been given on entanglement degradation in non-inertial frames and in the expanding universe, for example, \cite{FuentesSchuller:2004xp, Torres-Arenas:2018vei, Wang:2010qq, Wu:2021pja, Dong:2019jhs, Hwang:2001etg, Agullo:2009zza, Antoniadis:1986sb, Martin:2015qta, EE, bell:2017, vaccum EE for fermions, QC in deSitter, SSS, SHN:2020, Choudhury:2016cso, Choudhury:2017bou, MartinMartinez:2010ar, Fuentes:2010dt, Wu:2023pge, Maldacena:2015bha, Esmaelifar:2022gya, Ebadi:2014wva, Ueda:2021nln, Higuchi:2018tuk, Higuchi:2017gcd }. The study of entanglement in the context of quantum field theoretic scattering processes has been made earlier in \cite{Seki:2014cgq}. The entanglement between degrees of freedom associated with different momenta has also been investigated in \cite{Balasubramanian:2011wt}. As we have stated earlier, we wish to study quantum field theoretic entanglement in the Minkowski, the Rindler and the primordial inflationary cosmological spacetime in the presence of background electric and magnetic fields. Since these spacetimes are physically very well motivated, to the best of our knowledge and understanding, this study seems to be interesting in its own right, and it supposes to give us useful insights about the role of the magnetic field in relativistic quantum entanglement.

The rest of the Chapter is organised as follows. In \ref{Quantum field theory in curved spacetime}, we have outlined some very basic ingredients of quantum field theory in curved spacetime. In \ref{Review of relevant spacetime geometries}, we review the relevant spacetime geometries considered in this thesis and some interesting quantum phenomenon in them are also mentioned. In \ref{The Schwinger effect}, we review the Schwinger effect for a complex scalar field in the flat spacetime. In \ref{A short introduction to quantum information}, we have given a very short introduction to quantum information, including some measures of it to quantify correlations or entanglement. Finally, in \ref{Highlights of the thesis}, we have given a brief overview of the rest of the Chapters in this thesis. 

We shall take the mostly positive signature of the metric and will chiefly work in $(1+3)$-dimensions $(-,+,+,+)$. We shall set $c=\hbar=1$. We shall use Einstein's summation convention, i.e. if not otherwise stated repeated indices will always be summed over. The logarithms are understood as $\log_2$ in our numerical calculations.
 \noindent

\section{Field theory in non-trivial backgrounds}
\label{Quantum field theory in curved spacetime}
The general theory of relativity is a classical field theory of gravity. It predicts that matter's gravitational influence results in curving spacetime and is used to study the geodesics or trajectories of freely falling particles. The metric field $g_{\mu \nu}$ represents the curved spacetime line element. In contrast to the flat spacetime metric $\eta_{\mu\nu}$, $g_{\mu\nu}$'s components can depend on spacetime. 

The general theory of relativity is governed by the Einstein field equations. These equations can be derived from the Einstein-Hilbert action,
\begin{equation}
    \label{EHA}
    S_G=\int d^4x  \sqrt{-g} \Big[\frac{1}{16 \pi G}(R-2\Lambda)+\mathcal{L}_M\Big] 
\end{equation}
where $g$ is the determinant of spacetime metric $g_{\mu\nu}(x)$. $R$ is the Ricci curvature scalar given by
\begin{equation}
\label{RScalar}
R = R_{\mu\nu}g^{\mu\nu}
\end{equation}
where $R_{\mu\nu}$ is the Ricci tensor given by
\begin{equation}
    \label{Rtensor}
    R_{\mu\nu}=\partial_{\lambda}\Gamma^{\lambda}_{\mu\nu}-\partial_{\nu}\Gamma^{\lambda}_{\mu\lambda}+\Gamma^{\lambda}_{\mu \nu} \Gamma^{\rho}_{\lambda \rho}-\Gamma^{\lambda}_{\mu \rho}\Gamma^{\rho}_{\nu \lambda}
\end{equation}
where $\Gamma^{\lambda}_{\mu \nu}$ are the Christoffel connections given by
\begin{equation}
  \Gamma^{\lambda}_{\mu \nu}=\frac{1}{2}g^{\lambda \alpha}\Big(\partial_\nu g_{\alpha \mu}+\partial_\mu g_{\alpha \nu}-\partial_\alpha g_{ \mu \nu}\Big).
\end{equation}
$G$ and $\Lambda$ in \ref{EHA} are respectively the Newton constant and the cosmological constant. $\mathcal{L}_M$ stands for the Lagrangian density of any relevant matter field. The action \ref{EHA} is diffeomorphism invariant, i.e. invariant under any general coordinate transformation. The Einstein field equations can be derived by extremising \ref{EHA} with respect to $g_{\mu\nu}$, and is given as
\begin{equation}
    \label{EEM}
    R_{\mu\nu}-\frac{1}{2}g_{\mu\nu}R+\Lambda g_{\mu\nu}=8\pi G T_{\mu\nu}
\end{equation}
where $T_{\mu\nu}$ is the energy-momentum tensor defined as $T_{\mu\nu}=-\frac{2}{\sqrt{-g}}\frac{\delta (\sqrt{-g} \mathcal{L}_{M})}{\delta g^{\mu\nu}}$. 

The equation of motion of the matter field can be found by extremising $\mathcal{L}_M$ with respect to the corresponding field. If the field is quantum, \ref{EEM} is modified by replacing $T_{\mu\nu}$ with its expectation value with respect to some suitable vacuum state. These modified Einstein's equations, along with the field equations, will describe the propagation and dynamics of the quantum field in curved spacetime, for example, \cite{Parker:2009uva, Peskin:2018, QFTCS, QFTCS1, QFTCS2, QFTCS3, Hollands:2014eia}. In particular, as we have previously emphasised, in dynamically curved spacetimes, there can be pair creation from the vacuum of a quantum field theory. It has exciting impacts in cosmological and black hole spacetime backgrounds. 

 Let us briefly discuss now how one writes the action of a matter field in a diffeomorphism invariant way in a curved background. For example, we recall that for a real scalar field with no interactions, self or otherwise, the Poincare invariant action in the Minkowski spacetime reads
\begin{equation}
    \label{actionM}
    S=-\frac{1}{2}\int d^4x \Big(\eta^{\mu \nu}(\partial_\mu\phi)(\partial_\nu \phi)+m^2 \phi^2\Big)
\end{equation}
To generalise this action to a curved spacetime, $\eta_{\mu \nu}$ must be replaced by $g_{\mu\nu}$, and the covariant derivative must replace the ordinary derivative. Moreover, to make the volume element invariant, we replace $d^4x$ with $d^4x\sqrt{-g}$ to have
\begin{equation}
    \label{actionC}
    S=-\frac{1}{2}\int d^4x \sqrt{-g} \Big( g^{\mu\nu}(\nabla_\mu\phi)(\nabla_\nu \phi)+m^2 \phi^2\Big)
\end{equation}
which describes a real scalar field coupled to gravity minimally \cite{Sean}. In a more general case, a non-minimal interaction term of the form $\xi R \phi^2$ is also added, where $\xi$ is a constant. The Klein-Gordon equation corresponding to \ref{actionC} is given by
\begin{equation}
    \label{KGC}
    (\nabla_\mu \nabla^\mu -m^2)\phi=\Big[\frac{1}{\sqrt{-g}}\partial_\mu(\sqrt{-g}g^{\mu\nu}\partial_\nu)-m^2\Big]\phi=0
\end{equation}
Similarly, one can consider a fermionic field in curved spacetime. Its action in the flat spacetime reads
\begin{equation}
    \label{DactionM}
    S=\int d^4x \;\Bar{\psi}(x)(\imath \gamma^{\mu} \partial_\mu-m)\psi(x)
\end{equation}
where $\gamma^{\mu}$'s are the flat space gamma matrices and $\psi(x)$, $\Bar{\psi}(x)$ are the fermionic field and its adjoint ($\Bar{\psi}=\psi^{\dagger}\gamma^0$). These gamma matrices will satisfy the anticommutation relation $[\gamma^\mu,\gamma^\nu]_{+}=- 2 \eta^{\mu \nu} I_{4\times4}$, where $I_{4\times4}$ is a $4\times4$ identity matrix. The generalization of this action for curved spacetime is given by
\begin{equation}
    \label{DactionC}
     S=\int d^4x \;\sqrt{-g}\Bar{\psi}(x)(\imath \gamma^{\mu} D_\mu-m)\psi(x)
\end{equation}
where $\gamma^\mu$'s are the curved space gamma matrices and $\Bar{\psi}=\psi^{\dagger}\gamma^0$ as earlier \cite{QFTCS, QFTCS1}. One can relate flat and curved spacetime gamma metrics using the tetrads $e_{a}^{\mu}$; here, the Latin indices correspond to the local Lorentz frame metric, whereas the Greek indices correspond to curved spacetime. We have
\begin{equation}
    \label{GMR}
    \gamma^\mu=e^{\mu}_a \gamma^{a}
\end{equation}
the tetrads satisfy
\begin{equation}
    \label{metricrel}
    g^{\mu\nu}(x)=e^{\mu}_a(x) e^{\nu}_b(x) \eta^{ab}
\end{equation}
\ref{GMR} and \ref{metricrel} gives the curved space annicommutation relation.
\begin{equation}
    [\gamma^\mu,\gamma^\nu]_{+}=- 2 g^{\mu \nu} I_{4\times4}
\end{equation}
We refer our reader to \cite{Parker:2009uva, Peskin:2018} for a detailed discussion on the tetrads. The spin covariant derivative $(D_\mu)$ in \ref{DactionC} is defined as
\begin{equation}
    \label{covariant}
    D_\mu(x)=\partial_\mu+\Gamma_\mu(x) 
\end{equation}
where $\Gamma_\mu$'s are the spin connections given by,
\begin{equation}
\label{conn11}
\Gamma_\mu=-\frac{1}{8}e^{\mu}_a \left(\partial_{\mu}e_{b\nu}- \Gamma_{\mu\nu}^{\lambda} e_{b\lambda}\right)[\gamma^a, \gamma^b],
\end{equation}
The Dirac equation in curved spacetime reads
\begin{equation}
    \label{DECST}
    (\imath \gamma^{\mu} D_\mu -m)\psi(x)=0
\end{equation}
Since in this thesis, we are interested in studying quantum field theory with background electric and magnetic fields in flat as well as in curved spacetimes, we need to consider the gauge field coupling to the matter field, for which we need to replace 
\begin{equation}
\label{1}
     \nabla_\mu \to D_\mu=(\partial_\mu+\imath e A _\mu)
\end{equation}
in \ref{actionC} for a complex scalar field and  
\begin{equation}
\label{2}
 D_\mu \to D_\mu=(\partial_\mu+\imath e A_\mu+\Gamma_\mu) 
\end{equation}
for a fermionic field. In \ref{1} and \ref{2}, $A_\mu$ and $e$ denote the gauge field and coupling parameter, respectively. In the same spirit, one can write down the action of any matter field in a curved spacetime by following the minimal substitution prescription. We now wish to briefly discuss below the structure of certain spacetimes and some important quantum field theory phenomenon in them relevant to this thesis.
\noindent
\section{Review of the relevant spacetime geometries}
\label{Review of relevant spacetime geometries}
In \ref{The Minkowski spacetime}, we shall discuss the global structure of the Minkowski spacetime. In \ref{The de Sitter spacetime}, we shall discuss the global geometry of the de Sitter spacetime and the particle creation due to the gravitational field. In \ref{The Rindler spacetime}, we discuss the Rindler spacetime, i.e. the Minkowski spacetime seen by a uniformly accelerated non-inertial observer, and the associated thermal effect.
\noindent
\subsection{The Minkowski spacetime}
\label{The Minkowski spacetime}
\begin{figure}[ht]
    \centering
\includegraphics[scale=.47]{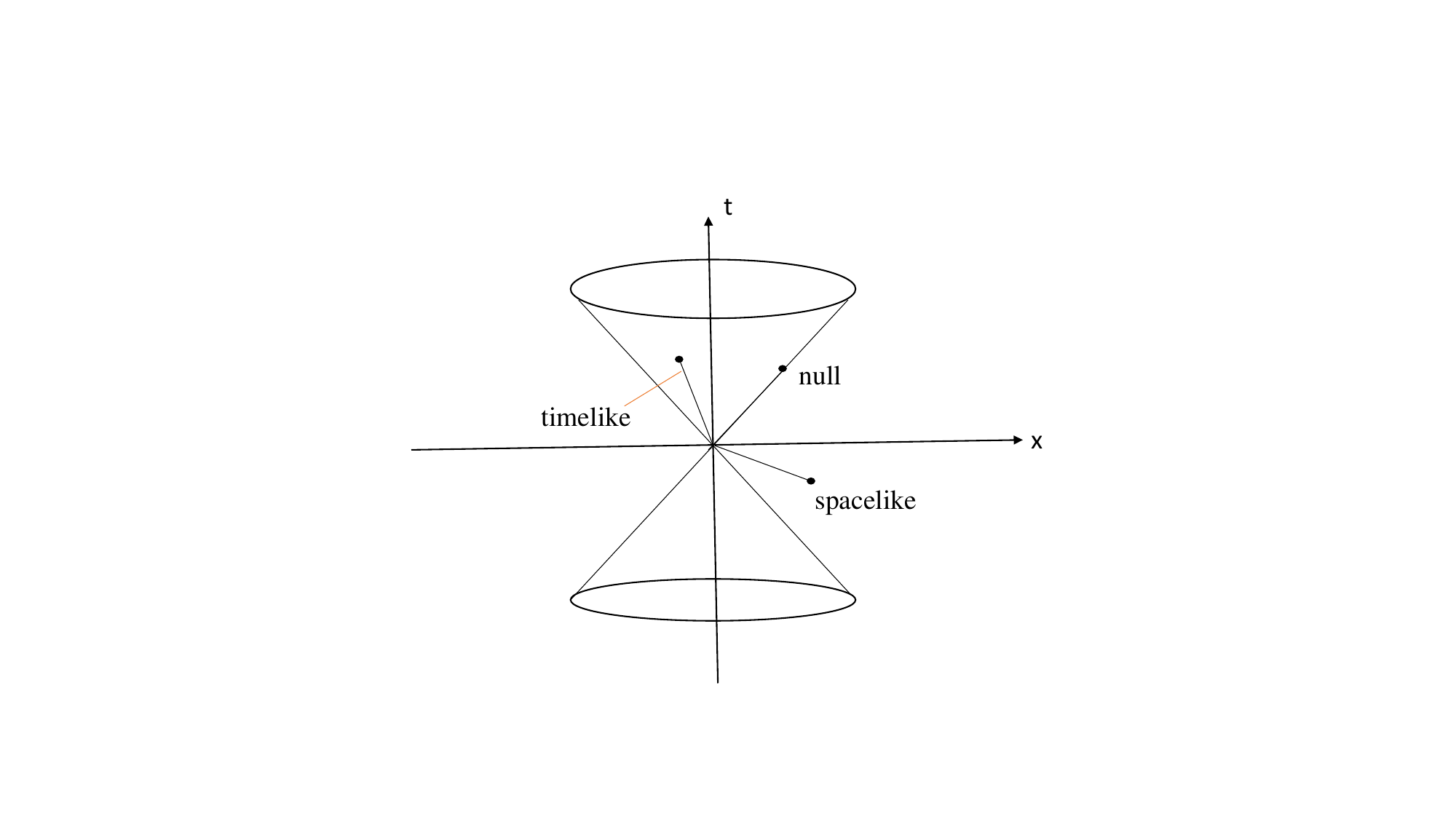}
    \caption{\it{\small This figure illustrates the Minkowski spacetime light cone, with the t and x axes shown at right angles and the y and z axes suppressed. The spacelike, timelike, and lightlike occurrences have been demonstrated.}}
    \label{lightcone}
\end{figure}

\begin{figure}[ht]
    \centering
    \includegraphics[scale=.47]{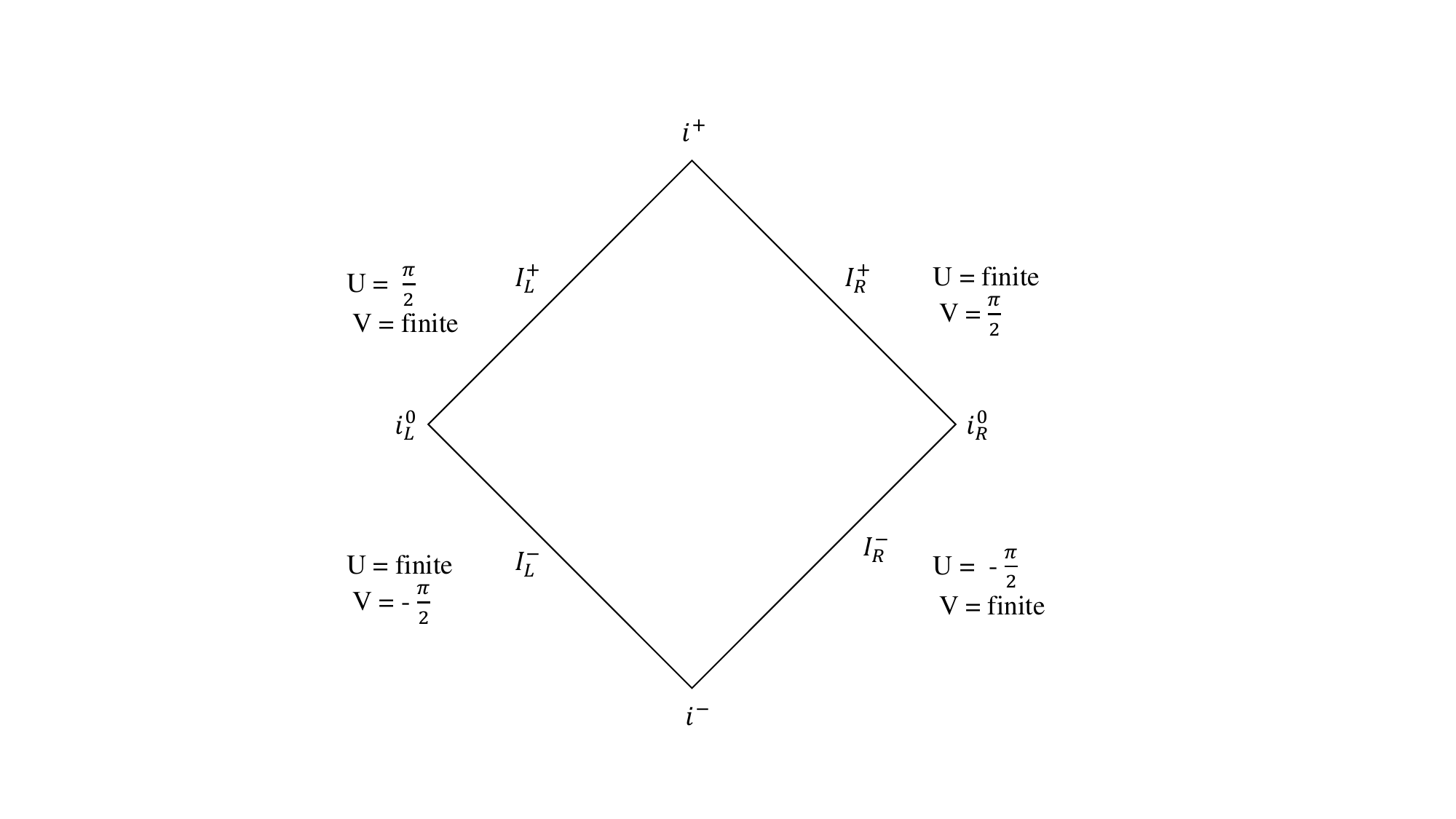}
    \caption{\it{\small The Penrose-Carter diagram of the $(1+1)$-dimensional Minkowski spacetime}}
    \label{Minkowskipenrose}
\end{figure}

The line element of the Minkowski spacetime is
\begin{equation}
    \label{MM}
    ds^2=-dt^2+dx^2+dy^2+dz^2
\end{equation}
The various intervals are classified as
\begin{equation}
 ds^2 \; \begin{cases}
      > \;0,\; \text{spacelike}\\
     =\; 0, \;\text{null or lightlike}\\
      <\; 0,  \;\text{timelike}
  \end{cases}  
\end{equation}
In \ref{lightcone}, where we consider the $t$ and $x$ of \ref{MM} and suppressed the $y$ and $z$ coordinates to consider a $2$-dimensional spacetime and represented these events with respect to the lightcone. The null or lightlike intervals are located on the surface of the lightcone, whereas the spacelike and timelike intervals are respectively located outside and inside of it. 
In order to understand the global structure of any spacetime, the Penrose-Carter diagrammatic technique is very useful \cite{Penrose:1962ij, Carter, Carter:1969zz}. This uses certain conformal transformations in order to bring the boundaries, which are located at spacetime infinities, to finite distances. Thus, the Penrose-Carter diagram allows us to `see' the entire spacetime on a piece of paper. For the $(1+1)$-dimensional Minkowski spacetime, the diagram is depicted in \ref{Minkowskipenrose}, in which the light rays are always at $\pm45^o$ with the vertical, can be derived as follows.

Let us consider the $t$ and $x$ parts of \ref{MM}. The ranges of these coordinates are $-\infty < x< \infty $ and $-\infty < t< \infty $. Introducing the retarded and advance null coordinates, $u=t-x$ and $v=t+x$, the line element becomes
\begin{equation}
\label{uvmetric}
    ds^2=-du dv
\end{equation}
Lines of constants $u$ and $v$, respectively, describe right-moving and left-moving light rays. To compactify the length of coordinates, we define
\begin{equation}
    \label{Compactcoord}
    u=\tan{U}\;,\;v=\tan{V}
\end{equation}
in terms of $U$ and $V$, \ref{uvmetric} becomes
\begin{equation}
    \label{UVmetric}
    ds^2=-\sec^2{U} \sec^2{V}dU\; dV
\end{equation}
The coordinates $U$ and $V$ now has finite ranges
\begin{equation}
    \label{UVrange}
    -\pi/2<U\leq V < \pi/2
\end{equation}
We give a conformal transformation to the \ref{UVmetric} to cast it to the form of the Minkowski metric \ref{uvmetric}. However, the only difference is that \ref{UVmetric} has finite coordinate ranges. Since these two metrics are conformally related, they have the same causal structure. 
The past timelike infinity, denoted as $i^-$, is the point from which timelike trajectories originate, and the future timelike infinity, denoted as $i^+$, is the point at which they terminate. The past null infinity, denoted as $I^-$, is the point from which light rays originate, and the future null infinity, denoted as $I^+$, is the point at which they terminate. The spacelike infinity is denoted by $i^0$. A $(3+1)$-dimensional Minkowski spacetime can also be represented in a similar manner. 
\noindent 
\subsection{The de Sitter spacetime}
\label{The de Sitter spacetime}

The idea that the universe is static has persisted for a long time. Einstein gave a model for a static universe \cite{QFTCS2}. Later, in the $1930$'s, Hubble discovered that the galaxies were drifting apart, by analyzing the redshift of the light rays coming from the distant stars. The study of the universe's evolution at a very large scale, as well as its properties and dynamics, is known today as cosmology or cosmological modelling, for example, \cite{Weinberg, Sean, Atkatz:1981tk}. These models, namely the Friedmann-Lemaitre-Robertson-Walker (FLRW) model, assume as per the observation that the universe is spatially homogeneous and isotropic at very large scales ($\gtrsim 300$ million lightyears) and solve the Einstein field equations to describe the dynamics of the universe. To solve these equations, one needs a suitable form of the energy-momentum tensor of the universe, and owing to the symmetries mentioned above, it is taken to be a spatially homogeneous and isotropic perfect fluid. The general ansatz for the metric of an FLRW universe is given by \cite{Weinberg,F,L,R,W} 
\begin{equation}
    \label{FRW}
    ds^2=-dt^2+a^2(t)\Big[\frac{dr^2}{1-Kr^2}+r^2(d\theta^2+\sin{\theta}^2d\phi^2)\Big]
\end{equation}
where $a(t)$ is the scale factor, a function of time that reflects the dynamics of the universe. $K$ is a constant representing the curvature of the three-space and it can take only three values owing to the aforementioned spatial isosymmetries
\begin{center}
$K=1$, for a closed universe\\
    $K=0$, for a flat universe\\
    $K=-1$, for an open universe
\end{center}
so that \ref{FRW} takes the form
\begin{equation}
\label{RW}
    ds^2=-dt^2+a^2(t)\begin{cases}
    d\psi^2+\sin^2\psi (d\theta^2+\sin^2\theta d\phi^2)\\
    dx^2+dy^2+dz^2\\
    d\psi^2+\sinh^2\psi (d\theta^2+\sin^2\theta d\phi^2)
    \end{cases}
\end{equation}
respectively correspond to the spatial geometries : a three-sphere, a Euclidean three space and a three hyperboloid. 
The de Sitter spacetime is the solution of the Einstein equation with a positive cosmological constant and without any matter field,
$$G_{\mu\nu}+\Lambda g_{\mu \nu}=0$$.
The corresponding metric can be put into the FLRW form 
\begin{equation}
    \label{dS}
    ds^2=-dt^2+a^2(t) \Big[dx^2+dy^2+dz^2\Big]
\end{equation}
where $a(t)=e^{Ht}$, with $H=\sqrt{\Lambda/3}$ being the Hubble rate. 
The above metric represents an accelerated expansion of our universe, relevant to the early primordial inflationary phase as well as to the current accelerating phase of our universe \cite{darkenergy}. 

\ref{dS} can be brought to a conformally flat form as
\begin{equation}
\label{Confltads}
ds^2=a^2(\eta)(-d\eta^2+dx^2+dy^2+dz^2)
\end{equation}
where $a(\eta)=-1/H\eta$ and the range of $\eta$ is $-\infty < \eta < 0^-$. 
$\eta=-e^{-Ht}/H$ is called the conformal time. The de Sitter spacetime is maximally symmetric like the Minkowski spacetime \cite{WaldGR}. Due to its maximal symmetries, many computations can be done exactly in the de Sitter spacetime. Let us now review below the particle creation due to the gravitational field in it.


\noindent
  \subsubsection{Particle creation in the de Sitter spacetime}
\label{Particles creation in de Sitter spacetime}
Let us consider a quantum field theory in the cosmological de Sitter spacetime. The non-zero curvature of the spacetime or the time-dependent gravitational field leads to particle creation from vacuum fluctuations, for example, \cite{Mottola:1984ar, Greenwood:2010mr, Anderson:2013ila }.
This particle creation process in the expanding universe results in the primordial variations of the cosmic microwave background. These primordial oscillations also impact the structure of the cosmos on a large scale, for example, \cite{PhysRevD.70.043502, PhysRevD.72.063516, PhysRevD.93.023505}. Studying quantum field theory in the cosmological de sitter spacetime will give us some insight into the physics of the early inflationary scenarios.\\

\noindent
Let us consider a massive scalar field moving in the background of \ref{Confltads}. Let us also ignore the backreaction due to the scalar field. Defining $\phi= \chi/a(\eta)$, the Klein-Gordon equation \ref{KGC} becomes \cite{Parker:2009uva}
\begin{equation}
    \label{DSKG}
    \chi^{\prime \prime}-\Vec{\nabla}^2 \chi-\frac{1}{\eta^2}\Big(2-\frac{m^2}{H^2}\Big)\chi=0
\end{equation}
where the $\prime$ denotes derivative with respect to $\eta$ once and $\Vec{\nabla}^2$ represents the spacial Laplacian operator. Assuming the ansatz
\begin{equation}
    \label{DSA1}
    \chi(x)= \chi_k (\eta) e^{-\imath \Vec{k}.\Vec{x}}
\end{equation}
and substituting this ansatz into \ref{DSKG}, we have
\begin{equation}
    \label{DFEeta}
    \chi_k^{\prime \prime}(\eta)+k^2\Big[1-\frac{1}{k^2 \eta^2}\Big(2-\frac{m^2}{H^2}\Big)\Big]\chi_k (\eta)=0
\end{equation}
where $k=|\Vec{k}|$. \ref{DFEeta} is the differential equation for the Bessel function, and its general solution is taken in terms of the Hankel functions of the first and second kinds \cite{AS}
\begin{equation}
    \label{chi}
    \chi_k (\eta)=\sqrt{-k\eta} \Big[C_1 \mathcal{H}^{(1)}_{\nu}(-k\eta)+C_2\mathcal{H}^{(2)}_{\nu}(-k\eta)\Big]
\end{equation}
where $C_1$ and $C_2$ are constants and $\nu =\sqrt{9/4-m^2/H^2}$. We assume that the positive frequency `in' modes are defined as $\eta \to - \infty$, read as
\begin{equation}
    \label{inDS}
    \chi^{\text{in}}_k(x)=C_1 \sqrt{-k \eta} \mathcal{H}^{(1)}_{\nu}(-k\eta) e^{-\imath \Vec{k}. \Vec{x}}
\end{equation}
Likewise, the positive frequency `out' modes are defined as $\eta \to 0^-$, read as
\begin{equation}
    \label{outDS}
    \chi^{\text{out}}_k(x)=C_2 \sqrt{-k \eta} \mathcal{H}^{(2)}_{\nu}(-k\eta) e^{-\imath \Vec{k} .\Vec{x}}
\end{equation}
The full, normalised positive frequency modes are given by
\begin{eqnarray}
\label{fullmodes}
   \phi^{\text{in}}_{+,k}(x)=-\frac{\sqrt{\pi}}{2} H (-\eta)^{\frac{3}{2}}\mathcal{H}^{(1)}_{\nu}(-k\eta) e^{-\imath \Vec{k}. \Vec{x}} \\
  \phi^{\text{out}}_{+,k}(x)=-\frac{\sqrt{\pi}}{2} H (-\eta)^{\frac{3}{2}} \mathcal{H}^{(2)}_{\nu}(-k\eta) e^{-\imath \Vec{k}. \Vec{x}}  
\end{eqnarray}
Likewise, the normalised negative frequency `in' and `out' modes are
\begin{eqnarray}
\label{fullmodesL}
   \phi^{\text{in}}_{-,k}(x)=-\frac{\sqrt{\pi}}{2} H (-\eta)^{\frac{3}{2}}(\mathcal{H}^{(1)}_{\nu}(-k\eta))^* e^{\imath \Vec{k}. \Vec{x}} \\
  \phi^{\text{out}}_{-,k}(x)=-\frac{\sqrt{\pi}}{2} H (-\eta)^{\frac{3}{2}} (\mathcal{H}^{(2)}_{\nu}(-k\eta))^* e^{\imath \Vec{k}. \Vec{x}}  
\end{eqnarray}
These modes satisfy the Klein-Gordon orthonormality conditions. Using the asymptotic expansion of `in' modes at $\eta \to 0^-$ and assuming $m/H\gg 3/2$ (i.e. $\nu\approx \frac{im}{H}$), the Bogoliubov relations are given by
\begin{equation}
    \label{BGDS}
    \phi^{\text{in}}_{+,k}(x)=\alpha_k \phi^{\text{out}}_{+,k}(x)+\beta_k \phi^{\text{out}}_{-,k}(x)
\end{equation}
where $\alpha_k$ and $\beta_k$ are the Bogoliubov coefficients and their forms can be found using the relation between the Hankel functions of first and second kind \cite{AS},
\begin{equation}
    \label{BCDS}
   | \alpha_k|^2=\frac{e^{\frac{2\pi m}{H}}}{e^{\frac{2\pi m}{H}}-1}, \;\;\;\;|\beta_k|^2=\frac{1}{e^{\frac{2\pi m}{H}}-1}
\end{equation}
which satisfy $\absN{\alpha_{k}}^2 - \absN{\beta_{k}}^2=1$.
At the operator level, the Bogoliubov transformation reads
\begin{equation}
    \label{BGDS11}
     a^{\text{in}}_{\Vec{k}}=\alpha_k b^{\text{out}}_{\Vec{k}}+\beta_k b^{\text{out}\dagger}_{-\Vec{k}}
\end{equation}
where
$a^{\text{in}}_{\Vec{k}}
	\ket{0 }^{\text{in}}
=0=
	b^{\text{out}}_{\Vec{k}}
	\ket{0 }^{\text{out}}
	$.
Using \ref{BGDS11} we compute the number density of `out' particles with respect to the `in' vacuum, given by
  \begin{equation}
      \label{NDDS}
      ^{\text{in}}\langle 0|a^{\text{out}\dagger}_{\Vec{k}} a^{\text{out}}_{\Vec{k}}|0\rangle^{\text{in}} = \frac{1}{e^{\frac{2\pi m}{H}}-1}
  \end{equation}
  which is a bosonic blackbody Planck spectrum with temperature $H/2\pi$. For $H \to 0$, the number density provided by \ref{NDDS} vanishes, which replicates the usual flat space result. For a more detailed review, we refer our reader to, for example, \cite{ QFTCS, Parker:1969au, Parker:1971pt, Parker:2012at}.
\noindent
\subsection{The Rindler spacetime}
\label{The Rindler spacetime}
The Rindler spacetime is the metric seen by a uniformly accelerated observer in the Minkowski spacetime. This metric is also physically very relevant because, for any non-extremal black hole, the
near horizon geometry takes the Rindler spacetime form. Studying quantum field theory from the perspective of such non inertial observer plays a vital role in understanding some exciting phenomenon that can occur in the presence of horizons, such as the Hawking radiation in black holes and its thermodynamics, we refer our reader to \cite{Unruh:1976db} for the first discussion on it.

\begin{figure}[ht]
   \centering
     \includegraphics[scale=.55]{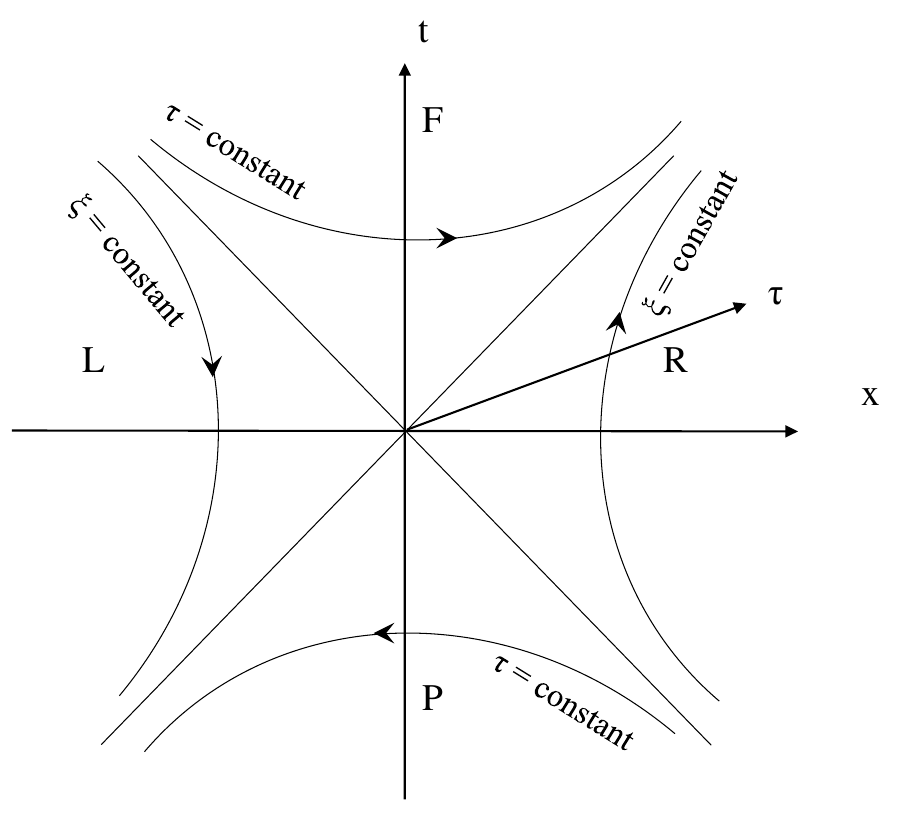}  
    \caption{\it{\small{The Rindler observer trajectory in the $(1+1)$-dimensional Minkowski spacetime.}}}
    \label{fighyperbola}
\end{figure}

The transformations between the Rindler $(\tau,\xi)$ and the Minkowski coordinates for the four regions right, left, future and past, denoted hereafter
by the labels $R$, $L$, $F$ and $P$, \ref{fighyperbola}, are given as
\begin{equation}
\label{coordtrans}
\begin{split}
&R \begin{cases}
  t= \frac{e^{a\xi_R}}{a}\sinh a \tau_R\\
  x= \frac{e^{a\xi_R}}{a}\cosh a \tau_R
\end{cases}\hspace{-5mm}\;
(x > 0, |t|< x)\;\quad L \;\begin{cases}
  t=- \frac{e^{a\xi_L}}{a}\sinh a \tau_L\\
  x=- \frac{e^{a\xi_L}}{a}\cosh a \tau_L 
\end{cases}(x < 0, |t|< x)\\
&F \begin{cases}
  t= \frac{e^{a\xi_F}}{a}\cosh a \tau_F\\
  x= \frac{e^{a\xi_F}}{a}\sinh a \tau_F
\end{cases}\hspace{-5mm}\;
(t > 0, |x|< t)\;\quad P \;\begin{cases}
  t=- \frac{e^{a\xi_{P}}}{a}\cosh a \tau_P\\
  x=- \frac{e^{a\xi_P}}{a}\sinh a \tau_P 
\end{cases}\hspace{-5mm}(t<0,|x|<|t|)   
\end{split}
\end{equation}
Each region's coordinates range from $-\infty \;\text{to}\; \infty$. The Rindler line element for region $R$ is
\begin{equation}
    \label{1+1rindler}
    ds^2=e^{2a\xi}(-d\tau^2+d\xi^2)+dy^2+dz^2
\end{equation}
The lines of constant $\xi$ describe a Rindler observer with proper acceleration $\alpha = ae^{-a \xi}$, where $a$ is a constant of dimension $\text{length}^{-1}$, known as the acceleration parameter. The Rindler observers are represented by a family of hyperbolae in the Minkowski spacetime given by
\begin{equation}
    \label{hyperbola}
    x^2-t^2=1/\alpha^2=e^{2a\xi}/a^2
\end{equation}
as shown in \ref{fighyperbola}. The region $R$ begins from $x\to\infty$ at $t\to-\infty$ (past null infinity for region $R$) and returns to $x\to\infty$ for $t\to\infty$ (future null infinity for region $R$). The acceleration of the Rindler observer varies from $\alpha \in (0,\infty)$. The lightcone surfaces $x=\pm t$ (i.e. $\xi \to -\infty$) in \ref{fighyperbola}, serve as the horizon for the Rindler observer. The regions $x>t$ and $-x>t$ in \ref{fighyperbola}, are the $F$ and $P$ regions, respectively, and the segments $x=t, \;t>0$ and $x=-t,\;t<0$ are called the future and past horizons of the $R$ region, respectively. Near the horizon, the acceleration $\alpha$ diverges and vanishes rapidly as we move away from it. Likewise, another branch of the hyperbola of \ref{hyperbola} lies in the $L$ region. It begins from $x \to -\infty$ at $t \to \infty$ (past null infinity for region $L$) and returns to $x \to -\infty$ at $t\to -\infty$ (future null infinity for region $L$). In terms of the Minkowski coordinates, trajectories in $R$ and $L$ are interchanged by the transformations: $x\to-x$. In the Rindler coordinates, $R$ and $L$ trajectories are interchanged by transforming $\xi \to -\xi$ and $\tau \to -\tau$.
 A Rindler observer confined in region $R$ will remain causally separated from all the other regions by the virtue of the horizon. Likewise, a Rindler observer located in region $L$ is causally disconnected from all the other regions. The region of interest for quantizing a field in the Rindler spacetime is the causally disconnected wedges, $R \cup L$. 
 In \ref{Schwinger effect and a uniformly accelerated observer}, the Penrose-Carter diagram for the Rindler spacetime will be presented for our purpose of studying the Schwinger-Unruh effect in it.
 
The Unruh effect is an interesting quantum field theory phenomenon predicted in the Rindler spacetime, and below, we wish to give a brief overview of it.

 \noindent
\subsubsection{The Unruh effect}
\label{The Unruh effect}
In $1973$ Fulling point out an ambiguity which arises in the quantum theory of fields when the background metric is not explicitly Minkowskian \cite{Fulling}. In $1975$ Davies discovered that a uniform gravitational field has effects equivalent to those of an accelerating observer \cite{Davies:1974th}. The Unruh effect was discovered in $1976$ \cite{Unruh:1976db} and describes how a uniformly accelerated detector moving in the Minkowski spacetime perceives the vacuum to be a thermally populated state at a temperature of $T=a/2\pi$, where $a$ is the acceleration parameter. For a detailed review and exhaustive list of references, we refer our reader to \cite{Higuchi:2017gcd, Crispino:2007eb, Global_1}. 
 
Let us consider a massless, Hermitian scalar field in the $(1+1)$-Minkowski spacetime \cite{Crispino:2007eb}. The Klein-Gordon equation and the canonical field quantization are given by
\begin{equation}
    \label{KGUE}
    (\partial^2/\partial t^2-\partial^2/\partial x^2)\phi(t,x)=0
\end{equation}
\begin{equation}
    \label{FE}
    \phi(t,x)=\int \frac{dk}{\sqrt{4\pi k}}(\hat{a}_k e^{i (k x-\omega_k t)}+\hat{a}^\dagger_k e^{i (-k x+\omega_k t)})
\end{equation}
where $\omega_k=|k|$ and $(\hat{a}_k$, $\hat{a}^\dagger_k)$ are the annihilation and creation operators, respectively. Likewise, the Klein-Gordon equation for $R$ region in the $(1+1)$-Rindler spacetime reads 
\begin{eqnarray}
    \label{KGUERS}
    (\partial^2/\partial \tau_R^2-\partial^2/\partial \xi_R^2)\phi(\tau_R,\xi_R)=0
\end{eqnarray}
and the canonical field quantization is
 \begin{equation}
     \label{FERS}
      \phi(\tau_R,\xi_R)=\int \frac{dk}{\sqrt{4\pi k}}\left(\hat{b}^R_k f^{(1)^*}_k+\hat{b}^{R^\dagger}_k f^{(1)}_k\right)
 \end{equation}
 where $f^{(1)}_k=e^{-i (\omega_k\tau_R-k\xi_R)}$ and $(\hat{b}^R_k$, $\hat{b}^{R^\dagger}_k)$ are the annihilation and creation operators corresponding to the Rindler observers in the $R$ region. Likewise, the field quantization for the region $L$ is
 \begin{equation}
     \label{FELS}
      \phi(\tau_L,\xi_L)=\int \frac{dk}{\sqrt{4\pi k}}(\hat{b}^L_k f^{(2)^*}_k+\hat{b}^{L^\dagger}_k f^{(2)}_k)
 \end{equation}
 where $f^{(2)}_k=e^{-i (\omega_k\tau_L-k\xi_L)}$. Since the Minowski and the Rindler temporal coordinates are different as noted in \ref{coordtrans}, one expects the corresponding vacua to be different as well. Thus one defines the Minkowski and the Rindler vacua, respectively as
 $$ \hat{a}_k|0_M\rangle =0$$
 $$\hat{b}^R_k|0_R\rangle =0=\hat{b}^L_k|0_L\rangle$$
We recall that the Minkowski modes are defined globally in $R\cup L$ whereas the Rindler modes are confined to their corresponding regions only. Thus, in order to compare the quantum field theory via these two different quantization schemes, one needs to construct the Rindler global modes, analytic in $R\cup L$.

In order to achieve this, we need to make linear combinations of the local modes on $R$ and $L$ wedges. However, before doing so, one must keep in mind that we cannot add them directly as they are not analytic at the origin with respect to the Minkowski coordinates, i.e., $t=x=0$, \ref{fighyperbola}. One uses the analytic continuation to overcome this ambiguity \cite{Unruh:1976db, Higuchi:2017gcd}. According to this, in order to by-pass the origin which serves as a branch point, we first note that 
$$a(-t+x)=e^{-a(\tau_R-\xi_R)},\quad a(-t+x)=-ae^{-a(\tau_L-\xi_L)}$$
Using the above relations and the local modes for $R$ and $L$ regions, we now make the following linear combination
\begin{equation}
    \label{comb}
    f^{(1)}_k+(-1)^{-i\omega_k/a}f^{(2)*}_{-k}=a^{i\omega_k/a}(-u)^{i\omega_k/a}
\end{equation}
where $u=t-x$ is the Minkowski retarded null coordinate. 
We note that the function $(-u)^{i\omega_k/a}$ is singular at the origin, and this ambiguity can be overcome by thinking of $u$ as a complex variable. 
Thus, \ref{comb} will represent a positive frequency Minkowski mode if and only if it is analytic and bounded in the lower half complex plane. Then on normalising the positive frequency global mode obtained from \ref{comb}, we have 
\begin{equation}
\label{gp}
    g^{(1)}_k=\frac{1}{\sqrt{2 \sinh{\frac{\pi \omega_k}{a}}}}\Big(e^{\frac{\pi \omega_k}{2a}} f^{(1)}_k+e^{-\frac{\pi \omega_k}{2a}} f^{(2)^*}_{-k}\Big)
\end{equation}
Similarly, one can analytically extend $f^{(2)}_k$ to the right region and the normalised positive frequency global mode obtained from it is given by
\begin{eqnarray}
    \label{RSGlobal}
    g^{(2)}_k=\frac{1}{\sqrt{2 \sinh{\frac{\pi \omega_k}{a}}}}\Big(e^{\frac{\pi \omega_k}{2a}} f^{(2)}_k+e^{-\frac{\pi \omega_k}{2a}} f^{(1)^*}_{-k}\Big)
\end{eqnarray}
The field quantization in terms of these global modes is
\begin{equation}
\begin{split}
\label{FEGM}
\phi(t,x)=\int \frac{dk}{\sqrt{2 \sinh{\frac{\pi \omega_k}{a}}}}\Big(\hat{d}^{(1)}_k g^{(1)}_k +\hat{d}^{(1)\dagger}_k g^{(1)^*}_k+  \hat{d}^{(2)}_k g^{(2)}_k +\hat{d}^{(2)\dagger}_k g^{(2)^*}_k\Big)
\end{split}
\end{equation}
The operators $\hat{d}_{k}$'s and $\hat{d}^{\dagger }_{k}$'s now annihilate the Minkowski vacuum
\begin{equation}
    \label{Gvacuum}
  \hat{d}^{(1)}_{k} |0_M\rangle=\hat{d}^{(2)}_{k} |0_M\rangle=0
\end{equation}
The field quantization of the Minkowski modes can be written in terms of the direct sum of \ref{FERS} and \ref{FELS}, e.g. \cite{Global_1}, and is given as
\begin{equation}
    \label{directsum}
    \phi(t,x)=\int \frac{dk}{\sqrt{4\pi k}}(\hat{b}^R_k f^{(1)^*}_k+\hat{b}^{R^\dagger}_k f^{(1)}_k+\hat{b}^L_k f^{(2)^*}_k+\hat{b}^{L^\dagger}_k f^{(2)}_k)
\end{equation}
Equating \ref{FEGM} and \ref{directsum} and taking the Klien-Gordon inner product with respect to $f^{(1)}_k$ on both sides, we obtain
\begin{equation}
    \label{BTG}
    \hat{b}^R_k=\frac{1}{\sqrt{2 \sinh{(\frac{\pi \omega_k}{a})}}}\Big(e^{\pi \omega_k/2a}\hat{d}^{(1)}_k+e^{-\pi \omega_k/2a}\hat{d}^{(1)\dagger}_{-k}\Big)
\end{equation}
representing the Bogoliubov transformation between the Minkowski and the Rindler observer's operators. The Bogoliubov relationships show us that a positive frequency mode with respect to the Minkowski observer will be a linear superposition of both positive frequency and negative frequency modes with respect to the Rindler observer. In particular, using this transformation, we can compute the number density of the Rindler particles with respect to the Minkowski vacuum, given by
\begin{equation}
    \label{Nop}
  \langle 0_M |  \hat{b}^{R\dagger}_k \hat{b}^R_k |0_M \rangle =\frac{1}{e^{2\pi \omega_k/a}-1}
\end{equation}
which is the well known bosonic blackbody Planck spectrum with temperature  $T=a/2\pi$, first shown by Unruh in \cite{Unruh:1976db}. This phenomenon is the celebrated Unruh effect. 
\section{The Schwinger effect}
\label{The Schwinger effect}

As we mentioned earlier, the Schwinger effect is the phenomenon of the creation of charged particle-antiparticle pair in the presence of a strong background or classical electric field \cite{Sauter, Schwinger}.
Let us briefly review the Schwinger effect by quantizing a complex scalar field in the $(1+1)$-dimensional Minkowski spacetime in the presence of a background electric field of constant strength, for example, \cite{Li:2016zyv, Ebadi:2014ufa}, 
\begin{equation}
    \label{C1KG}
    (\partial_\mu +i e A_\mu) (\partial^\mu +i e A^\mu) \phi(x) = m^2 \phi(x)
\end{equation}
Here the gauge field $A_\mu$ is classical and is not to be quantized, and any quantum fluctuation of it will be ignored. We make a gauge choice
\begin{equation}
    \label{C1GF}
    A_\mu =(Ez,0) 
\end{equation}
which gives a constant electric field of strength $E$ along $\hat{z}$. There will be two orthonormal sets of modes for particles and antiparticles at $z\to -\infty$ (the `in' modes) and $z\to \infty$ (the `out' modes). One could instead choose a time-dependent gauge as well, for example, $A_\mu = (0,-Et)$. In this case, the `in' and `out' modes lie at $t \to \pm \infty$. However, the final result, as desired, remains independent of the gauge choice.

The solution of \ref{C1KG} with gauge fixing is described in terms of the parabolic cylinder functions. The `incoming' particle and antiparticle modes are respectively given as
\begin{eqnarray}
\phi^{p\;\text{in}}_{\omega}=\frac{1}{N}D_{\frac{1}{2}\big(\frac{im^2}{eE}-1\big)}(\xi)e^{-i \omega t}e^{\frac{ieEtz}{2}}\;\;,\;\;
\phi^{a\;\text{in}}_{\omega}=\frac{e^{-\frac{3i \pi}{4}}}{N}D_{\frac{1}{2}\big(\frac{im^2}{eE}-1\big)}(-\xi)e^{i \omega t}e^{- \frac{ieEtz}{2}tz}
\end{eqnarray}
where $N$ is the normalisation constant, $\xi=e^{-\frac{3 i \pi}{4}}\sqrt{2eE}\Big(z+\frac{\omega}{eE}\Big)$ and `outgoing' modes are achieved by $\phi^{p\;\text{out}}_\omega(t,z)=(\phi^{p\;\text{in}}_\omega(-t,z))^*$ and $\phi^{a\;\text{out}}_\omega(t,z)=(\phi^{a\;\text{in}}_\omega (-t,z))^*$. Using the properties of the parabolic cylinder functions \cite{AS}, one finds out the Bogoliubov relations between these modes
\begin{equation}
  \label{modesa}
    \phi^{p\;\text{out}}_\omega=\alpha_\omega\; \phi^{\text{in}}_\omega-\beta_\omega^{*} \;\phi^{a\;\text{in}*}_\omega  
\end{equation}
\begin{equation}
    \label{modesa1}
     \phi^{a\;\text{out}*}_\omega=\alpha_\omega^* \;\phi^{a\;\text{in}*}_\omega-\beta_\omega\; \phi^{p\;\text{in}}_\omega
\end{equation}
where $\alpha_\omega$ and $\beta_\omega$ are the Bogoliubov coefficients given by 
$$\alpha_\omega=\frac{i\sqrt{2\pi}}{\Gamma(\frac{1}{2}+\frac{i m^2}{2eE})}e^{-\frac{\pi m^2}{4eE}}\;\;,\;\;\beta_\omega=i e^{-\frac{\pi m^2}{2eE}}$$
They satisfy the charge conservation $|\alpha_\omega|^2-|\beta_\omega|^2=1$. And the creation and annihilation operator level of \ref{modesa} and \ref{modesa1} are reflected as 
\begin{eqnarray}
    \label{C1BT}
    a^{\text{in}}_\omega=\alpha_\omega\; a^{\;\text{out}}_\omega-\beta_\omega^{*} \;b^{\text{out}\dagger}_\omega\\
    \label{C2BT}
 b^{\text{in}\dagger}_\omega=\alpha_\omega^* \;b^{\text{out}\dagger}_\omega-\beta_\omega\; a^{\text{out}}_\omega
\end{eqnarray}
where $ (a_\omega$, $a^\dagger_\omega )$ and $ (b_\omega$, $b^\dagger_\omega )$ are the annihilation and creation operators of particles and antiparticles, respectively. The `in' and `out' vacuum are defined as
\begin{eqnarray}
    a^{\text{in}}_{\omega}|0\rangle^{\text{in}}=0, \quad \quad  b^{\text{in}}_{\omega}|0\rangle^{\text{in}}=0,\quad \quad
     a^{\text{out}}_{\omega}|0\rangle^{\text{out}}=0 \quad \text{and}  \quad  b^{\text{out}}_{\omega}|0\rangle^{\text{out}}=0
\end{eqnarray}
On substituting \ref{C1BT} and \ref{C2BT} in the definition of `in' vacuum
one can write the `in' vacuum in terms of the `out' basis states as
\begin{equation}
    \label{C1SS}
    |0\rangle^{\text{in}} = \prod_\omega \frac{e^{\frac{\beta^*}{\alpha}a^{\text{out}\dagger}_\omega b^{\text{out}\dagger}_\omega}}{|\alpha_\omega|}  |0\rangle^{\text{out}}
\end{equation}
This explicitly represents that the `in' vacuum is filled with `out' particles. The number density of `out' particles in the `in' vacuum is 
\begin{equation}
    \label{C1ND}
    ^{\text{in}}\langle 0| a^{\text{out} \dagger}_\omega a^{\text{out}}_\omega|0\rangle^{\text{in}} = |\beta_\omega|^2 = e^{-\frac{\pi m^2}{eE}}
\end{equation}
The above formula illustrates that the mass of the created pairs suppresses the number density, whereas the electric field strength enhances it, as one expects intuitively.

The created particle-antiparticle Schwinger pairs turn out to be quantum entangled as well. Such entanglement features are also present in the Rinder, black hole and cosmological spacetimes \cite{Li:2016zyv, Wu:2020dlg, Ebadi:2014ufa, Fuentes:2010dt, Arias:2019pzy, EE, bell:2017, vaccum EE for fermions, QC in deSitter}. Here are a few instances of entanglement in pure states. Additionally, we are focusing on calculating correlations related to entanglement for certain mixed states obtained by tracing out some sectors from maximally entangled states created from two fields. Hence, we would like to briefly review below some basic features of quantum entanglement to be useful for our future purpose.


\noindent
\section{A quick review of quantum information theory}
\label{A short introduction to quantum information}
In this thesis, we are interested in studying the aspects of entanglement in quantum field theoretic systems. In this section, we have briefly discussed the quantum information theory. The entanglement, or correlations, are the fundamental characteristics of quantum mechanics. After experimental confirmation, this has been placed on firm physical grounds~\cite{Aspect1, Aspect2}. The three basic questions that come up naturally are `Whether a state is correlated (or entangled) or not?', `How much is this state correlated as compared to another state?' and `How can these correlations be quantified?'. The answers to all these questions can be given by various measures used to quantify such correlations, studied in a wide range of theoretical research, for example, \cite{ measure, Monogamy, Vidal:1998re, Horodecki:2009zz, Plenio:2007zz, Zyczkowski:1998yd, Martin, Vidal:2002zz, wang,  Plenio:2005, Calabrese:2012nk, Nishioka:2018khk, jmath} and references therein.
\noindent

Quantum information theory brings a new perspective to the quantum field theory. Perhaps the most interesting scenario to study quantum information properties in quantum field theory are when particle creation takes place. The concepts used in quantum mechanics for studying quantum information remain the same in quantum field theory.

In quantum mechanics, any state of a system is defined by a normalized vector in the Hilbert space. Let us consider two systems, $A$ and $B$, with corresponding Hilbert spaces $\mathcal{H}_A$ and $\mathcal{H}_B$, respectively. A vector $|\psi\rangle$ describes any state constructed by systems $A$ and $B$ in the product Hilbert space $\mathcal{H}_{AB}=\mathcal{H}_A \otimes \mathcal{H}_B$. For a two-dimensional Hilbert space with two orthogonal states $|0\rangle_{A/B}$ and $|1\rangle_{A/B}$, a particular state $|\psi \rangle$, can be written as 
\begin{equation}
    \label{psi}
    |\psi\rangle =\cos{\theta}|0\rangle_A \otimes |0\rangle_B + \sin{\theta} |1\rangle_A \otimes |1\rangle_B
\end{equation}
For any generic value of $\theta$, $|\psi\rangle$ represents an entangled state of systems $A$ and $B$, except for the values for which it can be written as the product state of systems $A$ and $B$. One can represent it by a density matrix that reads as
\begin{equation}
    \label{DM}
    \rho = |\psi \rangle \langle \psi |
\end{equation}
It represents a pure state density matrix. A classical mixture of different quantum states is known as a mixed state. Unlike a pure state, a mixed state cannot be represented by a normalized state vector in the Hilbert space. The density matrix for a mixed state is given as
\begin{equation}
    \label{DMmixed}
    \rho^{\text{mixed}}=\sum_j^N p_j |\psi_j\rangle \langle \psi_j| 
    \end{equation}
where each $|\psi_j\rangle$ represents a pure state, and the probability of finding the system in state $|\psi_j\rangle$ is $p_j$. $\rho^{\text{mixed}}$ can be entangled or separable and has classical and quantum correlations. A separable mixed state is a mixture of the product state of density matrices of subsystems, given as
$$\rho=\sum_i p_i \rho_{A,i}\otimes\rho_{B,i}$$
where $\rho_{A,i}$ and $\rho_{B,i}$ represents density matrix of subsystems $A$ and $B$ respectively.
One can also define the reduced density matrix of subsystem $A$ as $\rho_A = {\rm Tr}_B \rho_{AB}$, where the partial trace ${\rm Tr}_B$ is taken over the Hilbert space $\mathcal{H}_B$.

Let us now discuss some measures to quantify correlations pertaining to the entanglement for pure and mixed states. We refer our reader to \cite{NielsenChuang} for a detailed pedagogical discussion.
    \noindent
\subsection{Measures of correlations}
\label{Measures of correlations}
Following e.g.~\cite{Plenio:2007zz, NielsenChuang},
let us consider a bipartite system constituted by subsystems, $A$ and $B$, so that the Hilbert space can be decomposed  as $\mathcal{H}_{AB} = \mathcal{H}_{{A}} \otimes \mathcal{H}_{{B}}$.
Let $\rho_{AB}$ be the density matrix of states on $\mathcal{H}_{AB}$.
\noindent
\subsubsection{Entanglement entropy}
\label{Entanglement entropy}
\noindent
The entanglement entropy of $A$ is defined as the von Neumann entropy of $\rho_A$:
\begin{equation}
	S(\rho_A)
=	-
	{\rm Tr}_A
	\left(\rho_A
	\log \rho_A\right)
 \label{VNE}
\end{equation}
When $\rho_{AB}$ corresponds to a pure state, one has $S(\rho_A)=S(\rho_B)$, and it is zero when $\rho_{AB}$ is also separable.
The von Neumann entropies satisfy a subadditivity: $S(\rho_{AB}) \leq 	S(\rho_A) + S(\rho_B)$, where $S(\rho_{AB})$ is the von Neumann entropy of $\rho_{AB}$.
The equality holds if and only if $\rho_{AB}=\rho_A \otimes \rho_B  $ \cite{NielsenChuang}.

\noindent

\subsubsection{The mutual information}
\label{The mutual Information}
\noindent
The mutual information is a measure of  quantum as well as classical correlations between the subsystems $A$ and $B$. 
For the state $\rho_{AB}$, it is defined as
\begin{equation}
    I(A,B)
=
	S(\rho_{A})
	+
	S(\rho_{B})
	-
	S(\rho_{AB})
 \label{MIBP}
\end{equation}
The lower bound of the mutual information, $I(A,B) \geq 0$, is immediately obtained by the subadditivity of the entanglement entropy, where
the equality holds  only if $\rho_{AB} = \rho_A \otimes \rho_B$.

\noindent
\subsubsection{Entanglement negativity and logarithmic negativity}
\label{Entanglement negativity and logarithmic negativity}
\noindent
Even for mixed states, there is a measure of the entanglement of bipartite states~\cite{Zyczkowski:1998yd,Vidal:2002zz}, called the entanglement negativity, defined as
$
    \mathcal{N}(\rho_{AB})
=
	\frac{1}{2} \left(\lvert \lvert \rho_{AB}^{T_A} \rvert \rvert_1-1\right)
$,
where $\rho_{AB}^{\text{T}_A}$ is the partial transpose of $\rho_{AB}$ with respect to the subspace of $A$, i.e., $\left( \lvert i\rangle_{\! A} \hspace{-0.2ex} \langle n \rvert \otimes \lvert j \rangle_{\! B} \hspace{-0.2ex} \langle \ell \rvert \right)^{\text{T}_A}: = \lvert n \rangle_{\! A} \hspace{-0.2ex} \langle i \rvert \otimes \lvert j \rangle_{\! B} \hspace{-0.2ex} \langle \ell \rvert$.
Here, $\lvert \lvert \rho_{AB}^{T_A} \rvert\rvert_{1}$ is the trace norm, $\lvert \lvert\rho_{AB}^{T_A}\rvert \rvert_{1} = \sum_{i=1}^{\text{all}} \lvert \mu_i \rvert$, where $\mu_i$ is the $i$-th eigenvalue of $\rho_{AB}^{T_A}$. 
The logarithm of $\lvert \lvert \rho_{AB}^{T_A} \rvert \rvert_1$ is called the logarithmic negativity, which can be written as 
\begin{equation}
L_{N} (\rho_{AB})
=
	\log (1 + 2 \mathcal{N}(\rho_{AB}) )
    \label{L-N}
\end{equation}
These quantities are entanglement monotones which do not increase under local operations and classical communications.

These two quantities measure a violation of the positive partial transpose (PPT) in $\rho_{AB}$.
The PPT criterion can be stated as follows. If $\rho_{AB}$ is separable, the eigenvalues of $\rho_{AB}^{T_A}$ are non-negative.
Hence, if $\mathcal{N} \not = 0$ (i.e. $L_\mathcal{N} \not = 0$), $\rho_{AB}$ is an entangled state.
On the other hand, if $\mathcal{N} = 0$ (i.e. $L_\mathcal{N} = 0$), we cannot judge the existence of entanglement from this measure since there exist PPT and entangled states in general. Further discussions on it can be found in example ~\cite{Horodecki:2009zz}.
\noindent
\subsubsection{The Bell inequality violation}
\label{The Bell Inequality Violation}
Following \cite{NielsenChuang, bell:2017}, let us consider two pairs of non-commuting observables defined respectively over the Hilbert spaces ${\cal H}_{A}$ and  ${\cal H}_{B}$ : $({\mathit{O}_{1},\mathit{O}'_{1})\in{\cal H}_{A}} \; {\rm and}\; ({\mathit{O}_{2},\mathit{O}'_{2})\in {\cal H}_{B}} $. We assume that these are spin-$1/2$ operators along specific directions, such as $\mathit{O}=n_{i}\sigma_{i}$, $\mathit{O}'=n_{i}^{\prime}\sigma_{i}$, where $\sigma_{i}$'s are the Pauli matrices and $n_{i}$, $n_{i}^{\prime}$ are unit vectors on the $3-$dimensional  Euclidean space. The eigenvalues of each of these operators are $\pm1$. The Bell operator  ${\cal B}$, acting on $\in{\cal H}_{A}\otimes{\cal H}_{B}$ is defined as (suppressing the tensor product sign),
\begin{eqnarray}
\mathcal{B}=\mathit{O}_{1}\left(\mathit{O}_{2}+\mathit{O}'_{2}\right) +\mathit{O}'_{1}\left(\mathit{O}_{2}-\mathit{O}'_{2}\right)
\label{op2}
\end{eqnarray}
In theories with  classical local hidden variables, we have the so called Bell's inequality, $\langle\mathcal{B}^{2} \rangle \leq4$ and $|\langle\mathcal{B}\rangle|\leq2$. However, this inequality is violated in quantum mechanics as follows. We have from \ref{op2},
\begin{eqnarray}
\mathcal{B}^{2} = {\bf I} - [\mathit{O}_{1},\mathit{O}'_{1}][\mathit{O}_{2},\mathit{O}'_{2}],
\end{eqnarray}
where {\bf I} is the identity operator. Using the commutation relations for the Pauli matrices, one gets $|\langle \mathcal{B}\rangle| \leq\;2\sqrt{2}$, thereby obtaining a potential violation of Bell's inequality, where the equality is regarded as the maximum violation. The above construction can also be extended to multipartite systems with pure density matrices corresponding to squeezed states formed by mixing different modes. We refer our reader to~\cite{bell:2017} and references therein for details, including the construction of the Bell operator for scalars.

\noindent

Throughout the thesis, we will use these measures to quantify correlations or entanglement for the pure and mixed states. We choose these measures because of their computational simplicity and their suitability for calculating correlations pertaining to entanglement. There are other measures also, such as quantum discord, the entanglement of formations, distillable entanglement and many more, for which we refer our reader to \cite{ Vidal:2002zz, Plenio:2005, Calabrese:2012nk, Nishioka:2018khk}.

Having introduced and briefly reviewed the topics relevant to this thesis, we now wish to provide below a quick summary of the explicit issues we have considered in this thesis.
\section{Highlights of the thesis}
\label{Highlights of the thesis}
The main goal of this thesis is to investigate analytically and numerically the aspects of entanglement for quantum field theoretic systems in the presence of constant strength background electric and magnetic fields in various spacetime backgrounds. In particular, we wish to emphasize the role of the magnetic field in the presence of background electric and/or gravitational fields. Precisely, we are interested in computing various correlation measures between the particles and antiparticles for pure and mixed states.

\noindent
In \ref{Background magnetic field and quantum correlations in the Schwinger effect}, we consider the simplest scenario in which we take a complex scalar field coupled to background electric and magnetic fields of constant strength in the $(3+1)$-dimensional Minkowski spacetime. We investigate how the correlations between the created Schwinger pairs are impacted by the strength of the background fields. We examine these correlations pertaining to the entanglement for pure and some mixed states. Subsequently, we compute the vacuum entanglement entropy. Next, we consider some maximally entangled states for a two-scalar field system and compute the logarithmic
negativity and the mutual information corresponding to different particle-antiparticle
excitations. The qualitative differences between these results pertaining to the charge content
of the states are highlighted. 

\noindent
In \ref{Fermionic Bell violation in de Sitter spacetime with background electric and mangnetic fields}, we consider a more complex scenario where we take a fermionic field in the presence of background electric and magnetic fields of constant strength in the time-dependent cosmological de Sitter spacetime \ref{Confltads}. In this case, there are two sources of pair creation, i.e. the background electric field and the time-dependent gravitational field. We wish to investigate how the magnetic field affects the correlations pertaining to entanglement between the particles and antiparticles created for pure and mixed states subject to the Bunch-Davies vacuum. First, we compute the entanglement entropy and Bell's inequality violation for the Bunch-Davies vacuum. Next, we consider some maximally entangled initial states of two fermionic fields and compute Bell's inequality violation and mutual information for the final states corresponding to different particle-antiparticle excitations. Qualitative differences regarding the correlations for different maximally entangled initial states are discussed. Further extension of these results to the so-called de Sitter $\alpha$-vacua is also discussed. Perhaps this computation might give us some insight into the entanglement characteristics of the early universe scenarios where primordial electromagnetic fields could be present.

\noindent
In \ref{Schwinger effect and a uniformly accelerated observer}, we wish to study the Schwinger pair creation in the presence of a background magnetic field from the perspective of a uniformly accelerated or Rindler observer in the Minkowski spacetime. We consider a fermionic field in the presence of constant strength background electric and magnetic fields and solve for the orthonormal Dirac modes in the Rindler left and right regions \ref{1+1rindler}. There are two
sources of particle creation in this scenario : the Schwinger and the Unruh effects. Our chief aim is to investigate the role of the strength of the background electric and magnetic fields on the spectra of created particles and their entanglement properties. We note that in addition to the entanglement between the Schwinger pairs, there will be quantum correlations between the causally disconnected Rindler wedges as well. We compute the particle-antiparticle number densities with respect to the Minkowski vacuum for the Rindler number operator. We next compute the entanglement measures such as Bell inequality violation, logarithmic negativity and mutual information for various states. Since the near horizon geometry of any non-extremal black hole can be approximated as the Rindler spacetime, perhaps this analysis will provide us with a very crude insight into a charged quantum field's dynamics near an eternal non-extremal charged black hole. We also note that astrophysical black holes can be endowed with background electromagnetic fields due to the accreting plasmas onto them. However, such black holes do not have any past or white hole horizon, so they do not have any left-right Rindler wedge structure. Hence, particle creation in only one Rindler wedge will be relevant there. Moreover, in such practical scenarios, the electromagnetic field configuration might be quite complex, and there seems to be little hope of tackling such scenarios analytically.

Finally, the conclusive Chapter (\ref{Summary and Outlook}) contains a summary of the results discussed in the main chapters of the thesis. We also mention some future directions that might help us in gaining a deeper understanding of the correlations between particles and antiparticles created in the presence of background fields.

\chapter{Background electric and magnetic fields and quantum correlations in the Minkowski spacetime}
\label{Background magnetic field and quantum correlations in the Schwinger effect}

In this Chapter, we consider the simplest scenario. We first take a complex scalar field coupled to background electric and magnetic fields of constant strength in the $(1+3)$-dimensional Minkowski spacetime. We wish to investigate some measures of quantum correlations (namely, the vacuum entanglement entropy, the logarithmic negativity, and the mutual information for entangled states) in the context of the Schwinger pair creation mechanism~\cite{Parker:2009uva, Schwinger}. Various aspects of entanglement properties between created particle-antiparticle pairs in the Schwinger mechanism in Abelian or non-Abelian gauge theories can be seen in~\cite{Ebadi:2014ufa, Li:2016zyv, Agarwal:2016cir, Li:2018twv, Xia:2019ztf, Gavrilov:2019vyi, Karabali:2019oxq, Karabali:2019ucc, Dai:2019nzv}. We also refer our readers to, for example, ~\cite{Seki:2014cgq, Balasubramanian:2011wt, Momentum space entanglement} for interesting aspects of entanglement in the flat space quantum field theory and to~\cite{Ryu:2006bv, Ryu:2006ef} for holographic aspects of entanglement. 
However, the role of a background magnetic field in this scenario has not been addressed yet in the literature. 

It is well known that in quantum electrodynamics, a magnetic field itself cannot give rise to pair creation, for example, ~\cite{Karabali:2019oxq, Karabali:2019ucc}. However, we may expect, in general, that it would affect the pair creation rate if a background electric field is present. Thus, it seems interesting to ask : what will be the effect of a background magnetic field on the quantum correlations between the particle-antiparticle pairs? We may intuitively expect, {\it a priori} that the magnetic field will oppose the effect of the electric field. However, how do these correlations explicitly depend upon the magnetic field strength, for example, are they monotonic?
How do these behaviours differ subject to the charge content of the state we choose?
We wish to address these questions in this Chapter. We shall also discuss briefly a similar analysis for fermions and will point out some qualitative differences with that of the scalar.

The rest of the Chapter is organised as follows. We obtain the solution of the complex scalar's mode functions with the background electric and magnetic fields in \ref{sec:ComplexScalar}.
We compute the vacuum entanglement entropy for a single scalar field in \ref{sec:vacuum}, and the logarithmic negativity and the mutual information for maximally entangled states of a two scalar field system in \ref{sec:bipartite}. 
Finally, we summarise and discuss our results and related issues in \ref{sec:SD1}. In particular, we speculate that the well-known degradation of the quantum entanglement in an accelerated frame~for example,~\cite{MartinMartinez:2010ar}, can perhaps be restored for a charged field upon application of a `strong enough' magnetic field.

\noindent
\section{Complex scalar in background electric and magnetic field}
\label{sec:ComplexScalar}
\noindent
Let us now focus on a complex scalar field theory coupled to external, or background, electric and magnetic fields in the four-dimensional Minkowski spacetime.
Our analysis in this section is in parallel with~\cite{Ebadi:2014ufa, Gabriel:1999yz, Bavarsad:2017oyv}.
The Klein-Gordon equation reads  
\eq{
	\roundLR{
	D_\mu D^\mu - m^2 
	}
	\phi(t,\vec x)
&=
	0
,
\label{eq:KGeq}
}
where  $D_\mu = \partial_\mu - i e A_\mu$ is the gauge covariant derivative and $e$ stands for the electric charge of the field.
We consider the external gauge field as $A_\mu = (Ez,-By,0,0)$, which leads to parallel electric and magnetic fields of constant strengths, $E$ and $B$. One could instead choose a time-dependent gauge as well, for example, $A_\mu = (0,-By,0,-Et)$ which leads to perpendicular electric and magnetic fields. Throughout this thesis, we shall assume that $E$, $B$ and $e$ are positive.
On substituting the gauge field $A_\mu$ in \ref{eq:KGeq}, we have
\begin{equation}
\squareLR{
    -(\partial_t-ieEz)^2+(\partial_x+i e B y)^2+\partial_y^2+\partial_z^2-m^2}\phi(t,\Vec{x})=0
\end{equation}
We quantise the field as,
\eq{
	\phi(x)
&=
	\sum_{n_L}
	\int
	\frac{dk^0 dk^1}{\sqrt{4\pi k^0}}
	\squareB{
	a_{k,n_L}
	\phi^{(+)}_{k,n_L} 
    +
	b_{k,n_L}^\dagger
	\left(
	\phi^{(-)}_{k,n_L} 
	\right)^*
	}
\label{eq:phi}
}
where $k^0$ is restricted to be positive, and $a_{k,n_L}$ ($b_{k,n_L}^\dagger$) corresponds to the annihilation (creation) operator for the particle (antiparticle).
$n_L=0,1,2, \dots$ stands for the Landau level.
To shorten the notation, we shall suppress the label $n_L$. 
The mode functions $\phi^{(\pm)}_{k} $ are given by
\begin{equation}
	\phi^{(+)}_{k} 
=
    e^{-i(k^0t-k^1 x)}
	\phi^{-}_{k} (y,z)
, \qquad \qquad 
	\phi^{(-)}_{k} = e^{-i(k^0t-k^1 x)}
	\phi^{-}_{k} (y,z),
\label{phi+-}
\end{equation}
where the superscripts $+ (-)$ stand for particle (antiparticle). Substituting \ref{phi+-} into \ref{eq:KGeq} we have,
\begin{equation}
	\squareLR{
	\roundLR{
	k^0 \pm eEz
	}^2
	-
	\roundLR{
	k^1 \pm eBy
	}^2
	+
	\partial_y^2
	+
	\partial_z^2
	-
	m^2
	}
	\phi^{\pm}_{k} (y,z)
=
	0
\label{eq:EOM}
\end{equation}
Using now the decomposition $\phi^{\pm}_{k} (y,z)=h^{\pm}(y)g^{\pm}(z)$, we obtain
\begin{equation}
    \label{eqy}
     \left[\partial_y^2-(eBy\pm k^1)^2+s\right]h^{\pm}(y)=0
\end{equation}
\begin{equation}
  \left[\partial_z^2+(eEz\pm k^0)^2-m^2-s\right]g^{\pm}(z)=0
  \label{eqz}
\end{equation}
where $s$ is the separation constant. These equations clearly show that momentum $k^0$ and $k^1$ only serve to shift the origin in the $y$ and $z$ directions and have no effect on the energy of the particle or antiparticle mode. 
Let us first focus on the $y$ differential equation. We define the variable
 $${y}_\pm
=
	\sqrt{eB}
	\roundN{y \pm {k^1}/{eB}},$$
and with respect to this variable, it is easy to see that \ref{eqy} reduces to the Hermite differential equation, with the separation constant 
$s=2(n_L+1)eB$, where $n_L$ denotes the Landau levels. Thus the normalised solution of \ref{eqy} is,
\begin{equation}
  h^{\pm}(y)=\left(\frac{\sqrt{eB}}{2^{n_{L}+1}\sqrt{\pi}(n_L+1)!}\right)^{1/2}e^{-y_\pm^2/2}H_{n_L}(y_\pm)
\end{equation}
where $H_{n_L}(y_\pm)$ are the Hermite polynomials of order $n_L$. Now, let us focus on the differential equation for $z$, \ref{eqz}, for which we define the variable
$$
	\zeta_\pm
=
	e^{i \pi/4}
	\sqrt{2eE}
	\roundN{
	z \pm {k^0}/{eE}
	}
$$
The solution of the corresponding equations are given by the parabolic cylinder functions and their complex conjugations, $D_\nu(\zeta_\pm)$ and $(D_\nu(\zeta_\pm))^*$, where 
$
	\nu
=
	-(1+i\mu)/2
$,
and
\begin{eqnarray}
	\mu
=
	\frac{ m^2+eB(2n_L+1) }{eE}
\label{mu}
\end{eqnarray}
We consider a particle incoming in the $z$-direction at $\absN{z}\to\infty$. The independent solutions of \ref{eq:EOM} with this boundary condition are derived as
\begin{eqnarray}
	\phi^{(+) \text{in} }_{k} (y,z)
=
	N^{-1}
	e^{-{y}_+^2/2}
	H_{n_L}({y}_+)
	D_\nu (\zeta_+)\\
\qquad 
	\phi^{(-) \text{in}}_{k} (y,z)	
=N^{-1}
	e^{-{y}_-^2/2}
	H_{n_L}({y}_-)D_\nu (\zeta_-)
\end{eqnarray}
where $N$ is the normalisation constant. The above are the incoming particle and antiparticle modes. From now on, we consider that the variation of $\mu$ solely depends upon the magnetic field strength $B$, by keeping all the other parameters at some fixed values. 
In particular, a vanishing $\mu$ below will approximately imply a vanishing $B$, along with the restriction $m^2/e E \ll 1$, which can be achieved by applying a `sufficiently strong' electric field.

The incoming modes, $\phi^{(\pm) \text{in}}_{k} (x)$, satisfy the orthonormality conditions, 
defined via the Klein-Gordon inner product, $\angleN{\phi_1, \phi_2} = i \int d^3\vec x \roundN{\phi_1^* \partial_t \phi_2 - \phi_2 \partial_t \phi_1^*}$. Using the properties of the parabolic cylinder functions~\cite{AS}, we can check that
\begin{equation}
    \angleLR{
    \phi^{(+) \text{in}}_{k} (x)
    ,
    \phi^{(+) \text{in}}_{k'} (x)
    }
=
    \delta(k^0 - k^0{}')
    \delta(k^1 - k^1{}')
    \delta_{n_L n_L'}
\label{eq:Orthonormal}
\end{equation}
\begin{equation}
    \angleLR{
    \left(
    \phi^{(-) \text{in}}_{k} (x)
    \right)^*
    ,
   \left(
    \phi^{(-) \text{in}}_{k'} (x)
    \right)^*
    }
=
    -\delta(k^0 - k^0{}')
    \delta(k^1 - k^1{}')
    \delta_{n_L n_L'}
\end{equation}
\begin{equation}
    \angleLR{
    \phi^{(+) \text{in}}_{k} (x)
    ,
    \left(
    \phi^{(-) \text{in}}_{k'} (x)
    \right)^*
    }
=
    0
    \label{eq:Orthonormal1}
\end{equation}
Similarly, we find the outgoing modes for particles 
$
\phi^{(+) \text{out}}_{k} \sim
	e^{-{y}_+^2/2}
	H_{n_L}({y}_+)
\squareN{
	D_\nu(-\zeta_+)
	}^*$ and for antiparticles
$\phi^{(-)\text{out}}_{k} 	\sim
	e^{-{y}_-^2/2}
	H_{n_L}({y}_-)
	[D_\nu(-\zeta_-)]^*
$.
These modes also satisfy the orthonormality conditions in the same way as \ref{eq:Orthonormal} and \ref{eq:Orthonormal1}.
The incoming and the outgoing modes furnish two independent quantisations of the complex scalar, and they are related via the Bogoliubov transformation,
\eq{
	\phi^{(+) \text{in} }_{k} 
&=
	\alpha_k
	\phi^{(+) \text{out} }_{k} 
	+
	\beta_k
	\left(
	\phi^{(-) \text{out} }_{-k} 
	\right)^*
\label{eq:Bogoliuboftrf0}
}
where $\alpha_k$ and $\beta_k$ are the Bogoliubov coefficients. \ref{eq:Bogoliuboftrf0} yields,
\eq{
	D_\nu(\zeta_+)
&=
	\alpha_k
	\squareLR{
	D_\nu(-\zeta_+)
	}^*
	+
	\beta_k
	D_\nu(-\zeta_+)
}
where $
	\squareLR{
	D_\nu(-\zeta)
	}^*
=
	D_{-\nu-1}(i\zeta)
$.
Using the relation \cite{AS}, 
\eq{
	D_\nu(\zeta)
&=
	e^{-i\pi \nu}
	D_\nu(-\zeta)
	+
	\frac{\sqrt{2\pi}}{\Gamma(-\nu)}
	e^{-\frac{i\pi (\nu+1)}{2}}
	D_{-\nu-1}(i\zeta)
}
we obtain
\eq{
	\alpha_k
&=
	\frac{\sqrt{2\pi}}{\Gamma(-\nu)}
	e^{-\frac{i\pi (\nu+1)}{2}}=e^{-\frac{i\pi}{4}}\sqrt{1+e^{-\pi \mu}} 
,
\qquad
	\beta_k
=
	e^{-i\pi \nu}=e^{\frac{i\pi}{2}-\frac{\pi \mu}{2}}
\label{eq:defofalphabeta}
}
where $\nu$ is a function of $\mu$ defined in \ref{mu} and we have $\absN{\alpha_{k}}^2 - \absN{\beta_{k}}^2=1$, as a consistency. We note that the Bogoliubov coefficients are not only functions of the electric field strength but also magnetic field strength and the Landau levels. If we set $E\to0$ (i.e., $\mu \to \infty$) above, we have $|\alpha_k| \to 1$ and $|\beta_k|\to0$, irrespective of the value of $B$.
Thus in the presence of a background magnetic field alone, there is no pair creation or vacuum instability, as was shown in \cite{Parker:2009uva} by the effective action method. On the other hand if we set $E\to \infty$ (i.e., $\mu \to 0$) above, we have $|\alpha_k| \to \sqrt{2}$ and $|\beta_k|\to 1$, irrespective of the value of $B$. Thus at the large strength of the electric field, the number density of particles reaches its maximum. However, it is clear that in the presence of an $E$, the magnetic field would affect the pair creation rate, as was speculated before \ref{Quantum field theory in curved spacetime} of the preceding Chapter. We shall explicitly investigate pair creation in this scenario in the following section.

Substituting now the Bogoliubov relations \ref{eq:defofalphabeta} into the field quantization, we have the Bogoliubov relationship at the operator level
\eq{
	a^{\text{in}}_k
=
	\alpha_k
	a^{\text{out}}_k
	-
	\beta_{k}
	b^{\text{out} \dagger}_{-k}
 \\
\qquad \qquad 
	b^{\text{in}}_{-k}
=
	-
	\beta_{k}
	a^{\text{out} }_{k}{}^\dagger
	+
	\alpha_{k}
	b^{\text{out}}_{-k}
\label{eq:Bogoliuboftrf}
} 
Being equipped with this, we are now ready to investigate the correlation properties.
\noindent
\section{Entanglement entropy for the vacuum }
\label{sec:vacuum}
\noindent
We consider first the vacuum state of the incoming modes.
The state space $\mathcal{H}$ is constructed by the tensor product, $\mathcal{H} = \prod_k \mathcal{H}_k \otimes \mathcal{H}_{-k}$, where $\mathcal{H}_k$ and $\mathcal{H}_{-k}$ are the state spaces of the modes of the particle and the antiparticle, respectively.
The full `in' vacuum state $\ket{0}_{\text{in}}$ is decomposed as 
\eq{
	\ket{0}_{\text{in}}
&=
	\prod_{k,-k}
	\ket{0_k }_{\text{in}}
	\otimes
	\ket{0_{-k} }_{\text{in}}
\equiv
	\prod_{k,-k}
	\ket{0_k 0_{-k}}_{\text{in}}
}
where
\eq{
	a^{\text{in}}_k
	\ket{0_k }_{\text{in}}
&=
	b^{\text{in}}_{-k}
	\ket{0_{-k} }_{\text{in}}
=
	0
\label{eq:defofvacuum}
}
and likewise for the `out' states.
The  state $\ket{0_k 0_{-k}}_{\text{in}}$ can be expanded in terms of the `out' states as 
\eq{
	\ket{0_k 0_{-k}}_{\text{in}}
&=
	\sum_{n=0}^\infty
	C_{n_k}^0
	\ket{n_k n_{-k}}_{\text{out}}	
\label{eq:relabetinandout}
}
by using the Schmidt decomposition \cite{NielsenChuang}.
The normalisation, ${}_{\text{in}} \angleN{0_k 0_{-k}|0_k 0_{-k}}_{\text{in}}=1$, yields 
$\sum_{n=0}^\infty \absN{C_{n_k}^0}^2=1.$
%

The properties of $C_{n_k}^0$ and the Bogoliubov transformation \ref{eq:Bogoliuboftrf} yield the recurrence relation $ C_{n_k}^0 = \roundN{\beta_k / \alpha_k} C_{(n-1)_k}^0$, giving,
$
	C_{n_k}^0
=
	\roundN{
	{\beta_k}/{\alpha_k}
	}^n
	C_{0_k}^0
$
as discussed in~\cite{Ebadi:2014ufa} in the context of a pure electric field.
Using now this relation and \ref{eq:defofalphabeta},
we obtain
\eq{
	\absLR{C_{0_k}^0}
&=
	\frac{
	1
	}{
	\absLR{\alpha_k}
	}
=
	\frac{
	1
	}{
	\sqrt{2\pi}
	}
	\absLR{
	\Gamma(-\nu)
	e^{i \pi(1+\nu)/2}
	}
\label{eq:absC00}
}
We then derive
$$
    C_{n_k}^0
=
	\sqrt{1-\absLR{\gamma}^2}
	\gamma^n
	e^{i\theta_\nu^0}
$$
  where $\gamma
=
	\squareN{1+e^{\pi \mu}}^{-1/2}
	\exp \squareLR{i \roundLR{ {3\pi}/{4} + \rm{Arg}{\Gamma(-\nu)}}}
$ and
$
    \theta_\nu^0
=
	{\pi}/{4}
	+
	\rm{Arg}{\Gamma(-\nu)}
	+
	\phi_c
$,
where $\phi_c$ is a constant. Recalling that we have taken $E$, $B$ and $e$ to be positive (cf. \ref{sec:ComplexScalar}), we have $\mu > 0$.
Note that $\absN{\gamma} < 1/\sqrt{2}$ when $\mu > 0$, and hence $C_{n_k}^0$ decreases monotonically as the label $n$ increases.

Let us comment on other features of $C_{n_k}^0$.
Since, $C_{n_k}^0$ depends on only the variable $\mu$, \ref{mu}, 
$C_{n_k}^0$ reflects the charge and the mass but not the momentum $k^0$ and $k$ as the feature of the (anti)particle.
Second, when $\mu \to \infty$, $ \absN{ C_{n_k}^0 } \to \delta_{n0}$ since $\absN{\gamma}$ approaches $0$, in this case, and hence \ref{eq:relabetinandout} becomes $
	\ket{0_k 0_{-k}}_{\text{in}}
\to
	C_{0_k}^0
	\ket{0_k 0_{-k}}_{\text{out}}	
$, where the difference between the left- and right-hand side is just a phase factor.
Third, when $\mu \to 0$, $\absN{\gamma}$ approaches $1/\sqrt{2}$, and hence $\absN{ C_{n_k}^0 } \to 2^{-(n+1)/2}$.

The density matrix for the `in' vacuum state $\ket{0_k 0_{-k}}_{\text{in}}$
 is given by 
$
	\rho^{(\text{v})}
=
	\ket{0_k 0_{-k}}_{\text{in}}
	\hspace{-0.7ex}
	\bra{0_k 0_{-k}}
$, which is a pure state. To compute entanglement entropy we follow the definition of entanglement entropy given in \ref{VNE} of \ref{Entanglement entropy}. Employing \ref{eq:relabetinandout}, we first obtain the reduced density matrix for the particle as
$
	\rho_k
=
	\Tr_{-k}
	\rho^{(\text{v})}
=
	\sum_{n=0}^\infty
	\absN{C_{n_k}^{0}}^2
	\ket{n_k}_{\text{out}}
	\hspace{-0.6ex}
	\bra{n_k}
$. Since $\rho^{(\text{v})}$ is a pure state the reduced density matrix for antiparticles is the same as particles.
Hence the entanglement entropy, defined in \ref{VNE},
is given by 
\eq{
	S_k
&=
	-
	\Tr_k
	\rho_k
	\log \rho_k
=
	-
	\absLR{\beta_{k}}^2
	\log \absLR{\beta_{k}}^2
	+
	\roundLR{
	1
	+
	\absLR{\beta_{k}}^2
	}
	\log
	\roundLR{
	1
	+
	\absLR{\beta_{k}}^2
	}
\label{eq:EEforrhovk}
}
where $
    {}_{\text{in}}\braN{0_k} 	
    a^{\text{out}}_k{}^\dagger
    a^{\text{out}}_k
	\ketN{0_k }_{\text{in}}
=
    {}_{\text{in}}\braN{0_{-k}} 	
    b^{\text{out}}_{-k}{}^\dagger
    b^{\text{out}}_{-k}
	\ketN{0_{-k} }_{\text{in}}
= 
    \absN{\beta_{k}}^2 
=
    \exp\squareN{-\pi \mu}
$, is the density of created particles.
We are dealing with a pure state, and hence the entanglement entropies for the particle and antiparticle sectors satisfy 
$S_k=S_{-k}$.
\begin{figure}[ht]
	\centering
  \includegraphics[scale=.63]{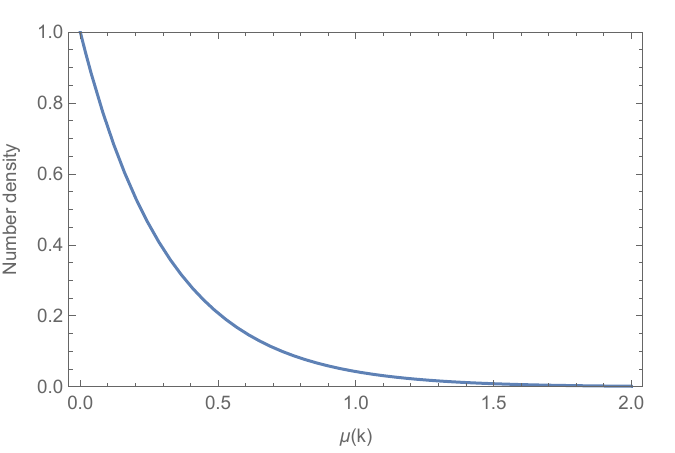}\hspace{1.0cm}
		\includegraphics[scale=.50]{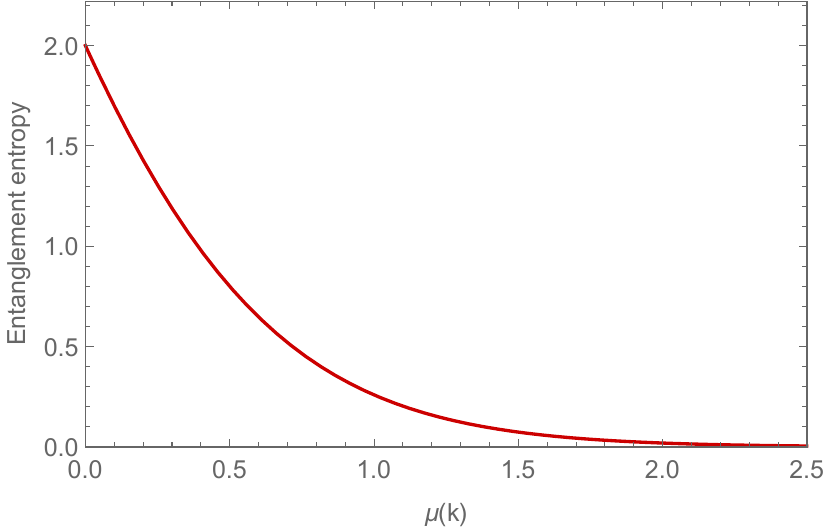}
		\caption{\it{\small The density of created particles (left) and 
		the entanglement entropy $S_k$ (right) of the vacuum state $\ket{0_k 0_{-k}}$ vs. the parameter $\mu$ given by \ref{mu}.
		Considering $m^2$ and $E$ to be fixed, $S_k$ is maximum in the small $\mu$ $(\absN{B}, n_L \ll \absN{E}$, along with our initial condition, $m^2/|qE|\ll1)$, where the number of created particles is maximum, and vanishing for large $\mu$ $(E \ll n_L, B)$, where the number density of the created particles is zero. Both of these plots show the re-stabilisation of the vacuum state with the increasing magnetic field strength. See main text for discussions.}
		}
		\label{fig:EEforVacuum1}
\end{figure}
We obtain the $\mu$-dependence of $S_k$ and the density of created particles as shown in \ref{fig:EEforVacuum1}.
Thus $S_k$ decreases as $\mu$ increases, and it is maximum, $S_k=2$, in the limit $\mu \to 0$, where $\absN{\beta_k}^2$ becomes unity. This means that at $\mu \to 0$ $\left(m^{2} \ll E\right.$ and $\left.|B| \ll E\right)$, the state $\rho^{(\mathrm{v})}$ is the maximally entangled state of the outgoing particles and antiparticles. On the other hand, $S_k \to 0$ in the limit $\mu \to \infty$, where the reduced density matrix $\rho_k$ returns to the incoming pure state and $\absN{\beta_k}^2$ vanishes.
This corresponds to the suppression of pair creation due to the stabilisation of the vacuum with increasing $B$.
\noindent
\section{Mutual information and logarithmic negativity in systems of two scalar fields}
\label{sec:bipartite}
\noindent
Let us now consider systems which are constructed by two complex scalar fields (not interacting with each other).
There are two species of (anti)particles, which do not interact with each other.
The total state space $\mathcal{H}$ is given by, $\mathcal{H} = \prod_{s,k} \mathcal{H}_s \otimes \mathcal{H}_{-s} \otimes \mathcal{H}_k \otimes \mathcal{H}_{-k}$, where $ s$ and $ k$ stand for the two species of scalar fields. We assume that these two scalar fields have the same charge to preserve the local gauge transformation, but different masses are allowed.

We shall focus on the maximally entangled states for the incoming states of the (anti)particles. Now, the gauge transformation properties of the wave function of a charged field in quantum electrodynamics put a constraint on how one can prepare those states, as follows. The wave functions corresponding to two states with different charge content will have different transformation properties under the local gauge transformation. Hence if we add two or more states to construct an entangled state, we must ensure that the charge content of each of these states are the  {\it same}, so that the wave function for the full state has a definite transformation property. This will be reflected in the states \ref{charge1} and \ref{eq:PAstate}
we work with. 

\noindent
\subsection{Single-charge state}
\label{sec:1particle}
\noindent
Based upon the above argument, we consider a maximally entangled single-charge state,
$
	\rho^{(1)}
=
	\ketN{\psi_{sk}^{(1)}}
	\braN{\psi_{sk}^{(1)}}
$, which is a pure state, with
\eq{
	\ket{\psi_{sk}^{(1)}}
&=
	\frac{\ket{0_s 0_{-s}; 1_k 0_{-k}}_{\text{in}}+\ket{1_s 0_{-s}; 0_k 0_{-k}}_{\text{in}}}{\sqrt{2}}
\label{charge1}
}
In our notation, the first (second) pair of entries appearing in the kets stands for the first (second) scalar. For a specific pair, the first (second) entry represents particle (antiparticle).  

Using the expansion of the incoming vacuum, \ref{eq:relabetinandout}, we rewrite $
	\ket{ 1_k 0_{-k}}_{\text{in}}
=
	{ a^{\text{in}}_k }^\dagger
	\ket{ 0_k 0_{-k}}_{\text{in}}
$
by the outgoing states as
\eq{
	\ket{ 1_k 0_{-k}}_{\text{in}}
&=
	\sum_{n=0}^\infty
	C_{n_k}^1
	\ket{(n+1)_k n_{-k}}_{\text{out}}	
\label{eq:10byout}
}
where the coefficient $C_{n_k}^1$ is given by
\eq{
	C_{n_k}^1
=
	\frac{\sqrt{n+1}}{\alpha_k}
	C_{n_k}^0
=
	\roundLR{1-\absLR{\gamma_k}^2}
	\gamma_k^n
	e^{i\theta_\nu^1}		
,
\qquad \qquad 
	\theta_\nu^1 
=
	2 
	\roundLR{
	\frac{\pi}{4} 
	+ 
	\arg{\Gamma(-\nu)}) 
	}
	+ 
	\phi_c
\label{eq:C1nkandtheta1}
}
Here, we are using the label $k$ for $\gamma_k$, as it depends on the mass and the charge of the particle with  momentum $k^1$.
The features of $C_{n_k}^1$ are given in parallel with that of  $C_{n_k}^0$, discussed in the preceding section.

The coefficients $C_{n_k}^1$ depend on only the variable $\mu$, and hence $C_{n_k}^1$ reflects the charge and the mass but not the momentum $k^0$ and $k^1$.
When $\mu \to \infty$, we obtain $ \absN{ C_{n_k}^1 } \to \delta_{n0}$, and hence \ref{eq:10byout} becomes $
	\ket{ 1_k 0_{-k}}_{\text{in}}
\to
	C_{0_k}^1
	\ket{1_k 0_{-k}}_{\text{out}}	
$, where the difference between the left- and right-hand side is just a phase factor.
When $\mu \to 0$, we have $\absN{ C_{n_k}^1 } \to \sqrt{n+1}/ 2^{(n+2)/2}$.


Also, using the relations \ref{eq:relabetinandout} and \ref{eq:10byout}, the single-particle `in' state $\ketN{\psi_{sk}^{(1)}}$ can be written in terms of the `out' states, necessary to make the squeezed state expansion
\eq{
	\ket{\psi_{sk}^{(1)}}
&=
	\frac{1}{\sqrt{2}}
	\sum_{\ell,n=0}^\infty
	\squareB{
	C_{\ell_s}^0
	C_{n_k}^1
	\ket{\ell_s \ell_{-s} (n+1)_k n_{-k}}_{\text{out}}
	+
	C_{\ell_s}^1
	C_{n_k}^0
	\ket{(\ell+1)_s \ell_{-s} n_k n_{-k}}_{\text{out}}
	}
\label{eq:rho1-ssk-k}
}
In the limit of large $\mu$, the state behaves as
\begin{equation}
    \left|\psi_{s k}^{(1)}\right\rangle \stackrel{\mu \rightarrow \infty}{\longrightarrow} \frac{1}{\sqrt{2}}\left[C_{0_{s}}^{0} C_{0_{k}}^{1}\left|0_{s} 0_{-s} 1_{k} 0_{-k}\right\rangle_{\text {out }}+C_{0_{s}}^{1} C_{0_{k}}^{0}\left|1_{s} 0_{-s} 0_{k} 0_{-k}\right\rangle_{\text {out }}\right]
    \label{largemu}
\end{equation}
which can differ from the original incoming state \ref{charge1} by phase factors.
\noindent
\subsubsection{ Mutual information}
\noindent
Here we compute the  mutual information defined in \ref{MIBP} of \ref{The mutual Information},
corresponding to the state in \ref{charge1}. We shall focus on two reduced density matrices that characterise the particle-particle and also the particle-antiparticle correlations between the two scalar fields.

Let us begin with the particle-particle correlation.
The reduced density matrix is given by $
	\rho_{s,k}^{(1)}
=
	\Tr_{-s,-k}
	\rho^{(1)}
$ and is written in terms of the `out' states as 
\eq{
	\rho_{s,k}^{(1)}
&=
	\frac{1}{2}
	\sum_{n,\ell=0}^\infty
	\roundB{
	C_{\ell_s}^0
	C_{n_k}^1
	\ket{\ell_s (n+1)_k}_{\text{out}}
	+
	C_{\ell_s}^1
	C_{n_k}^0
	\ket{(\ell+1)_s n_k}_{\text{out}}
	}
	\times
	\roundLR{
	\text{h.c.}
	}
 \label{eq:rho1sk}
}
where $\roundN{\text{h.c.}}$ stands for the Hermitian conjugate of the first parenthesis. The reduced density matrix is no longer in a pure state since the partial trace is performed. In the limit of large $\mu(k)$ and $\mu(s)$, the state $\rho_{s,k}^{(1)}$ behaves as 
\eq{
	\rho_{s,k}^{(1)}
&\stackrel{\mu \rightarrow \infty}{\longrightarrow}
	\frac{1}{2}
	\roundB{
	C_{0_s}^0
	C_{0_k}^1
	\ket{0_s 1_k}_{\text{out}}
	+
	C_{0_s}^1
	C_{0_k}^0
	\ket{1_s 0_k}_{\text{out}}
	}
	\times
	\roundLR{
	\text{h.c.}
	}
\label{eq:rho1sk'}
}
which is a pure and maximally entangled state.

The reduced density matrices, given by $\rho_{s}^{(1)}=\operatorname{Tr}_{k} \rho_{s k}^{(1)}$ and $\rho_{k}^{(1)}=\operatorname{Tr}_{s} \rho_{s k}^{(1)}$, are obtained as
\begin{eqnarray}
\rho_{s}^{(1)} =\frac{1}{2} \sum_{\ell=0}^{\infty}\left(\left|C_{\ell_{s}}^{0}\right|^{2}+\left|C_{(\ell-1)_{s}}^{1}\right|^{2}\right)\left|\ell_{s}\right\rangle_{\text {out }}\left\langle\ell_{s}\right| \\
\rho_{k}^{(1)} =\frac{1}{2} \sum_{n=0}^{\infty}\left(\left|C_{n_{k}}^{0}\right|^{2}+\left|C_{(n-1)_{k}}^{1}\right|^{2}\right)\left|n_{k}\right\rangle_{\text {out }}\left\langle n_{k}\right|
\label{reducedsk}
\end{eqnarray}
When the two scalar fields have the same mass, the eigenvalues of $\rho_{s}^{(1)}$ and $\rho_{k}^{(1)}$ are the same, and hence the entanglement entropies satisfy $S\left(\rho_{s}^{(1)}\right)=S\left(\rho_{k}^{(1)}\right)$. In the limit of large $\mu(k)$ and $\mu(s)$,
 $$
 	\rho_{s,k}^{(1)}
 \to
 	\frac{1}{2}
 	\squareN{
 	C_{0_s}^0
 	C_{0_k}^1
 	\ketN{0_s 1_k}_{\text{out}}
 	+
 	C_{0_s}^1
 	C_{0_k}^0
 	\ketN{1_s 0_k}_{\text{out}}
 	}
	\times
 	\squareN{
 	\text{h.c.}
 	}
 $$
and the von Neumann entropy
$S(\rho_{s,k}^{(1)})$ vanishes.

\begin{figure}[ht]
	\centering
		\includegraphics[scale=.5]{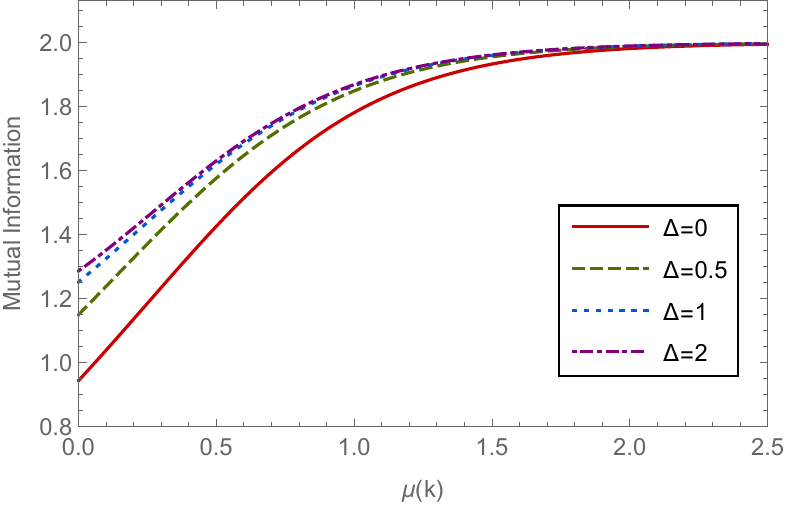}
		\caption{\it{\small The mutual information of $\rho_{s,k}^{(1)}$ vs. $\mu(k)$ (i.e., the particle-particle sector) for each value of $\Delta$, corresponding to the single charge state in \ref{charge1}. All lines approach $S(\rho_{s}^{(1)})+S(\rho_{k}^{(1)})=2$ as $\mu(k)$ increases. This asymptotic value corresponds to the initial state itself, see text for discussion.}}
		\label{fig:MIForRhoskForOP1}
\end{figure}

The mutual information is defined as $
	I(\rho_{s,k}^{(1)})
=
	S(\rho_{s}^{(1)})
	+
	S(\rho_{k}^{(1)})
	-
	S(\rho_{s,k}^{(1)})
$.
The summation in \ref{eq:rho1sk} converges rapidly and hence for numerical purposes, we replace the infinity with a finite but large $n$- and $\ell$-value.
We thus obtain the $\mu$-dependence of $I(\rho_{s,k}^{(1)})$, shown in \ref{fig:MIForRhoskForOP1}.
Here we have defined 
 $$\Delta \equiv \mu(s)-\mu(k)$$
reflecting, e.g., the mass difference between the fields. 
Moreover, as we consider various parametric values of $\Delta$, we assume to vary $\mu(s)$ keeping $\mu(k)$ fixed.

\ref{fig:MIForRhoskForOP1} shows that $I(\rho_{s,k}^{(1)})$ approaches its maximum value, $S(\rho_{s}^{(1)})+S(\rho_{k}^{(1)})=2$, as $\mu(k)$ increases, showing
the correlation of the particle-particle sector is maximum
for the large $\mu$ limit of \ref{eq:rho1sk}.
This behaviour comes from the fact that the eigenvalues become independent of $\mu$ in the large $\mu$ limit, 
which resembles the reduced density matrix of $\ketN{\psi_{sk}^{(1)}}$ traced out by the incoming states.
When $\mu$ is small, the lines for the different values of $\Delta$ split, e.g., the mass difference of the two scalar fields can be estimated with fixed $E$ and $B$ in that region. 
\ref{fig:MIForRhoskForOP1} also implies that the pair-creation disturbs the correlation, which initially exists in terms of the incoming modes.
\\


%

Next, we consider correlations of the particle-antiparticle and antiparticle-antiparticle sector.
We write the reduced density matrices,
$
	\rho_{\pm s,-k}^{(1)}
=	
	\Tr_{\mp s,k}
	\rho^{(1)}
$, in terms of  the outgoing modes as
\eq{
	\rho_{s,-k}^{(1)}
&=
	\frac{1}{2}
	\sum_{n,\ell=0}^\infty
	\roundB{
	C_{\ell_s}^0
	C_{(n-1)_k}^1
	\ket{\ell_s (n-1)_{-k}}_{\text{out}}
	+
	C_{\ell_s}^1
	C_{n_k}^0
	\ket{(\ell+1)_s n_{-k}}_{\text{out}}
	}
\times	\roundLR{
	\text{h.c.}
	}
\label{eq:rho1s-k}
}
\eq{
\rho_{-s,-k}^{(1)} 
&=
	\frac{1}{2}
	\sum_{n,\ell=0}^\infty
	\roundB{
	C_{\ell_s}^0
	C_{(n-1)_k}^1
	\ket{\ell_{-s} (n-1)_{-k}}_{\text{out}}
	+
	C_{(\ell-1)_s}^1
	C_{n_k}^0
	\ket{(\ell-1)_{-s} n_{-k}}_{\text{out}}
	}
\times	\roundLR{
	\text{h.c.}
	}
\label{eq:rho1-s-k}
}
with the requirement $C^1_{(-1)_k}=0$.
We also define 
$
	\rho_{-k}^{(1)}
=
	\Tr_{s}
	\rho_{s,-k}^{(1)}
$ and 
$
	\rho_{-s}^{(1)}
=
	\Tr_{-k}
	\rho_{-s,-k}^{(1)}
$. The reduced density matrix $\rho_{s}^{(1)}$ is given by \ref{reducedsk}, and $\rho_{-k}^{(1)}=\operatorname{Tr}_{s} \rho_{s,-k}^{(1)}$ is given by

$$
\rho_{-k}^{(1)}=\frac{1}{2} \sum_{n=0}^{\infty}\left(\left|C_{n_{k}}^{0}\right|^{2}+\left|C_{n_{k}}^{1}\right|^{2}\right)\left|n_{k}\right\rangle_{\text {out }}\left\langle n_{k}\right|
$$

The eigenvalues of $\rho_{s}^{(1)}$ and $\rho_{-k}^{(1)}$ are different even if the fields have the same mass and hence the entanglement entropies of them are also different, unlike the case of $\rho_{k}^{(1)}$. 
Note that $\rho_{\pm s,-k}^{(1)}$ becomes a product state
$
	\rho_{\pm s}^{(1)}
	\otimes
	\rho_{-k}^{(1)}
$ 
in the limit of large $\mu(k)$ and $\mu(s)$, and consequently, the mutual information becomes zero, as discussed in \ref{The mutual Information}. 

\begin{figure}[ht]
	\centering
		\includegraphics[scale=.65]{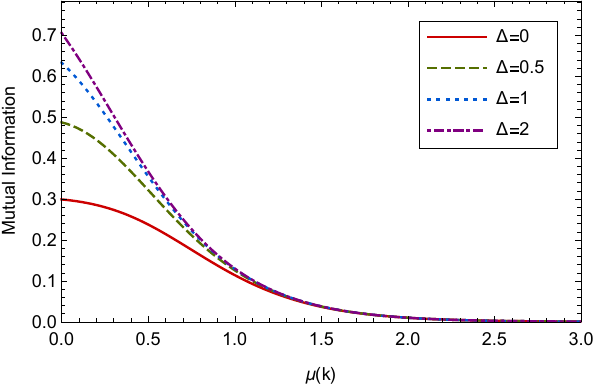}\hspace{1.0cm}
		\includegraphics[scale=.5]{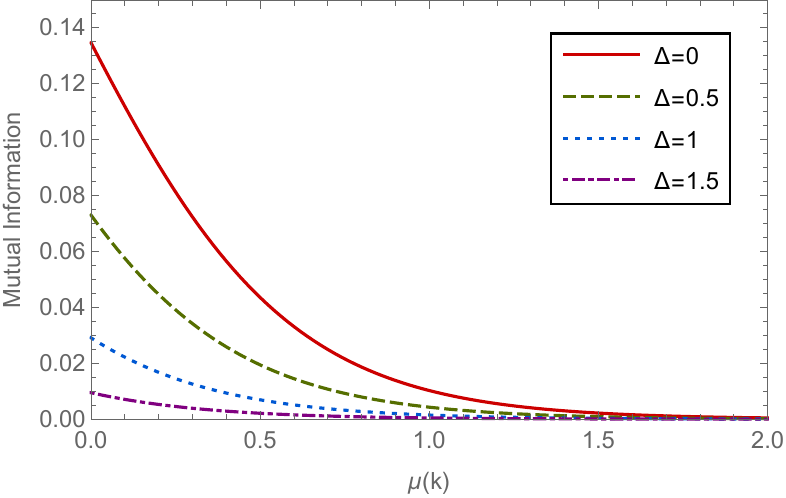}
		\caption{\it {\small The mutual information of $\rho_{s,-k}^{(1)}$ (left) and of $\rho_{-s,-k}^{(1)}$ (right) vs. $\mu(k)$ for each value of $\Delta$ (i.e., the particle-antiparticle and antiparticle-antiparticle sector), corresponding to the single charge state in \ref{charge1}. 
		All lines approach zero, since $S(\rho_{\pm s,-k}^{(1)})=S(\rho_{\pm s}^{(1)})+S(\rho_{-k}^{(1)})$ in the limit of large $\mu(k)$. Note the	qualitative difference from \ref{fig:MIForRhoskForOP1}.} }
		\label{fig:MIForRhosmkForOP1}
\end{figure}

\ref{fig:MIForRhosmkForOP1} shows that the mutual information of $\rho_{\pm s,-k}^{(1)}$ approaches zero as $\mu(k)$ increases. This corresponds once again to the fact,  that for large $\mu(k)$ values, the Bogoliubov transformation becomes trivial, and the `out' and `in' states coincide modulo some trivial phase factors, as discussed in \ref{sec:1particle}. However, \ref{charge1} has no antiparticle content in it, resulting in a vanishing mutual information between the particle-antiparticle (antiparticle-antiparticle) sector in this limit.
On the other hand, for smaller $\mu$ values, the lines split as \ref{fig:MIForRhoskForOP1}.
In addition, we observe that $\rho_{- s,-k}^{(1)}$ shows the inverted hierarchy of $\Delta$ compared to $\rho_{s,\pm k}^{(1)}$.



%
%

\subsubsection{Logarithmic negativity}
\noindent
Let us now compute the logarithmic negativity, 
first for the particle-particle sector, $\rho_{s,k}^{(1)}$.
 To obtain the logarithmic negativity, as discussed in \ref{L-N} of \ref{Entanglement negativity and logarithmic negativity}, we need the partial transpose of $\rho_{s, k}^{(1)}$, which is given by
$$
\begin{aligned}
\left(\rho_{s, k}^{(1)}\right)^{\mathrm{T}_{s}}=\frac{1}{2}[ & \left|C_{\ell_{s}}^{0} C_{n_{k}}^{1}\right|^{2}\left|\ell_{s}(n+1)_{k}\right\rangle_{\text {out }}\left\langle\ell_{s}(n+1)_{k}\right| \\
& +C_{\ell_{s}}^{1} C_{n_{k}}^{0}\left(C_{\ell_{s}}^{0} C_{n_{k}}^{1}\right)^{*}\left|\ell_{s} n_{k}\right\rangle_{\text {out }}\left\langle(\ell+1)_{s}(n+1)_{k}\right| \\
& +\left(C_{\ell_{s}}^{1} C_{n_{k}}^{0}\right)^{*} C_{\ell_{s}}^{0} C_{n_{k}}^{1}\left|(\ell+1)_{s}(n+1)_{k}\right\rangle_{\text {out }}\left\langle\ell_{s} n_{k}\right| \\
& \left.+\left|C_{\ell_{s}}^{1} C_{n_{k}}^{0}\right|^{2}\left|(\ell+1)_{s} n_{k}\right\rangle_{\text {out }}\left\langle(\ell+1)_{s} n_{k}\right|\right]
\end{aligned}
$$

Note that $\left(\rho_{s, k}^{(1)}\right)^{\mathrm{T}_{s}}$ approaches the partial transpose of the pure and maximally entangled state in the limit of large $\mu(s)$ and $\mu(k)$, as discussed in the case of the mutual information. 

\begin{figure}[ht]
	\centering
		\includegraphics[scale=.65]{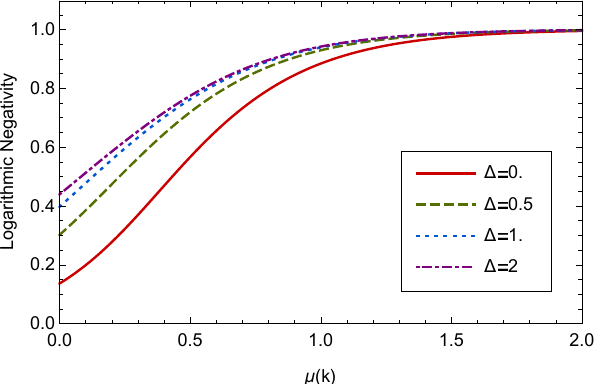}
		\caption{
		\it{\small The logarithmic negativities of $\rho_{s,k}^{(1)}$ vs. $\mu(k)$  (i.e. the particle-particle sector) for each value of $\Delta$, corresponding to the single charge state in \ref{charge1}. In the limit of large $\mu(k)$, the logarithmic negativities approach unity. Also, we have not plotted the particle-antiparticle sector, where the logarithmic negativity turns out to be vanishingly small for all $\mu(k)$ values.
		In the region where $\mu$ is small, the lines for the different values of $\Delta$ split.}
		}
		\label{fig:LNPskmu50}
\end{figure}

Employing the definition of the logarithmic negativity, the $\mu(k)$-dependence of the logarithmic negativity of $\rho_{s,k}^{(1)}$ is shown in 
\ref{fig:LNPskmu50} for different values of $\Delta$. The logarithmic negativity increases as $\mu(k)$ increases and for large $\mu(k)$-values, all the lines converge to unity. This is because $\rho_{s,k}^{(1)}$ approaches the maximally entangled pure state for large $\mu(k)$. 
This behaviour implies that the entanglement of $\rho_{s,k}^{(1)}$ is disturbed by the pair creation.
%
For the particle-antiparticle and antiparticle-antiparticle sector, however, 
we find that the logarithmic negativities are vanishingly small, $\mathcal{L}_N(\rho_{\pm s, -k}^{(1)}) \lesssim{\cal O}(10^{-15})$, for all $\mu(k)$ values, showing once again the qualitative differences of these sectors with the particle-particle sector.

We shall consider  another example of entangled state below and will see the differences between the information quantities associated with it and those of \ref{charge1}.

\subsection{Zero-charge state}
\label{sec:zerocharge}
\noindent
Keeping in mind the discussion made at the end of \ref{sec:bipartite},
we now consider a pure state $
	\rho^{(\text{PA})}
=
	\ketN{\psi_{sk}^{(\text{PA})}}
	\braN{\psi_{sk}^{(\text{PA})}}
$, where 
\eq{
	\ket{\psi_{sk}^{(\text{PA})}}
&=
	\frac{
	1
	}{
	\sqrt{2}
	}
	\roundB{
	\ket{1_s 0_{-s}; 0_k 1_{-k}}_{\text{in}} 
	+ 
	\ket{0_s 1_{-s}; 1_{k} 0_{-k}}_{\text{in}} 
	}
\label{eq:PAstate}
}
has zero net charge content. Let us compute the same measures as earlier in order to see the qualitative differences.
\subsubsection{ Mutual information}
\noindent

Using the earlier techniques, we derive the reduced density operator for the particle-particle sector,
$
	\rho_{s,k}^{(\text{PA})}
=
	\Tr_{-s,-k}
	\rho^{(\text{PA})}
$, which is written in terms of the `out' states as 
\eq{
	\rho_{s,k}^{(\text{PA})}
&=
	\frac{1}{2}
	\sum_{\ell,n=0}^\infty
	\roundB{
	C_{\ell_s}^1
	C_{(n-1)_k}^1
	\ket{(\ell+1)_s (n-1)_k}_{\text{out}}
	+
	C_{(\ell-1)_s}^1
	C_{n_k}^1
	\ket{(\ell-1)_s (n+1)_k}_{\text{out}}
	}
	\times
	\roundLR{
	\text{h.c.}
	}
\label{eq:rhoPAsk}
}
\begin{figure}[htbp]
	\centering
		\includegraphics[scale=.5]{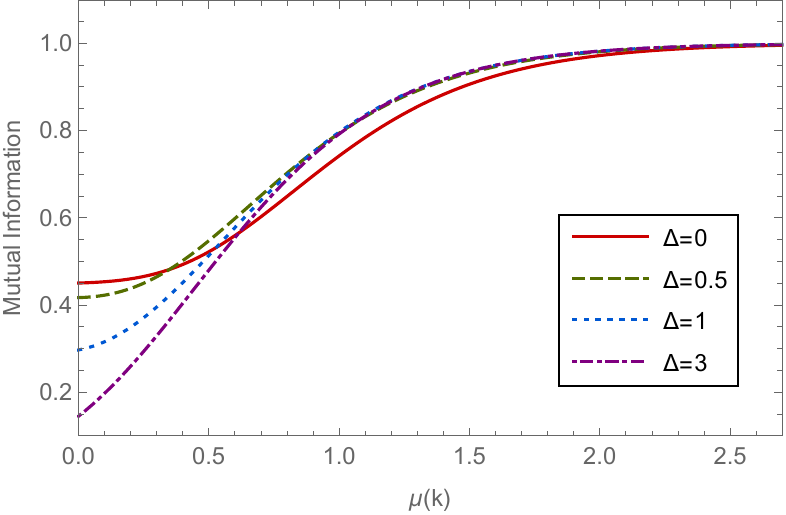}\hspace{1.0cm}
  \includegraphics[scale=.48]{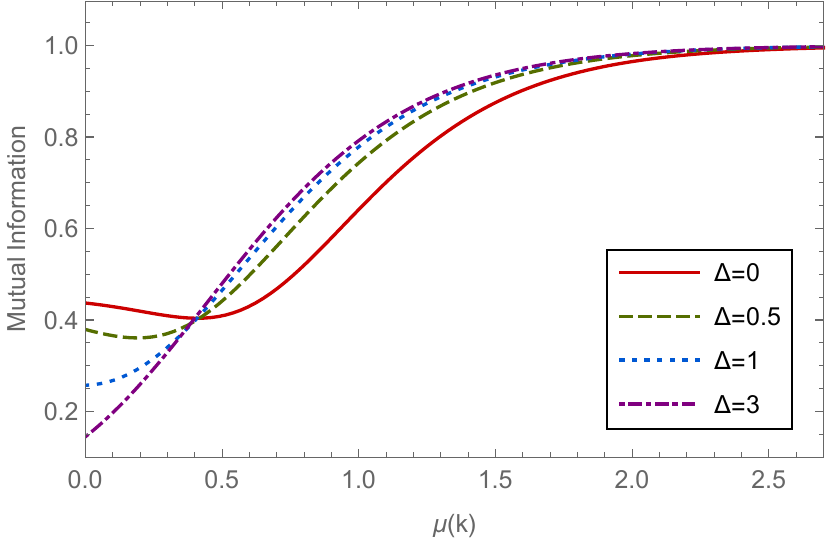}
		\caption{\it{\small The mutual information of $\rho_{s,k}^{(\text{PA})}$ vs. $\mu(k)$ (i.e., the particle-particle sector)(left) and $\rho_{s,-k}^{(\text{PA})}$ vs. $\mu(k)$ (i.e., the particle-antiparticle sector) (right) for each value of $\Delta$, corresponding to the particle-antiparticle state in \ref{eq:PAstate}.}}
		\label{fig:MIForRhoskForPA}
\end{figure}
The reduced density matrices, given by $\rho_{s}^{(PA)}=\operatorname{Tr}_{k} \rho_{s,k}^{(PA)}$ and $\rho_{k}^{(PA)}=\operatorname{Tr}_{s} \rho_{s,k}^{(PA)}$, are obtained as
\begin{eqnarray}
\rho_{s}^{(PA)} =\frac{1}{2}\sum_{\ell=0}^{\infty}(|C_{\ell_{s}}|^2+|C_{(\ell+1)_{s}}|^2)|\ell_s\rangle_{\text{out}} \langle \ell_s| \\
\rho_{k}^{(PA)} =\frac{1}{2}\sum_{n=0}^{\infty}(|C_{n_{k}}|^2+|C_{(n+1)_{k}}|^2)|n_k\rangle_{\text{out}} \langle n_k|
\label{reducedsk1}
\end{eqnarray}
On the other hand, the reduced density operator for the  particle-antiparticle sector is given by
\eq{
	\rho_{s,-k}^{(\text{PA})}
&=
	\frac{1}{2}
	\sum_{n,\ell=0}^\infty
	\roundB{
	C_{\ell_s}^1
	C_{n_k}^1
	\ket{(\ell+1)_s (n+1)_{-k}}_{\text{out}}
	+
	C_{(\ell-1)_s}^1
	C_{(n-1)_k}^1
	\ket{(\ell-1)_s (n-1)_{-k}}_{\text{out}}
	}
	\times
	\roundLR{
	\text{h.c.}
	}
\label{eq:rhoPAs-k}
}
%
The reduced density matrix $\rho^{(PA)}_{-k}$ is equal to $\rho^{(PA)}_{k}$ given by \ref{reducedsk1}.
The $\mu(k)$-dependence of the mutual information corresponding to the particle-particle sector, \ref{eq:rhoPAsk}, and the particle-antiparticle sector, \ref{eq:rhoPAs-k}, for different values of $\Delta$ is depicted respectively in \ref{fig:MIForRhoskForPA}. We note the overall qualitative similarity between them, although the right plot of \ref{fig:MIForRhoskForPA} shows a minimum for $\Delta \to 0$. 
We also note that the mutual information for both of them has the same numerical orders, unlike those of the single-charge state.
These similarities should correspond to the symmetry in particle and antiparticle number of the initial state, \ref{eq:PAstate}. Due to the same reason, the antiparticle-antiparticle sector of this state yields exactly the same correlations as those of the particle-particle sector, and we do not pursue it any further. 
We also note some qualitative differences in the behaviour of figures \ref{fig:MIForRhoskForOP1} and \ref{fig:MIForRhoskForPA} :
the left plot of \ref{fig:MIForRhoskForOP1} has simple features such as the monotonicity and unchanging hierarchy with respect to $\Delta$, while the right plot of \ref{fig:MIForRhoskForOP1} and \ref{fig:MIForRhoskForPA} shows interchanges of the hierarchy and the non-monotonic behaviour in the right plot of \ref{fig:MIForRhoskForPA}.
 This implies that the slightly complicated behaviour of the zero-charge state originates from the existence of both particle and antiparticle in the incoming state.
 \noindent
\subsubsection{Logarithmic negativity}

 Finally, we compute the logarithmic negativity for the particle-particle and particle-antiparticle sectors by following the earlier methods. The partial transposed matrix for these sectors is given by
 \begin{equation}
      \begin{split}
\left(\rho_{s,k}^{(\text{PA})}\right)^{\mathrm{T}_{s}}=\frac{1}{2}\Big[ & \left|C_{\ell_{s}}^{1} C_{(n-1)_{k}}^{1}\right|^{2}\left|\ell_{s}(n+1)_{k}\right\rangle_{\text {out }}\left\langle\ell_{s}(n+1)_{k}\right| \\
& +C_{\ell_{s}}^{1} C_{(n-1)_{k}}^{1}\left(C_{(\ell-1)_{s}}^{1} C_{n_{k}}^{1}\right)^{*}\left|\ell_{s} n_{k}\right\rangle_{\text {out }}\left\langle(\ell+1)_{s}(n+1)_{k}\right| \\
& +(C_{\ell_{s}}^{1} C_{(n-1)_{k}}^{1})^*C_{(\ell-1)_{s}}^{1} C_{n_{k}}^{1}\left|(\ell+1)_{s}(n+1)_{k}\right\rangle_{\text {out }}\left\langle\ell_{s} n_{k}\right| \\
& +\left|C_{\ell_{s}}^{1} C_{n_{k}}^{0}\right|^{2}\left|(\ell+1)_{s} n_{k}\right\rangle_{\text {out }}\left\langle(\ell+1)_{s} n_{k}\right|\Big]\\
     \end{split}
 \end{equation}
  \begin{equation}
     \begin{split}
\left(\rho_{s,-k}^{(\text{PA})}\right)^{\mathrm{T}_{s}}=\frac{1}{2}\Big[& \left|C_{(\ell+1)_{s}}^{1} C_{(n+1)_{-k}}^{1}\right|^{2}\left|(\ell+1)_{s}(n+1)_{-k}\right\rangle_{\text {out }}\left\langle(\ell+1)_{s}(n+1)_{-k}\right| \\
& +C_{\ell_{s}}^{1} C_{n_{-k}}^{1}\left(C_{(\ell-1)_{s}}^{1} C_{(n-1)_{-k}}^{1}\right)^{*}\left|(\ell-1)_{s} (n+1)_{-k}\right\rangle_{\text {out }}\left\langle(\ell+1)_{s}(n-1)_{-k}\right| \\
& +(C_{\ell_{s}}^{1} C_{n_{-k}}^{1})^*C_{(\ell-1)_{s}}^{1} C_{(n-1)_{-k}}^{1}\left|(\ell+1)_{s} (n-1)_{-k}\right\rangle_{\text {out }}\left\langle(\ell-1)_{s}(n+1)_{-k}\right| \\
& +\left|C_{(\ell-1)_{s}}^{1} C_{(n-1)_{-k}}^{0}\right|^{2}\left|(\ell-1)_{s} (n-1)_{-k}\right\rangle_{\text {out }}\left\langle(\ell-1)_{s} (n-1)_{-k}\right|\Big]
     \end{split}
 \end{equation}


 \begin{figure}[h]
	\centering
		\includegraphics[scale=.45]{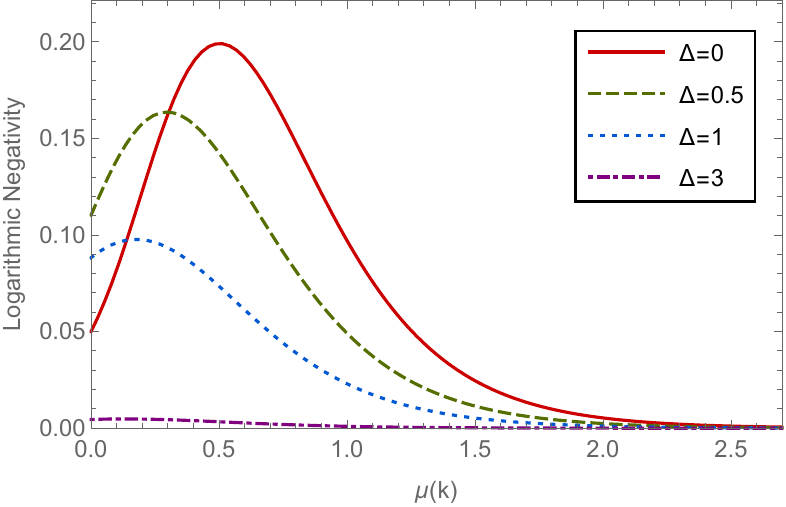}\hspace{1.0cm}
  \includegraphics[scale=0.45]{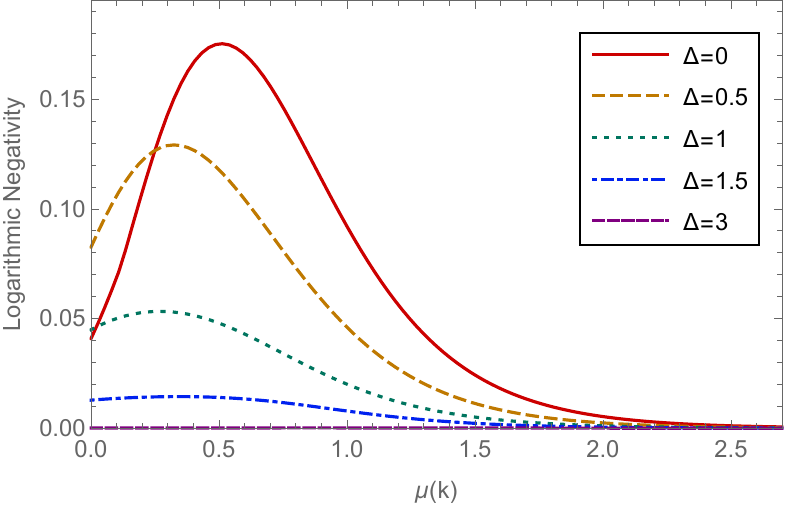}
		\caption{\it{\small The logarithmic negativity of $\rho_{s,k}^{(\text{PA})}$ (left) and of $\rho_{s,-k}^{(\text{PA})}$ (right) vs. $\mu(k)$ for each value of $\Delta$ (i.e., the particle-particle sector and the particle-antiparticle sector), corresponding to the particle-antiparticle state in \ref{eq:PAstate}.}}
		\label{fig:LNForRhoskForPA}
\end{figure}
 Using the definition of logarithmic negativity discussed in \ref{Entanglement negativity and logarithmic negativity}, we computed the logarithmic negativity for both sectors. They have respectively been plotted in  \ref{fig:LNForRhoskForPA}. We note their overall qualitative similarity, owing once again to the symmetry in the number of particles and antiparticles in \ref{eq:PAstate}. The asymptotically vanishing values in the plots for large $\mu(k)$  just as earlier correspond to the fact that in this limit we reach the initial state, \ref{eq:PAstate}, and it  has no logarithmic negativity, as can be checked easily. 
The antiparticle-antiparticle sector behaves in the same manner as the particle-particle sector.
We also note the qualitative differences of these plots with that of \ref{fig:LNPskmu50}, which arises from the existence of both particle and antiparticle in the incoming zero-charge state.
\section{Summary and outlook}
\label{sec:SD1}
\noindent
We now summarise our results. The chief motivation of this work was to quantify the effect of a background magnetic field on the entanglement or correlations between the Schwinger pairs, for a complex scalar field. We have studied the vacuum entanglement entropy in \ref{sec:vacuum}, and the mutual information and logarithmic negativity for maximally entangled  states with different electric charge content, respectively, in \ref{sec:1particle} and \ref{sec:zerocharge}. We have emphasised the qualitative differences in the behaviour of the information quantities between these  states. 
Extending these results to the Rindler and to the inflationary cosmological backgrounds seems interesting. 

Finally, we note that in all the plots, the various information quantities converge to some specific points for sufficiently large $\mu$ values. Assuming, for example, that the mass, the electric field, and the Landau level are fixed, a large $\mu$ corresponds to large values of the magnetic field, \ref{mu}.
In this limit,  the Bogoliubov transformation becomes trivial,  and an `out' state coincides with the `in' state, modulo some trivial phase factor (cf., the discussion below \ref{eq:absC00}).

Keeping in mind that the background electric field is analogous to the acceleration parameter in a non-inertial frame as far as the particle creation is concerned, it then seems possible that the degraded quantum correlation between two entangled states in such frames, for example, ~\cite{MartinMartinez:2010ar}, might be restored (for charged fields) via the application of a background magnetic field, as follows. As we discussed in \ref{Motivation and Overview}, the magnetic Lorentz force, $e \vec{v}\times \vec{B} $, acts in the same direction for the particle and antiparticle initially moving in the opposite direction, just after the pair creation. An electric field renders just the opposite effect by moving the created pairs away. 
Thus, to the best of our understanding, it seems interesting to see how the background magnetic field will impact the particle-antiparticle pair creation, causing the entanglement degradation in the Rindler frame. 
As an example, we may consider \ref{fig:LNForRhoskForPA}. The logarithmic negativity, which is a measure of entanglement for a mixed ensemble, grows and reaches its maximum for certain $\mu$ values. After this point, however, the particle creation becomes too weak, and the squeezed state expansion coincides with that of the initial state, which itself has vanishing logarithmic negativity. Depending upon the characteristics of the initial state, analogous arguments can be made for all the cases we have investigated in this Chapter. Such an effect can, in particular, can also be relevant for a non-extremal black hole endowed with a strong magnetic field on its exterior.

It would be interesting to extend this analysis to the more realistic case of Dirac fermions, which we briefly sketch below. The Dirac equation coupled to background electric and magnetic fields is
$$ \left(i \gamma^{\mu} \partial_{\mu}-q \gamma^{\mu} A_{\mu}-m\right) \Psi(x)=0$$
where $\gamma^{\mu}$ are the Dirac matrices. We consider the external gauge field the same as that of the earlier case, i.e., $A_\mu = (Ez,-By,0,0)$.
The orthonormal `in' and `out' modes for this case are also found to be given in terms of the Hermite polynomial and the parabolic cylinder function. We find after some algebra
\begin{equation}
\label{modef1}
     U^{\text{in}}_{k,s,n_L}(x)  =\frac{1}{M_s}\left(i \gamma^{\mu} \partial_{\mu}-e \gamma^{\mu} A_{\mu}+m\right) e^{-i(k^0t-k^1 x)}e^{-{y}_+^2/2}
	H_{n_L}({y}_+)
	D_{\nu_s} (\zeta_+)\epsilon_{s} \quad (s=1,2)
\end{equation}
\begin{equation}
\label{modef2}
V^{\text{in}}_{k,s,n_L}(x)  =\frac{1}{M_s}\left(i \gamma^{\mu} \partial_{\mu}-e \gamma^{\mu} A_{\mu}+m\right) e^{i(k^0t-k^1 x)}e^{-{y}_-^2/2}
	H_{n_L}({y}_-)[D_{\nu_s} (\zeta_+)]^* \epsilon_{s} \quad (s=3,4)    
\end{equation}
\begin{equation}
\label{modef3}
 U^{\text{out}}_{k,s,n_L}(x)  =\frac{1}{M_s}\left(i \gamma^{\mu} \partial_{\mu}-e \gamma^{\mu} A_{\mu}+m\right) e^{-i(k^0t-k^1 x)}e^{-{y}_+^2/2}
	H_{n_L}({y}_+) D_{\nu_s}(-\zeta_+)\epsilon_{s} \quad (s=1,2)
\end{equation}
\begin{equation}
\label{modef4}
   V^{\text{out}}_{k,s,n_L}(x)  =\frac{1}{M_s}\left(i \gamma^{\mu} \partial_{\mu}-e \gamma^{\mu} A_{\mu}+m\right) e^{i(k^0t-k^1 x)}e^{-{y}_-^2/2}
	H_{n_L}({y}_-)
	[D_{\nu_s}(-\zeta_-)]^* \epsilon_{s} \quad (s=3,4) 
\end{equation}

where $U$ and $V$ represents particle and antiparticle respectively and $M$ stands for the normalisation.
Also, $\epsilon_s$ are the four orthonormal spinors given by
$$
\begin{aligned}
&  \epsilon_{1}=\left(\begin{array}{l}
1 \\
0 \\
1 \\
0
\end{array}\right), \quad \epsilon_{2}=\left(\begin{array}{c}
0 \\
1 \\
0 \\
1
\end{array}\right),
& \epsilon_{3}=\left(\begin{array}{c}
0 \\
1 \\
0 \\
-1
\end{array}\right), \quad
 \epsilon_{4}=\left(\begin{array}{c}
1 \\
0 \\
-1 \\
0
\end{array}\right).
\end{aligned}
$$
The coefficient $\nu_s$ of the parabolic cylinder function in \ref{modef1}, \ref{modef2}, \ref{modef3} and \ref{modef4} is defined as
$\nu_s
=
	-(1+i\mu_s)/2$, and
 \begin{eqnarray}
	\mu_s
=
	\frac{ m^2+S_s }{eE}
\label{muf}
\end{eqnarray}
where $S_s$ are given by
 $$S_1=S_3=(2n_L+1)eB   \quad \text{and} \quad S_2=S_4=2n_LeB$$ 
where $n_L=0,1,2,3,...$ are the Landau levels as earlier. We find these modes are related via the Bogoliubov transformation,
\begin{equation*}
   U^{\text{in}}_{k,s,n_L}(x)
=
	\alpha_{s}
	 U^{\text{out}}_{k,s,n_L}(x)
	+
	\beta_{s}
	\left(
	 V^{\text{out}}_{-k,s,n_L}(x)
	\right)^*  
\end{equation*}
where $\alpha_{s}$ and $\beta_{s}$ are the Bogoliubov coefficients given by
\begin{eqnarray*}
 \alpha_{s}=\sqrt{\frac{\mu_s}{\pi}} \Gamma\left(\frac{i \mu_s}{2}\right) \sinh \left(\frac{\pi \mu_s}{2}\right) e^{-\frac{\pi \mu_s}{4}}, \quad \quad
 \beta_{s}=e^{-\frac{\pi\mu_s}{2}}    
\end{eqnarray*}
which satisfy the consistency relation $|\alpha_{s}|^{2}+|\beta_{s}|^{2}=1$. It is worth noting that there exists a sign change in the Bogoliubov consistency condition compared to scalars. This distinction arises from the fact that fermions adhere to an anticommutation relation, while scalars follow a commutation relation. It is easy to see from the expression for $\mu_s$ that is similar to the scalar field theory, if we let $E\to0$, we have $\beta_s \to 0$ as well, no matter how strong $B$ is. Also, $\beta_s$ decreases monotonically as $B$ increases with $E$ kept fixed.

Using these Bogoliubov coefficients, we can construct the squeezed state expansion as earlier,  keeping, however, in mind Pauli's exclusion principle. Using this, we may further compute the desired quantum correlations. The details of these calculations and results will be reported in future publications.
\noindent

\chapter{Fermionic entanglement in cosmological de Sitter spacetime with background electric and magnetic fields}
\label{Fermionic Bell violation in de Sitter spacetime with background electric and mangnetic fields}

There are a couple of distinct non-trivial spacetime backgrounds where entanglement properties of quantum fields emerge very naturally due to the creation of entangled particle pairs. One of them is the near horizon region of a non-extremal black hole spacetime, or the Rindler spacetime, where the entanglement of quantum fields between two causally disconnected spacetime wedges have been investigated, for example in ~\cite{BKAY:2017, FuentesSchuller:2004xp, degradation:2015, nper:2011, degradation:2015n}. Another one is the cosmological background, where the vacuum in the asymptotic future (the out vacuum) is related to that of the asymptotic past (the in vacuum) via squeezed state expansion due to pair creation. We refer our reader to 
for example, ~\cite{Fuentes:2010dt, Arias:2019pzy, bell:2017, Maldacena:2015bha, EE, vaccum EE for fermions, SSS, SHN:2020, QC in deSitter, Choudhury:2016cso, Choudhury:2017bou, Bhattacharya:2019zno} and references therein for discussions on various measures of scalar and fermionic entanglement in different coordinatisations of the de Sitter spacetime. 

A very important and useful measure of quantum entanglement is the violation of the Bell inequality~\cite{Bell, CHSH} (see also~\cite{NielsenChuang} for an excellent pedagogical discussion), which has been confirmed experimentally~\cite{Aspect1, Aspect2}. Such violation rules out the so called classical hidden variable theories and establishes the probabilistic and (for entangled states) the non-local characteristics associated with the quantum measurement procedure, e.g.~\cite{bell_3, bell1:2003, bell_1} (also references therein). The Bell inequality was originally designed for pure bipartite states, which was later extended to multipartite systems, altogether known as the Bell-Mermin-Klyshko  inequalities (or the  Clauser-Horne-Shimony-Holt inequality),~\cite{CHSH, NHB:1998zz, Mermin, Klyshko}.

 Studying correlations in the context of the early inflationary era can give us insight into the state of a quantum field in the early universe. Such investigations should not be regarded as mere academic interests, as attempts have also been made to predict their possible observational signatures. Specifically, entanglement generated in the early universe can affect the cosmological correlation functions or the cosmic microwave background (CMB). For example, the fermionic entanglement may break the scale invariance of  the inflationary power spectra~\cite{Boyanovsky:2018soy}. It was argued in~\cite{Rauch:2018rvx}
by studying the violation of the Bell inequality by the photons coming from certain high redshift quasars that they are entangled, indicating the existence of entangled quantum states in the early universe. We also refer our reader to~\cite{Morse:2020mdc} and references therein for discussion on the signature of Bell violation in the CMB and its observational constraints pertaining to the Bell operators and some course graining parameters.

In cosmological spacetimes, particle creation occurs due to the background spacetime curvature, for example,~\cite{Parker:2009uva}. However, if background electric and magnetic fields are also present, particle creation can be affected. A particularly interesting scenario is the early inflationary spacetime endowed with primordial electromagnetic fields. Computations on the Schwinger effect for both scalar and fermion in the de Sitter spacetime and its possible connection to the observed residual magnetic field in the inter-galactic spaces (i.e., the so called galactic dynamo problem,~\cite{Subramanian:2015lua}), can be seen in for example,~\cite{vilenkin, Xue:2017, Xue:2017cex, Xue:2017ecx, yoko:2016tty, Bavarsad:2017oyv}.   

In this Chapter, we consider a fermionic field in the presence of constant strength background electric and magnetic fields in the $(1+3)$-dimensional cosmological de Sitter spacetime.
We note that compared to the case we studied in \ref{Background magnetic field and quantum correlations in the Schwinger effect}, here the spacetime is itself dynamical and hence it would also create particles.  As earlier,
we wish to see how the magnetic field affects the correlations between particles and antiparticles created by the background electric field and the time-dependent gravitational field. We shall compute the Bell inequality violation to estimate the entanglement and the mutual information to estimate the total correlation. Previous studies on cosmological Bell violation without any background electric and magnetic fields can be seen in for example,~\cite{Maldacena:2015bha, bell:2017, SHN:2020}.

As we have seen for the flat spacetime in \ref{Background magnetic field and quantum correlations in the Schwinger effect}, the Schwinger pair creation process ceases upon the application of a strong enough magnetic field.
Accordingly, the degradation of correlation or information between entangled states due to particle creation would also cease. In our present scenario of a dynamical spacetime background, we may ask, will the magnetic field be able to stop the particle creation due to the gravitational field? The intuitive answer is No, as follows. In a pure gravitational background, a created particle pair will follow geodesics and become observable in spacetimes like the de Sitter due to the geodesic deviation~\cite{Mironov:2011hp}. Such deviations happen even for initially parallel trajectories. Thus as the particle-antiparticle pair created in the presence of geometric curvature propagate, they are expected to get separated irrespective of the presence of the Lorentz force imparted by the background magnetic field, even though that force is acting in the same direction for both of them. This also indicates that in the absence of an electric field, the magnetic field may not affect particle creation due to the gravitational field at all. We shall check these intuitive guesses explicitly in the next section. Our goal here is to study the effect of the background magnetic field strength on the Bell violation as well as the mutual information. 
 
Apart from this, a physical motivation behind this study comes from the possible connection between the primordial electromagnetic fields and the aforementioned galactic dynamo problem, for example,~\cite{Xue:2017} studied the effect of the Schwinger process on the generation of magnetic field in an inflationary de Sitter spacetime. We wish to consider fermions instead of a complex scalar, as the former is more realistic.  Let us speculate about some possible observational consequences of the model we study. For example, one can compute the power spectra by tracing out the fermionic degrees of freedom (interacting with the inflaton or gravitational excitations) and check the breaking of scale invariance as of~\cite{Boyanovsky:2018soy}. Likewise, if we also consider the quantum part of the electromagnetic sector, it should carry information about the entangled fermionic states once we trace out the fermionic degrees of freedom, originating from the photon-fermion interaction.  Thus one can expect that the photons coming from the distant past undergoing the Bell test as of~\cite{Rauch:2018rvx} will carry information about such entangled fermionic states. Since these states are defined in the presence of the primordial background electromagnetic fields, the Bell test might also carry information about those background fields. This can possibly be used to constrain the corresponding field strengths and test the proposition 
of~\cite{Xue:2017}. With this motivation, we shall compute below the fermionic  Bell violation and mutual information in the cosmological de Sitter spacetime in the presence of background electric and magnetic fields as a viable measure of correlations.

The rest of the chapter is organised as follows. In \ref{S2} and \ref{A}, we compute the orthonormal in and out Dirac modes in the cosmological de Sitter spacetime in the presence of constant background electric and magnetic fields. The Bogoliubov coefficients and the squeezed state relationship between the in and out vacua are also found. Using this, we compute the vacuum entanglement entropy in~\ref{EE}. The Bell inequality violations for the vacuum and also two maximally entangled initial states are computed in \ref{BV}. The mutual information for two maximally entangled initial states is computed in \ref{MI}. All these results are further extended in \ref{alph} to the so called one parameter fermionic $\alpha$-vacua. Finally, we conclude in \ref{sec:SD}. We shall assume either the field is heavily massive or equivalently, the electric field strength is very high (with respect to the Hubble constant).
\noindent


\section{The in and out Dirac  modes}\label{S2}

For our purpose, we first need to solve the Dirac equation in the cosmological de Sitter spacetime coupled to the constant strength background electric and magnetic fields. It is worth noting that with the inclusion of these fields, the cosmological de Sitter spacetime ceases to satisfy the Einstein equation as it did before. The following will be an extension of the solutions found earlier in the same spacetime but in the absence of any magnetic field~\cite{Xue:2017cex, yoko:2016tty}.

The Dirac equation in a general curved spacetime reads \cite{Parker:2009uva},
\begin{equation}
\label{diraceqincurve}
 (i\gamma^{\mu}D_{\mu}-m)\psi(x)=0
\end{equation}
where  the gauge cum spin covariant derivative reads, 
$$D_\mu\equiv \partial_\mu+ieA_\mu+\Gamma_\mu$$
The spin connection matrices are given by,
\begin{eqnarray}
\label{conn1}
\Gamma_\mu=-\frac{1}{8}e^{\mu}_a \left(\partial_{\mu}e_{b\nu}- \Gamma_{\mu\nu}^{\lambda} e_{b\lambda}\right)[\gamma^a, \gamma^b]
\end{eqnarray}
where the Latin indices represent the local inertial frame and  $e^{\mu}_a$ are the tetrads.

The expanding de Sitter metric in $(3+1)$-dimensions reads,
\begin{eqnarray}
\label{FLRW}
ds^2=\frac{1}{H^2\eta^2}\left(-d\eta^2+dx^2+dy^2+dz^2\right)
\end{eqnarray}
where $H$ is the Hubble constant  and the conformal time $\eta$ varies from $-\infty <\eta < 0^-$. Choosing now
$e^{a}_\mu=a(\eta)\delta^a_\mu$, we have from \ref{conn11},
\begin{equation}
\Gamma_{\mu}=\frac{1}{2}\gamma^{\mu}\gamma^0 a^\prime(\eta) a(\eta) \delta_{\mu}^i,\;\;\;i=1,2,3    
\label{conn2}
\end{equation}
where the prime denotes differentiation once with respect to  $\eta$.

Defining a new variable in terms of the scale factor $a(\eta)= -1/H\eta$, as
\begin{equation} 
 \label{conf}
 \xi=a^{\frac32}\psi
 \end{equation}
and using \ref{conn2}, the Dirac equation \ref{diraceqincurve}  becomes,
\begin{equation}
\label{dirac_3}
\left(ie^\mu_a  \gamma^a \partial_\mu-eA_\mu e^\mu_a \gamma^a-m\right)\xi(\eta,\vec{x})\;=\;0
\end{equation}%
Substituting next
\begin{equation}
\xi(\eta,\vec{x})=\left(ie^\mu_a \gamma^a\partial_\mu-eA_\mu e^\mu_a \gamma^a+m\right)\zeta(\eta,\vec{x})
\label{de0}
\end{equation}
 into \ref{dirac_3}, we obtain the squared Dirac equation
\begin{eqnarray}
\label{sq_dirac}
\left[\left(\partial_\mu+ieA_\mu\right)^2-m^2a^2+i\left(ma^\prime a e^0_0 \gamma^0-\frac{e}{2}a^2 e^\mu_a \gamma^ae^\nu_a \gamma^b F_{\mu \nu}\right)\right]\zeta(\eta,\vec{x})\;=\;0
\end{eqnarray}
We choose the gauge  to obtain constant electric and magnetic fields in the $z$-direction in \ref{FLRW} as, 
\begin{eqnarray}
\label{pot1}
A_\mu=By\delta^{x}_\mu-\frac{E}{H}\left(a-1\right)\delta^z_\mu
\end{eqnarray}
where $E$, and $B$, respectively, stand for electric and magnetic fields. As of the preceding Chapter, we shall also assume here that $e$, $E$ and $B$ are positive quantities.

Making now the ansatz $\zeta(\eta,\vec{x})=e^{-ieEz/H}e^{i\vec{k_\slashed{y}}\cdot\vec{x}}\zeta_s(\eta,y)\omega_s$, where $\vec{k_\slashed{y}}=(k_x,0,k_z)$ are momenta orthogonal to the background fields and substituting it into \ref{sq_dirac},  we have
\begin{equation}
\label{dirac41}
\begin{split}
\Bigg[\left(\partial_y^2-\left(k_x+eBy\right)^2\right)-\partial_0^2-k_z^2+2HLak_z-H^2L^2a^2-m^2a^2\\+i Ha^2\left(M\gamma^0+L\gamma^0\gamma^3 +\frac{eB}{H a^2}\gamma^1\gamma^2\right)\Bigg]\zeta_{s}(\eta,y)\omega_s=0\\
\end{split}
\end{equation}
where 
$$M=\frac{m}{H} \qquad {\rm and} \qquad  L=\frac{eE}{H^2}$$
 are dimensionless mass and electric field strengths. Note also in \ref{dirac41} that the matrices $(M\gamma^0+L\gamma^0\gamma^3) $ and $\gamma^1\gamma^2$ commute and hence we may treat $\omega_s$ to be their simultaneous eigenvectors. Thus \ref{dirac41} becomes
\begin{eqnarray}
\label{sq_dirac2}
\left[\left(\partial_y^2-\left(k_x+eBy\right)^2\right)+\left(-\partial_0^2-\omega_k^2+i\lambda_s\sigma(\eta)+ie\beta_s B\right)\right]\zeta_{s}(\eta,y) =0
\end{eqnarray} 
where $\lambda_s=\pm 1$, $\beta_s=\pm i$ and  we have abbreviated,
\begin{eqnarray}
\label{const}
\omega_k^2=k_z^2-2HLak_z+H^2a^2\left(L^2+M^2\right),\;\sigma(\eta)=a^2H^2\sqrt{L^2+M^2}
\end{eqnarray} 
The explicit expressions for the four orthonormal eigenvectors $\omega_s$ are given in \ref{A}.
Substituting now for the variable separation, $\zeta_{s}(\eta,y)=\varsigma_s(\eta) h_s(y)$ into \ref{sq_dirac2}, we obtain the variable decoupled equations,
\begin{eqnarray}
\left(\partial_\eta^2+\omega_k^2-i\lambda_s\sigma(\eta)+S_s\right){\varsigma_{s}(\eta)}=0,\; \left(\partial_y^2-\left(k_x+eBy\right)^2+S_s+i\beta_s eB \right) h_s(y)=0
\label{dirac9}
\end{eqnarray}
where $S_s$ is the separation constant. Clearly, we can have four sets of such pair of equations corresponding to the different choices of $\lambda_s =\pm 1,\,\, \beta_s = \pm i$. For example, for $\lambda_s=1$, $\beta_s =-i$ and 
$\lambda_s=1$, $\beta_s =i$, we respectively have,
\begin{eqnarray}
\left(\partial_\eta^2+\omega_k^2-i\sigma(\eta)+S_1\right){\varsigma_{1}(\eta)}=0, \;  \left(\partial_y^2-\left(k_x+eBy\right)^2+S_1+ eB \right) h_1(y)=0 \nonumber\\ \left(\partial_\eta^2+\omega_k^2-i\sigma(\eta)+S_2\right){\varsigma_{2}(\eta)}=0, \; \left(\partial_y^2-\left(k_x+eBy\right)^2+S_2- eB \right) h_2(y)=0 
\label{dirac9'}
\end{eqnarray}
Let us first focus on the spatial equations. In terms of the variable
 $$\overline{y}=\left(\sqrt{eB}y+\frac{k_x}{\sqrt{eB}}\right)$$
 it is easy to see that the spatial differential equations of \ref{dirac9'} reduce to the Hermite differential equation, with the separation constants,
 $$S_1=2n_L eB \quad \text{and} \quad S_2=2(n_L+1)eB$$ 
 where $n_L=0,1,2...$ denotes the Landau levels. Thus we have the normalised solutions,
$$h_1(y)=h_2(y)=\left(\frac{\sqrt{eB}}{2^{n_L+1}\sqrt{\pi}(n_L+1)!}\right)^{1/2}e^{-\overline{y}^2/2}\mathcal{H}_{n_L}(\overline{y})=h_{n_L}(\overline{y})~({\rm say})$$
where $\mathcal{H}_{n_L}(\overline{y})$ are the Hermite polynomials of order $n$.

For the two temporal equations in \ref{dirac9'}, we introduce the variables,
$$z_1=-\frac{2i\sqrt{k_z^2+S_1}}{aH}\quad \text{and}\quad  z_2=-\frac{2i\sqrt{k_z^2+S_2}}{aH}$$ 
so that they respectively become,
\begin{eqnarray}
\label{zcom1}
\left(\partial^2_{z_{1}}-\frac{1}{4}+\frac{\kappa_1}{z_1}+\frac{(1/4-\mu^2)}{z_1^2}\right){\varsigma_1(z_1)}=0,\left(\partial^2_{z_{2}}-\frac{1}{4}+\frac{\kappa_2}{z_2}+\frac{(1/4-\mu^2)}{z_2^2}\right){\varsigma_2(z_2)}=0
\end{eqnarray}
where we have abbreviated,
\begin{eqnarray}
\label{coeff}
\kappa_1=-\frac{ik_z L}{\sqrt{k_z^2+S_1}}, \qquad
\kappa_2=-\frac{ik_z L}{\sqrt{k_z^2+S_2}}, \qquad 
\mu=\left(\frac{1}{2}+i\sqrt{M^2+L^2}\right)
\end{eqnarray}
Note that $\kappa_{1,2}$ depend upon the sign of $k_z$. From now on we shall only focus on the situation $(M^2+L^2)\gg 1$, for which 
$$\mu\approx i\sqrt{M^2+L^2} \approx i |\mu|$$
 in \ref{coeff}.  This corresponds to either a very strong electric field or a  highly massive fermion, compared to the Hubble constant, or both.   Then the general solutions for \ref{zcom1} are given by,
\begin{eqnarray}
\label{sols1}
{\varsigma_1(z_1)}=C_1 W_{\kappa_1,i|\mu|}(z_1)+D_1M_{\kappa_1,i|\mu|}(z_1),\;\;
{\varsigma_2(z_2)}=C_2 W_{\kappa_2,i|\mu|}(z_2)+D_2M_{\kappa_2,i|\mu|}(z_2)
\end{eqnarray}
where $W$ and $M$ are the Whittaker functions~\cite{AS} and $C_1,\,C_2,\,D_1,\,D_2$ are constants.

Let us now find out the positive frequency `in' modes, i.e. the mode functions whose temporal part behaves as positive frequency plane waves as $\eta \to -\infty$. In this limit we have~\cite{AS},
$$W_{\kappa_1,i|\mu|}(z_1)\sim e^{-2i\eta \sqrt{k_z^2+S_1}}\eta^{\kappa_1}$$
Thus for such modes we must set $D_1=0=D_2$ in \ref{sols1}. Putting things together, we write the two positive frequency `in' mode functions as,
\begin{equation}
\label{in_comp}
\zeta^{\text{in}}_{s,n_L}({\eta,\vec{x}})= e^{-iHLz}e^{i\vec{k_\slashed{y}}\cdot\vec{x}}W_{\kappa_s,i|\mu|}(z_s)h_{n_L}(\overline{y})\,\omega_s \qquad (s=1,2)
\end{equation}
Likewise, since as $\eta \to 0^-$~\cite{AS} $M_{\kappa,i|\mu|}(z_1)\sim \eta^{i|\mu|+1/2},$
the positive frequency `out' modes can be defined with respect to the cosmological time, $t$ $(t=-\ln H\eta/H)$, and we choose them to be 
\begin{equation}
\label{out_comp}
\zeta^{\text{out}}_{s,n_L}({\eta,\;\vec{x}})= e^{-iHLz}e^{i\vec{k_\slashed{y}}\cdot\vec{x}}M_{\kappa_s,i|\mu|}(z_s)h_{n_L}(\overline{y})\,\omega_s \qquad (s=1,2)
\end{equation}
However, recall that the $\zeta$'s appearing in \ref{in_comp} and \ref{out_comp} are not the original Dirac modes, as of  \ref{conf},  \ref{sq_dirac}. We thus have the complete set of positive and  negative frequency in and out modes,
\begin{eqnarray}
\label{mainmodes}
U_{s,n_L}^{\rm \text{in}}&=&\frac{1}{N_s a^{3/2}} \hat{D}\zeta^{\text{in}}_{s,n_L},\qquad V_{s,n_L}^{\rm \text{in}}\;=\; \mathcal{C}(U_{s,n_L}^{\rm \text{in}})^*,\nonumber\\
U_{s,n_L}^{\rm \text{out}}&=&\frac{1}{M_s a^{3/2}}\hat{D}\zeta^{\text{out}}_{s,n_L},\qquad V_{s,n_L}^{\rm \text{out}}\;=\;\mathcal{C}(U_{s,n_L}^{\rm \text{out}})^*, \qquad (s=1,2)
\end{eqnarray}
where $\mathcal{C}= i \gamma^2$ is the charge conjugation matrix. Hence the $V$-modes appearing above are the negative frequency modes. The normalisation constants appearing above are given by
\begin{equation}
     N_1 = e^{\pi |\kappa_1|{\rm sgn}(k_z)/2}, \qquad N_2 = e^{\pi |\kappa_2|{\rm sgn}(k_z)/2}, \qquad M_1=M_2= \sqrt{2|\mu|} e^{\pi |\mu|/2}
\label{nc}
\end{equation}
where the sign dependence of the normalisation constants originates from the sign dependence  of $k_z$ appearing in the parameters $\kappa_s$, \ref{coeff}. 
The explicit form of the mode functions in \ref{mainmodes} and the evaluation of the normalisation constants are discussed in \ref{A}.
It can be checked that these mode functions satisfy the orthonormality relations,
\begin{eqnarray}
&&(U_{s,n_L}^{\rm \text{in}}(x;\vec{k}_{\slashed{y}}),U_{s^{\prime},n_L^{\prime}}^{\text{in}}(x;\vec{k^{\prime}_{\slashed{y}}}))= (V_{s,n_L}^{\rm \text{in}}(x;\vec{k}_{\slashed{y}}),V_{s^,n_L^{\prime}}^{\rm \text{in}}(x;\vec{k^{\prime}_{\slashed{y}}})) = \delta^{2}(\vec{k}_{\slashed{y}}-\vec{k'_{\slashed{y}}})\delta_{n_L n_L^{\prime}}\delta_{ss^{\prime}}\nonumber\\
&&(U_{s,n_L}^{\rm \text{out}}(x;\vec{k}_{\slashed{y}}),U_{s^{\prime},n_L^{\prime}}^{\rm  \text{out}}(x;\vec{k^{\prime}_{\slashed{y}}}))= (V_{s,n_L}^{\rm  \text{out}}(x;\vec{k}_{\slashed{y}}),V_{s^{\prime},n_L^{\prime}}^{\rm  \text{out}}(x;\vec{k^{\prime}_{\slashed{y}}})) = \delta^{2}(\vec{k}_{\slashed{y}}-\vec{k'_{\slashed{y}}})\delta_{n_L n_L^{\prime}}\delta_{s,s^{\prime}}\nonumber
\end{eqnarray}
with all the other inner products vanishing. In terms of these orthonormal modes, we now make the field quantisation, 
\begin{eqnarray}
\psi(\eta,\vec{x}) &&= \sum_{n_L; s=1,2}\int\frac{d^{2}\vec{k}_{\slashed{y}}}{2\pi a^{3/2}} \Bigg[a_{\rm \text{in}}(\vec{k}_{\slashed{y}},s,n_L)U_{s,n_L}^{\rm \text{in}}(x;\vec{k}_{\slashed{y}})+b^{\dagger}_{\rm \text{in}}(\vec{k}_{\slashed{y}},s,n_L)V_{s,n_L}^{\rm \text{in}}(x;\vec{k}_{\slashed{y}})\Bigg] \nonumber\\
&&=\sum_{n_L; s=1,2}\int\frac{d^{2}\vec{k}_{\slashed{y}}}{2\pi a^{3/2}}\Bigg[a_{\rm \text{out}}(\vec{k}_{\slashed{y}},s,n_L)U_{s,n_L}^{\rm \text{out}}(x;\vec{k}_{\slashed{y}})+b^{\dagger}_{\rm \text{out}}(\vec{k}_{\slashed{y}},s,n_L)V_{s,n_L}^{\rm \text{out}}(x;\vec{k}_{\slashed{y}})\Bigg]\nonumber
\label{field}
\end{eqnarray}
where the creation and annihilation operators are assumed to satisfy the usual canonical anti-commutation relations.
Using now the relations between the Whittaker functions~\cite{AS}, 
\begin{eqnarray}
\label{Whitidentity}
  && W_{\kappa,i|\mu|}(z) = \frac{\Gamma(-2i|\mu|)}{\Gamma(1/2-i|\mu|-\kappa)}M_{\kappa,i|\mu|}(z) + \frac{\Gamma(2i|\mu|)}{\Gamma(1/2+i|\mu|-\kappa)}M_{\kappa,-i|\mu|}(z)\nonumber\\
 &&   M_{\kappa,i|\mu|}(z)= 
    -i e^{\pi |\mu|} M_{-\kappa,i|\mu|}(-z)
\end{eqnarray}
into \ref{in_comp}, \ref{out_comp} and \ref{mainmodes}, we find
\begin{equation}
\label{bogo1}
U_{s,n_L}^{\rm \text{in}}(x;\vec{k}_{\slashed{y}})= \frac{M_s}{N_s} \frac{\Gamma(-2i|\mu|)}{\Gamma(1/2-i|\mu|-\kappa_s)} U_{s,n_L}^{\rm \text{out}}(x;\vec{k}_{\slashed{y}}) + i e^{-|\mu|\pi}\frac{M_s}{N_s}  \frac{\Gamma(2i|\mu|)}{\Gamma(1/2+i|\mu|-\kappa_s)} V_{s,n_L}^{\rm \text{out}}(x;\vec{k}_{\slashed{y}})
\end{equation}
Substituting this into \ref{field}, we obtain the Bogoliubov relations
\begin{eqnarray}
\label{in_out}
&& a_{\rm \text{out}}(\vec{k}_{\slashed{y}},s,n_L)=\alpha_s a_{\rm \text{in}}(\vec{k}_{\slashed{y}},s,n_L)-\beta_{s}^{*} b_{\rm \text{in}}^{\dagger}(-\vec{k}_{\slashed{y}},s,n_L)\nonumber\\
&& b_{\rm \text{out}}(\vec{k}_{\slashed{y}},s,n_L)=\alpha_s b_{\rm \text{in}}(\vec{k}_{\slashed{y}},s,n_L)+\beta_{s}^{*} a^{\dagger}_{\rm \text{in}}(-\vec{k}_{\slashed{y}},s,n_L)\qquad (s=1,2)
\end{eqnarray}
The canonical anti-commutation relations ensure,
$|\alpha_s|^2+|\beta_s|^2=1$, for $s=1,2$.

Recalling we are working with $|\mu|\gg1 $, we have 
\begin{equation}
\label{complxbv}
\beta_s\approx  i e^{-|\mu|\pi}\left\vert\frac{M_s}{N_s}\right\vert   \frac{\Gamma(2i|\mu|)}{\Gamma(i|\mu|-\kappa_s)} \qquad (s=1,2)
\end{equation}
We find from the above after using some identities of the gamma function~\cite{AS}, the spectra of pair creation 
\begin{equation}
\label{number_demsity}
|\beta_s|^2_{\pm}\approx e^{-\pi(|\mu|\pm |\kappa_s|)}\frac{\sinh\pi(|\mu|\pm|\kappa_s|)}{\sinh2\pi |\mu|}\qquad (s=1,2)
\end{equation}
where the $\pm$ sign correspond respectively to $k_z > 0$ and  $k_z < 0$, originates from the fact that $\kappa_s$ depend upon the sign of $k_z$, \ref{coeff}. The above expression is formally similar to the case where only a background electric field  is present \cite{Xue:2017cex}. 
The contribution to the particle creation 
from the magnetic field comes solely from the coefficients $\kappa_s$ and there is no contribution of it (i.e., $\kappa_s=0$) if either the electric field is vanishing, Or, the magnetic field strength is infinitely large. Note  also  that if we set $E=0$ in \ref{number_demsity}, we reproduce the well known fermionic blackbody spectra of created particles  with temperature $T_H=H/2\pi$, e.g.~\cite{SHN:2020},
\begin{equation}
|\beta_s|^2_{\pm}\approx \frac{1}{e^{2\pi |\mu|}+1}\qquad (s=1,2)
\label{x}
\end{equation}
where $|\mu| =M=m/H$. The above discussions show that in the absence of an electric field, the magnetic field cannot alter the particle creation rate, as we intuitively anticipated towards the middle of the introduction of this chapter. Finally, we also note from \ref{number_demsity} that since $|\mu|=(M^2+L^2)^{1/2}$, for $E\neq 0$, and even if $B \to \infty$, the particle creation due to the electric field does not completely vanish, unlike that of the flat spacetime~\ref{Background magnetic field and quantum correlations in the Schwinger effect}. Once again, this should correspond to the fact that the mutual separation of the pairs created  by the electric field as they propagate is also happening here due to the expanding gravitational field of the de Sitter, upon which the magnetic field has no effect whatsoever.

Since the parameters $M$ and $L$ denote the dimensionless rest mass and the strength of the electric field  (cf., discussion below \ref{dirac4}), let us consider in the following two qualitatively distinct cases, keeping in mind $(M^2+L^2)\gg 1$.\\

\noindent
\textbf{Case $1$}: $M^2\gg1\quad {\rm and}\quad  M^2\gg L^2$.  Hence in this case particle creation is happening chiefly due to the background spacetime curvature. We have from \ref{number_demsity} in this limit,
	\begin{equation}
	\label{case1}
	|\beta_s|^2_{\pm} \approx e^{-2 \pi M}\left(1- e^{- 2 \pi M}e^{\mp 2\pi|\kappa_s|}+{\cal O}(e^{-4\pi M})\right) \,	\end{equation}
Thus $|\beta_s|^2_+> |\beta_s|^2_-$. Although the electromagnetic field is weak here and hence they would have little effects on the particle creation, note in particular from the expression of $\kappa_s$ that if we keep the electric field strength and $k_z$ fixed,  $|\beta_s|^2_+$ decreases whereas $|\beta_s|^2_-$ increases with the magnetic field strength, and for extremely high $B$-value, the particle creation rate coincides to that of only due to the spacetime curvature. \\

\noindent
	\textbf{Case $2$}: $L^2\gg1\quad {\rm and}\quad  L^2 \gg M^2$. Hence in this case particle creation is happening chiefly due to the background electric field. We have from \ref{number_demsity},
	\begin{equation}
	\label{case2}
	|\beta_s|^2_{\pm} \approx e^{-2 \pi L} \left(1-e^{-2 \pi L}e^{\mp 2 \pi |\kappa_s|}+ {\cal O}(e^{-4\pi L})\right) \,	\end{equation}
Thus in this case  also $|\beta_s|^2_+>|\beta_s|^2_-$, and $|\beta_s|^2_+$ decreases whereas $|\beta_s|^2_-$ increases with the magnetic field strength, while the other parameters are held fixed. We wish to focus only on   $|\beta_s|^2_-$ in the following. In our computation, we shall often encounter  the complex $\beta_{s}$ value. Hence instead of using \ref{case1} or \ref{case2}, we shall work with  \ref{complxbv}, by taking numerical values of the parameters  appropriate for the particular case.

Subject to the field quantisation in \ref{field}, the `in' and the `out' vacua are defined as,
$$a_{\rm \text{in}}(\vec{k}_{\slashed{y}},s,n_L) |0\rangle^{\rm \text{in}}=0 = b_{\rm in}(\vec{k}_{\slashed{y}},s,n_L) |0\rangle^{\rm in}\quad {\rm and }\quad a_{\rm \text{out}}(\vec{k}_{\slashed{y}},s,n_L) |0\rangle^{\rm \text{out}}=0 = b_{\rm \text{out}}(\vec{k}_{\slashed{y}},s,n_L) |0\rangle^{\rm \text{out}}$$

Thus the Bogoliubov relations  \ref{in_out} imply a (normalised) squeezed state expansion between the `in' and `out' states for a given momentum, 
\begin{eqnarray}
\label{vaccum1}
|0_{k}\rangle^{\rm \text{in}}\;
= \left(\alpha_1 |0_{k}^{(1)}\,0_{-k}^{(1)}\rangle^{\rm \text{out}} +\beta_{1} |1_{k}^{(1)}\,1_{-k}^{(1)}\rangle^{\rm \text{out}}\right )\otimes  \left(  \alpha_2 |0_{k}^{(2)}\,0_{-k}^{(2)}\rangle^{\rm \text{out}} +\beta_{2} |1_{k}^{(2)}\,1_{-k}^{(2)}\rangle ^{\rm \text{out}}\right)
\end{eqnarray}
As we have discussed above, since we shall be working only with the `$-$' sign of \ref{number_demsity}, $\beta_s$ and $\alpha_s$ appearing above are understood as  $\beta_{1-}$, $\beta_{2-}$ and $\alpha_{1-}$ and $\alpha_{2-}$, respectively.

The excited `in' states can be written in terms of the `out' states by applying the in-creation operators on the left hand side of \ref{vaccum1} and then using the Bogoliubov relations  \ref{in_out} on its right hand side.

Finally we note that  the $s=1,2$ sectors are factorised in \ref{vaccum1}, leading to 
$$|0_{k}\rangle^{\rm \text{in}}  = |0_{k}^{(1)}\rangle^{\rm \text{in}}\otimes |0_{k}^{(2)}\rangle^{\rm \text{in}}$$
Thus for simplicity, we can work only with a single sector, say $|0_{k}^{(1)}\rangle^{\rm \text{in}}$, of the in-vacuum. 

Being equipped with these, we are now ready to go to the computation of the Bell violation. However, before we do that, we wish to compute the entanglement entropy associated with the vacuum state.

\noindent

\section{Entanglement entropy of the vacuum}
\label{EE}
We wish to compute the entanglement entropy for the state $|0_{k}^{(1)}\rangle^{\rm \text{in}}$, defined in \ref{VNE} of \ref{Entanglement entropy}.  The density matrix  corresponding to this state is pure, $\rho_0= |0_{k}^{(1)}\rangle^{\rm \text{in}\,\,\text{in}}\langle0_{k}^{(1)}|$. Using \ref{vaccum1}, we write down $\rho_0$ in terms of the out states, which contain both $k$ and $-k$ degrees of freedom. The reduced density matrix corresponding to the $k$ sector (say, particle) is given by, 
$
	\rho_k
=
	\text{Tr}_{-k}
	\rho_0
=
	|\alpha_1|^{2} |0_{k}^{(1)}\rangle^{\rm \text{out}\,\, \text{out}}\langle 0_{k}^{(1)}| + |\beta_1|^{2} |1_{k}^{(1)}\rangle^{\rm \text{out}\,\, \text{out}}\langle 1_{k}^{(1)}|
$, and hence the entanglement entropy
is given by 
\begin{equation}
    \label{entropy}
	S_k =
	-
	\text{Tr}_k
	\rho_k
	\ln \rho_k
=
	-   \left[ \ln	(1-|\beta_1|^{2})    +|\beta_1|^2  \ln \frac{|\beta_1|^{2}}{1-|\beta_1|^2}\right]
\end{equation}
We also have $S_k=S_{-k}$, as we are dealing with a pure state.

Since we are chiefly interested here in the effect of the magnetic field strength, let us extract a dimensionless quantity from \ref{coeff},
$$Q=\frac{2en_LB}{k^2_z}$$
\begin{figure}[h]
	\centering
		\includegraphics[scale=.53]{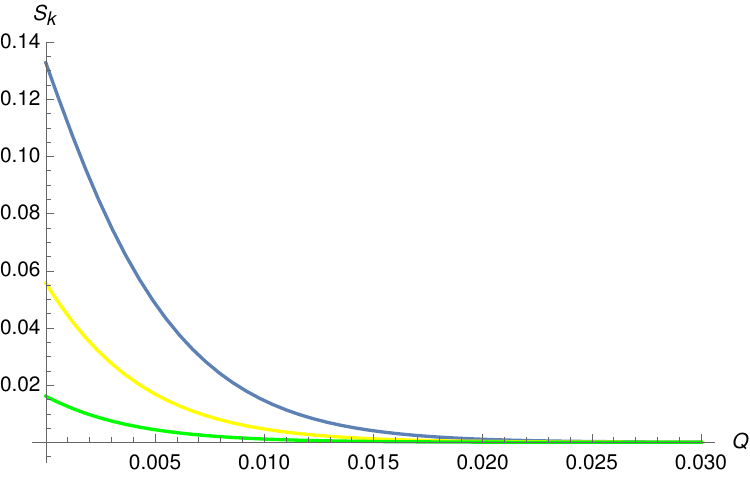}\hspace{1.0cm}
		\includegraphics[scale=.53]{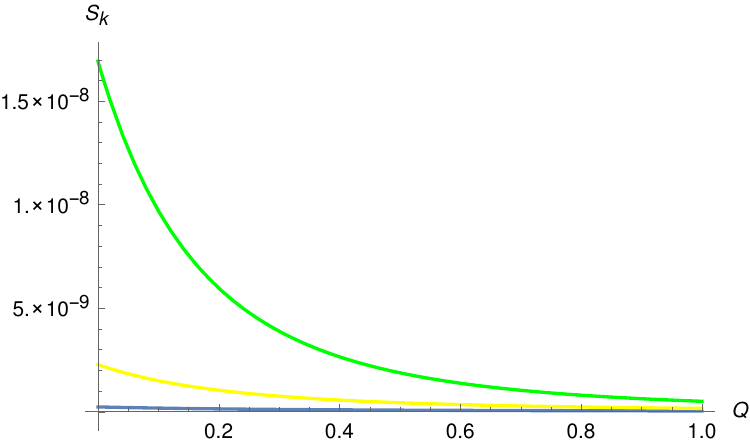} 
		\caption{\it{ \small We have plotted entanglement entropy \ref{entropy} corresponding to the vacuum state with respect to the parameter $Q=2n_LeB/k^2_z$. The left plot corresponds to the case $L^2\gg M^2$, where we have taken $L=100$ and different curves correspond to different $M$ values (blue $M=10$, yellow $M=12$ and green $M=16$). The right plot corresponds to $M^2\gg L^2$, where we have taken $M=5$, and different curves correspond to different $L$ values (blue $L=1$, yellow $L=1.5$ and green $L=2$). For a given mode, the entanglement entropy decreases monotonically with increasing $B$ for both cases due to the decrease in particle creation. See main text for details.} 
		}
		\label{fig:EEforVacuum11}
\end{figure}
The $Q$-dependence of $S_k$ is depicted in \ref{fig:EEforVacuum11}, for the two cases (`large' ($L^2\gg M^2$) and `small' ($L^2 \ll M^2$) electric fields), discussed in the preceding section. For a given mode (i.e. $n_L,\,k_z$ fixed), thus, the increase in $Q$ corresponds to the increase in the $B$ value. As shown in the figure, the entanglement entropy decreases monotonically with the increase in the magnetic field strength. This corresponds to the fact that the vacuum entanglement entropy originates from the pair creation, which decreases  with increasing $B$ for both the cases we have considered. This result is qualitatively similar to that of the flat spacetime, \ref{fig:EEforVacuum1}, discussed in the preceding Chapter. However, we note also that, unlike the previous case, the entanglement entropy does not vanish here, no matter how large $B$ is. This corresponds to the fact that the magnetic field cannot affect the pair creation due to the gravitational field at all and hence, $\beta_s \neq 0$ always (cf., the discussion below \ref{number_demsity}). A non-vanishing pair creation always refers to a non-vanishing entanglement entropy.

\noindent
\section{The violation of the Bell  inequalities}\label{BV}

We wish to investigate below Bell's inequality violation for the vacuum as well as some maximally entangled initial states. In order to do so, we will use the prescription described in \ref{The Bell Inequality Violation}. 
\noindent
\subsection{Bell violation for the vacuum state}
We wish to find out the expectation value of $\mathcal{B}$, defined in \ref{op2} of \ref{The Bell Inequality Violation}, with respect to the vacuum state $|0_{k}^{(1)} \rangle^{\rm  \text{in}}$, given at the end of \ref{S2}. In order to do this, one usually introduces the pseudospin operators, measuring the parity in the Hilbert space along different axes, for example, ~\cite{bell:2017} and references therein. These operators for fermionic systems with eigenvalues $\pm 1$ are defined as,
\begin{equation}
\label{unitvector}
\mathbf{\hat{n}}.\mathbf{S} = S_{z}\cos\theta + \sin\theta(e^{i\phi}S_{-}+e^{-i\phi}S_{+}), 
\end{equation}
where $\mathbf{\hat{n}} = (\sin\theta \cos\phi,\sin\theta \sin\phi,\cos\phi)$ is a unit vector in the Euclidean $3$-space. The action of the operators $S_z$ and $S_{\pm}$ are defined on the `out states' as,
\begin{equation}
\label{spin}
S_z | 0\rangle = - | 0\rangle, \quad S_z | 1\rangle =  | 1\rangle, \quad S_+| 0\rangle =| 1\rangle, \quad S_+| 1\rangle =0, \quad S_-| 0\rangle =0, \quad S_-| 1\rangle =|0\rangle
\end{equation}
Without any loss of generality, we take the operators to be confined to the $x-z$ plane so that we may set  $\phi = 0 $ in \ref{unitvector}. We may then take in \ref{op2}, $\mathit{O_{i}}=\;\mathbf{\hat{n}}_i\cdot\mathbf{S}$  and $\mathit{O}'_i=\mathbf{\hat{n}'}_i\cdot\mathbf{S}$ 
with $i=1,2$. Here $\mathbf{\hat{n}}_i$ and $\mathbf{\hat{n}'}_i$ are two pairs of unit vectors in the Euclidean $3$-space, characterised by their angles with the $z$-axis, $\theta_{i}$, $\theta'_{i}$ (with $i=1,2$) respectively. 

\begin{figure}[h]
\centering
\includegraphics[scale=0.65]{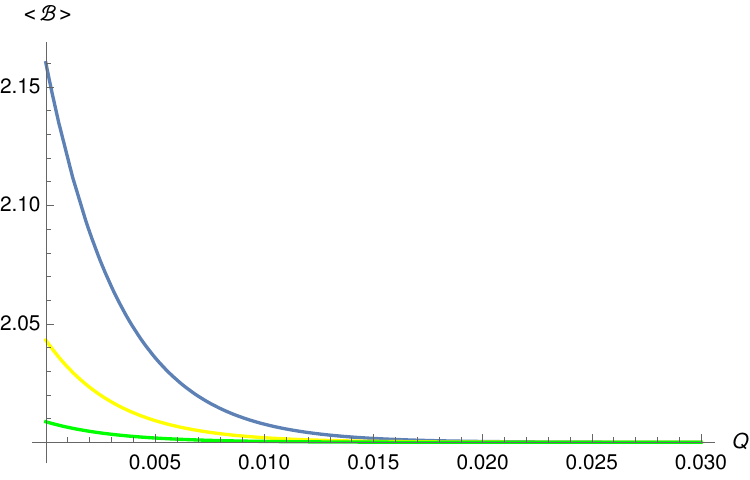}\hspace{1.0 cm}
\caption{\it{\small We have plotted Bell's violation $\langle\mathcal{B}\rangle_{\rm \text{max}}$ \ref{bvpm}, corresponding to the vacuum state, with respect to the parameter   $Q=2en_LB/k_z^2$. We have plotted only the case $L^2\gg M^2$, for the other case ($M^2\gg L^2$) does not show any significant violation or numerical variation.   We have taken $L=100$, and different curves correspond to different $M$ values (blue $M=10$, yellow  $M=12$ and green  $M=16$). }}
\label{fig:bellvofvacuum}
\end{figure}

Using the above constructions, and the squeezed state expansion of \ref{vaccum1}  and also the operations \ref{spin} defined on the out states,  the desired expectation value is given by,
\begin{equation}
\label{B2}
^{\rm  \text{in}}\langle 0^{(1)}_{k}|\mathcal{B}|0_{k}^{(1)}\rangle^{\rm  \text{in}} = [E(\theta_{1},\theta_{2})+E(\theta_{1},\theta'_{2})+E(\theta'_{1},\theta_{2})-E(\theta'_{1},\theta'_{2})]
\end{equation}
where, $\mathit{O}_i$ and $\mathit{O}'_i$ are assumed to operate respectively on the $k$ and $-k$ sectors of the out states in \ref{vaccum1}, and 
$$E(\theta_{1},\theta_{2})
=
\cos\theta_{1} \cos\theta_{2} + 2|\alpha_{1}\beta_{1}| \sin\theta_{1} \sin\theta_{2}$$
Choosing now for example, $\theta_{1}=0,\,\theta'_{1}=\pi/2$ and $\theta_{2}=-\theta'_{2}$, we have from \ref{B2},
\begin{equation}
^{\rm \text{in}}\langle 0_{k}^{(1)}|\mathcal{B}|0_{k}^{(1)}\rangle^{\rm  \text{in}} = 2(\cos\theta_{2}+ 2|\alpha_{1}\beta_{1}|\sin\theta_{2})
\label{bvpm}
\end{equation}
The above expression maximises at $\theta_{2} =\tan^{-1}(2|\alpha_{1}\beta_{1}|)$, so that the above expectation value becomes
$$\langle\mathcal{B}\rangle_{\rm \text{max}} = 2\left(1+ 4|\alpha_1 \beta_1|^2 \right)^{1/2}$$
Thus $\langle\mathcal{B}\rangle_{\rm \text{max}}\geq2$, and hence there is always Bell inequality violation as long as $\beta_1 \neq 0$, which is always the case in the de Sitter spacetime (cf., the discussion below \ref{number_demsity}).
We have plotted $\langle\mathcal{B}\rangle_{\rm \text{max}}$ in~\ref{fig:bellvofvacuum}  with respect to the parameter $Q=2en_LB/k_z^2$ as earlier.  We have considered only the case of a strong electric field, $L^2\gg M^2$, for the other case, the violation is tiny and also it does not show any significant numerical variation. As of the vacuum entanglement entropy, \ref{fig:EEforVacuum11}, the Bell violation decreases monotonically with the increasing magnetic field (for a given mode) and asymptotes to values close to $2$ due to the suppression of the particle creation by the magnetic field.

Note that the vacuum state is pure. Instead of the vacuum, if we consider a pure but maximally entangled initial state, make its squeezed state expansion, and then trace out some parts of it in order to construct a bipartite subsystem, the resulting density matrix turns out to be mixed. The above construction is valid for pure ensembles only, and one requires a different formalism to deal with mixed ensembles, for example,~\cite{bell_3, nper:2011}. We wish to study two such cases below in order to demonstrate their qualitative differences with the vacuum case. 
\noindent
\subsection{Bell violation for maximally entangled initial states} \label{bvme}

We wish to consider maximally entangled initial states (corresponding to two fermionic fields) in the following.   For computational simplicity, we assume that both the fields have the same rest mass, and we consider modes in which their momenta along the $z$-direction  and the Landau levels are the same.

The density matrix corresponding to the initial state can be expanded into the out states via \ref{vaccum1}, and then any two degrees of freedom are traced out in order to construct a bipartite system. The resulting reduced density matrix turns out to be mixed. For such a system, a procedure to compute the Bell inequality has been proposed  in~\cite{seprability}. According to this, any density matrix ($\rho$) in a Hilbert-Schmidt basis can be decomposed as
\begin{equation}
    \label{decomrho}
\rho=\frac{1}{4}\Big(I\otimes I+\mathit{O_i}\otimes I+I\otimes \mathit{O}'_i+\sum_{i,j=1}^{3}T_{ij}\sigma_i \otimes \sigma_j\Big)
\end{equation}
where $I$ stands for the identity operator and $\sigma_i$'s are the Pauli matrices. Here, $\mathit{O_{i}}=\;\mathbf{\hat{n}}_i\cdot\mathbf{S}$  and $\mathit{O}'_i=\mathbf{\hat{n}'}_i\cdot\mathbf{S}$  are the operators along specific directions in the $3$-dimensional Euclidean space as defined earlier in \ref{The Bell Inequality Violation}. In the decomposition \ref{decomrho}, the middle two terms describe the local behaviour of the state, and the last term describes the correlations of the state. 

We wish to compute the maximum average value of the Bell operator defined in \ref{op2} with respect to the density matrix $\rho$, \ref{decomrho}. Following \cite{bellmix}, one can choose
\begin{equation}
    \left(\mathbf{\hat{n}}_2+\mathbf{\hat{n}'}_2\right)= 2 \cos{\theta} \mathbf{\hat{e}},\qquad \qquad\left(\mathbf{\hat{n}}_2-\mathbf{\hat{n}'}_2\right)= 2 \cos{\theta} \mathbf{\hat{e}'}
\end{equation}
where $\mathbf{\hat{e}}$ and $\mathbf{\hat{e}'}$ are a pair of orthonormal vectors, and $\theta \in [0,\pi/2]$. Next, we choose $\mathbf{\hat{n}}_1$ and $\mathbf{\hat{n}'}_1$ in the direction of $T\mathbf{\hat{e}}$ and $T\mathbf{\hat{e}'}$, respectively. $T$ is called the correlation matrix for the general decomposition of $\rho$, \ref{decomrho}, and it can be easily checked that its elements are given by  $T_{ij} \equiv {\rm Tr}[\rho \sigma_{i}\otimes \sigma_{j}]$. On substituting all these choices of unit vectors in \ref{op2}, the expectation value of the Bell operator becomes
\begin{equation}
    \label{Btheta}
   \langle \mathcal{B}\rangle=2(||T\mathbf{\hat{e}}||\cos \theta+||T\mathbf{\hat{e}'}||\sin \theta)
\end{equation}
where $||\;||$ denotes the Euclidean norm in $3-$dimensional space and defined as $||T\mathbf{\hat{e}}||=(\mathbf{\hat{e}},\;T^T T\mathbf{\hat{e}})$, where the round brackets correspond to the Euclidean scalar product and $T^T$ stands for transpose of $T$. On maximising \ref{Btheta} with respect to $\theta$, we have
\begin{equation}
    \label{Btheta1}
\langle \mathcal{B}\rangle=2\sqrt{||T\mathbf{\hat{e}}||^2+||T\mathbf{\hat{e}'}||^2}
\end{equation}
Let us now maximize the above average with respect to the choices of the unit vectors, $\mathbf{\hat{e}}$ and $\mathbf{\hat{e}'}$. We choose them as the eigenvectors corresponding to the maximum eigenvalues of the matrix $U$, defined as
$$U=(T)^{\rm T}T$$
which is a symmetric matrix and can be diagonalised.  Having all these parameters fixed, thus the maximum value of the Bell operator is given by
\begin{equation}
    \langle \mathcal{B}_{\rm \text{max}}\rangle\;=\; 2\sqrt{\lambda_{1}+\lambda_{2}},
    \label{bv}
\end{equation}
where $\lambda_{1}$ and $\lambda_{2}$ are the two largest eigenvalues of the $(3\times 3)$ matrix $U$. Here also, the violation of the Bell inequality as earlier will correspond to $ \langle \mathcal{B}_{\rm max}\rangle>2$ in \ref{bv}. \\

We begin by considering the initial state,
\begin{eqnarray}
\label{Bell-1}
|\psi\rangle&=&\frac{|0_p 0_{-p}0_k0_{-k}\rangle^{\text{in}}+|1_p0_{-p}0_{k}1_{-k}\rangle^{\text{in}}}{\sqrt{2}}
\end{eqnarray}
In the four entries of a ket above, the first pair of states corresponds to one fermionic field, whereas the last pair corresponds to the other. The $\pm$-sign in front of the momenta stands respectively for the particle and antiparticle degrees of freedom.  

Recall that we are assuming the created particles have the same rest mass, and we are working with modes for which  the Landau levels and the  $k_z$ values for both fields are coincident. Using \ref{in_out}, \ref{vaccum1} and keeping in my the Pauli exclusion principle,  we re-express \ref{Bell-1} in the out basis as
\begin{eqnarray}
|\psi\rangle=\frac{(\alpha_{1}|0_p0_{-p}\rangle^{\rm \text{out}}+\beta_{1}|1_p1_{-p}\rangle^{\rm \text{out}})(\alpha_{1}|0_k0_{-k}\rangle^{\rm \text{out}}+\beta_{1}|1_k1_{-k}\rangle^{\rm \text{out}})+|0_p1_{-p}\rangle^{\rm \text{out}}|1_{k}0_{-k}\rangle^{\rm \text{out}}}{\sqrt{2}}\nonumber\\
\end{eqnarray}

We shall focus below only on the correlations between the particle-particle and the particle-antiparticle sectors corresponding to the density matrix of the above state. Accordingly, tracing out first the antiparticle-antiparticle degrees  of freedom of the density matrix $\rho^{(0)}\;=\;|\psi\rangle \langle \psi|$, we construct the reduced density matrix for the particle-particle sector,  
\begin{eqnarray}
\rho_{k,p}^{0}=\;\text{Tr}_{-k,-p}(\rho^{(0)})\;=\;\frac{1}{2}\left(\begin{array}{cccc}
|\alpha_{1}|^{4}&0&0&0\\
0&|\alpha_1\beta_1|^2&(\alpha_1\beta_1)^* &0\\
0&\alpha_1\beta_1&|\alpha_1\beta_1|^2+1&0\\
0&0&0&|\beta_{1}|^{4}\\
\end{array}\right)
\label{pp'}
\end{eqnarray}
Likewise, we obtain the reduced density matrix for the particle-anti-particle sector,
\begin{eqnarray}
\rho_{p,-k}^{0}\;=\;\text{Tr}_{-p,k}(\rho^{(0)})=\frac{1}{2}\left( {\begin{array}{cccc}
|\alpha_{1}|^{4}&0&0&(\alpha_{1}^{*})^2\\
0&|\alpha_1\beta_1|^2&0&0\\
0&0&|\alpha_1 \beta_1|^2&0\\
\alpha_{1}^2&0&0&|\beta_{1}|^{4}+1\\
\end{array}}\right)
\label{ap'}
\end{eqnarray}

The correlation matrices corresponding to \ref{pp'} and  \ref{ap'} are respectively given by,
\begin{eqnarray}
\label{correlation1}
T(\rho_{k,p}^{0})=\left( {\begin{array}{cccc}
	{\rm Re}(\alpha_1 \beta_1)&0&0\\
	0&-{\rm Re}(\alpha_1 \beta_1)&0\\
	0&0&\frac{1}{2}\left(|\alpha_{1}|^{4}-2|\alpha_1|^2|\beta_1|^2-1+|\beta_{1}|^4\right)\\
	\end{array}}\right)
\end{eqnarray}
and
\begin{eqnarray}
\label{correlation2}
T(\rho_{p,-k}^{0})=\left( {\begin{array}{cccc}
	{\rm Re}(\alpha_1^2)&0&0\\
	0&{\rm Re}(\alpha_1^2)&0\\
	0&0&\frac{1}{2}\left(|\alpha_{1}|^{4}-2|\alpha_1|^2|\beta_1|^2 +1+|\beta_1|^4\right)\\
	\end{array}}\right)
\end{eqnarray}
Using \ref{correlation1} and \ref{correlation2}, we compute the matrices $U(\rho_{k,p}^{0})=(T(\rho_{k,p}^{0}))^{\rm T}T(\rho_{k,p}^{0})$, and  $U(\rho_{p,-k}^{0})=(T(\rho_{p,-k}^{0})^{\rm T}T(\rho_{p,-k}^{0})$.   \ref{bv} yields then after a little algebra 
\begin{equation}
\label{Bell0kp}
     \langle \mathcal{B}^{0}_{kp}\rangle_{\rm \text{max}} = 2\sqrt{2}\,{\rm Re}(\alpha_1 \beta_1)
\end{equation}
and
\begin{equation}
\label{Bell0p-k}
     \langle \mathcal{B}^{0}_{p-k}\rangle_{\rm \text{max}} = 2\sqrt{({\rm Re}(\alpha_1^2))^2+\left(1-2|\alpha_1\beta_1|^2\right)^2}
\end{equation}

\begin{figure}[h]
\centering
\includegraphics[scale=0.45]{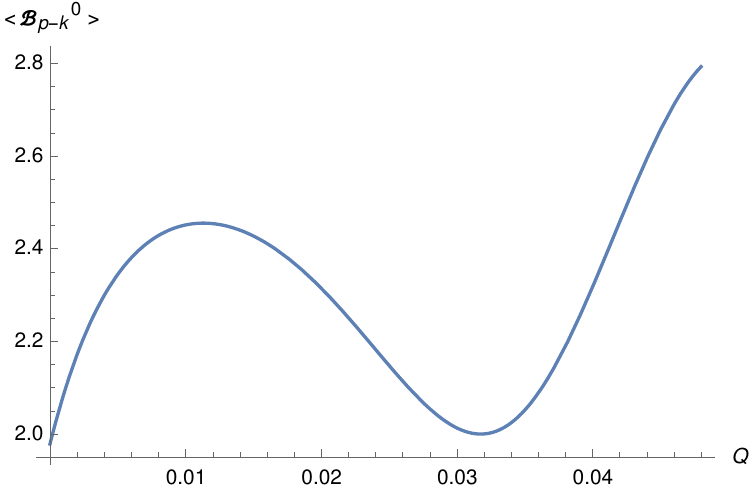}\hspace{1.0cm}
\includegraphics[scale=0.45]{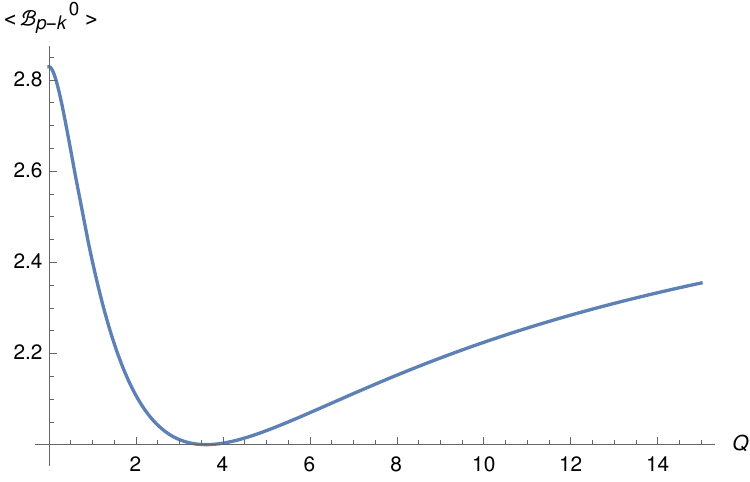}
\caption{\it{\small Bell violation for the particle-antiparticle sector $\rho_{p,-k}^{0}$, \ref{ap'}, corresponding to the initial state in \ref{Bell-1}.  We have plotted  \ref{Bell0p-k} with respect to the parameter $Q=2n_LeB/k_z^2$.  The left plot corresponds to $L^2\gg M^2$  ($L=100$ and $M=10$), whereas the right one corresponds to  $M^2\gg L^2$ ($M=5$ and $L=1$). $\langle \mathcal{B}^{0}_{p-k}\rangle_{\rm max}>2$ corresponds to the Bell violation. The choice of the fixed parameters $L$ and $M$ totally determines the dip in the above figures, and the non-monotonic behaviour is caused by the mixed structure of the density matrix.}}

\label{fig:rhop-k0}
\end{figure}

 We have plotted $\langle \mathcal{B}^{0}_{p-k}\rangle_{\rm \text{max}}$ in~\ref{fig:rhop-k0} with respect to the parameter $Q=2neB/k_z^2$ as earlier, depicting the Bell violation ($\langle \mathcal{B}^{0}_{p-k}\rangle_{\rm \text{max}}>2$) for both strong and weak electric fields.  $\langle \mathcal{B}^{0}_{kp}\rangle_{\rm \text{max}}$, on the other hand, does not show any such violation. This one again depicts the fact that for a very heavy rest mass field, the pair production is suppressed and as a consequence, nothing very non-trivial occurs. 

\bigskip
We next consider another maximally entangled state given by,
\begin{eqnarray}
\label{Bell-2}
|\chi\rangle=\frac{|1_p 0_{-p}0_k0_{-k}\rangle^{\text{in}}+|0_p0_{-p}1_{k}0_{-k}\rangle^{\text{in}}}{\sqrt{2}}
\end{eqnarray}

Following similar steps as described above, by partially tracing out  the original density matrix $\rho^{(1)}=|\chi\rangle \langle \chi |$, we have the mixed bipartite density matrices respectively for the particle-particle and the particle-antiparticle sectors,
\begin{equation}
\rho_{k,p}^{1}\;=\;\text{Tr}_{-k,-p}(\rho^{(1)}) \qquad {\rm and} \qquad 
\rho_{p,-k}^{1}\;=\;\text{Tr}_{-p,k}(\rho^{(1)}),
\label{z}
\end{equation}
which respectively yield the expressions,
\begin{equation}
     \langle \mathcal{B}^{1}_{kp}\rangle_{\rm \text{max}} = 2\sqrt{2}|\alpha_1|^2 \qquad 
{\rm and} \qquad 
     \langle \mathcal{B}^{1}_{p-k}\rangle_{\rm \text{max}} = 2\sqrt{2}\,{\rm Re}(\beta_1\alpha^{*}_1)
     \label{Bell1p-k}
\end{equation}

\begin{figure}[h]
	\begin{center}
		\includegraphics[scale=.60]{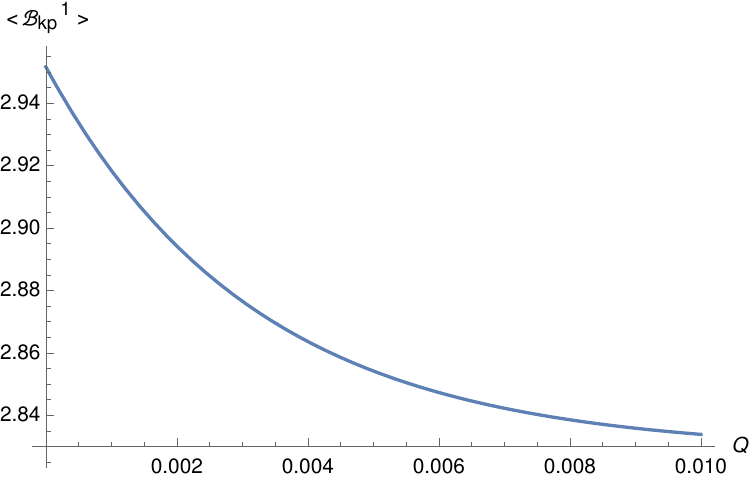}
		\caption{ \it{\small Bell violation for $\rho_{k,p}^{1}$, \ref{z}, corresponding to the initial state in \ref{Bell-2}. We have plotted $\langle \mathcal{B}^{1}_{kp}\rangle$, \ref{Bell1p-k},  with respect to the parameter $Q=2n_LeB/k_z^2$, for $L^2\gg M^2$ ($L=100$ and $M=10$). The other case, $M^2 \gg L^2$, does show Bell violation, but there is no significant numerical variation. Note that in contrast to \ref{fig:rhop-k0}, the behaviour is monotonic here, and qualitatively, it rather resembles the vacuum case, \ref{fig:bellvofvacuum}.
		}}
		\label{fig:rhokp1}
		\end{center}
\end{figure}

We have plotted $ \langle \mathcal{B}^{1}_{kp}\rangle_{\rm \text{max}}$ in \ref{fig:rhokp1} with respect to the parameter $Q$ for strong electric field, $L^2 \gg M^2$. For $M^2\gg L^2$, we also have a violation; however, it does not show any significant numerical variation. This again corresponds to the fact when the field rest mass is very high, the pair creation is suppressed, and hence nothing very non-trivial happens. This again relates to the fact that pair production is diminished when the field rest mass is very high, so nothing very non-trivial occurs.

 On the other hand, we find no violation for the particle-anti-particle sector, $\langle \mathcal{B}^{1}_{p-k}\rangle_{\rm \text{max}}$.
\noindent
\subsection{Mutual information for maximally entangled initial states} \label{MI}
\noindent
Following the definition given in \ref{MIBP} of \ref{The mutual Information}, we compute the mutual information for the different sectors corresponding to the states \ref{Bell-1} and \ref{Bell-2}.

\begin{figure}[h]
    \centering
\includegraphics[scale=0.65]{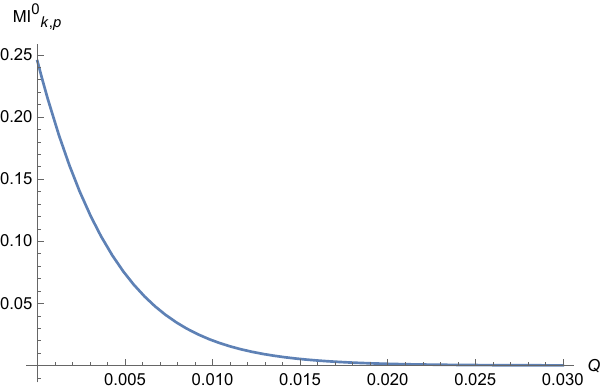}\hspace{1.0cm}
\includegraphics[scale=0.65]{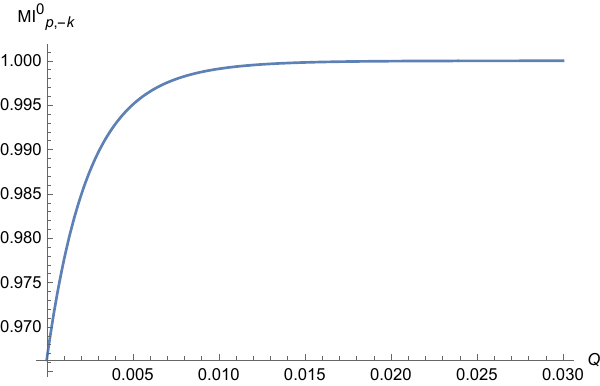 }
    \caption{\it{\small Mutual information for the particle-particle sector $\rho_{k,p}^{0}$ (left) and  particle-antiparticle sector $\rho_{p,-k}^{0}$ (right), corresponding to the initial state in \ref{Bell-1}.  We have plotted the mutual information for $\rho_{k,p}^{0}$ and $\rho_{p,-k}^{0}$ ($MI^{0}_{k,p}$ and $MI^{0}_{p,-k}$ respectively) concerning to the strength of parameter $Q=2n_LeB/k_z^2$. These plots correspond to $L^2\gg M^2$ ($L=100$ and $M=10$). For the other case, $M^2 \gg L^2$, the mutual information does not show any significant numerical variation with respect to parameter $Q$.}}
    \label{MIlarge0}
\end{figure}

The mutual information for particle-antiparticle and particle-particle sectors represented by density matrices $\rho^0_{p,-k}$ and $\rho^0_{k,p}$, respectively, are given by
\begin{equation}
    \begin{split}
    MI^{0}_{p,-k}=-| \alpha_1 | ^2 \log _2\frac{| \alpha_1 | ^2}{2}-| \beta_1 | ^2 \log _2\frac{| \beta_1 | ^2}{2}| +|\alpha_1 | ^2 | \beta_1 | ^2 \log _2\frac{| \alpha_1 | ^2 | \beta_1 | ^2}{2}\\+\frac{1+| \alpha_1 | ^4+| \beta_1 | ^4-\sqrt{| \alpha_1 | ^8+\left(| \beta_1 | ^4+1\right)^2-2 | \alpha_1 | ^4 \left(| \beta_1 | ^4-1\right)}}{4}\\\times\log_2\left(\frac{1+| \alpha_1 | ^4+| \beta_1 | ^4-\sqrt{| \alpha_1 | ^8+\left(| \beta_1 | ^4+1\right)^2-2 | \alpha_1 | ^4 \left(| \beta_1 | ^4-1\right)}}{4}\right)\\+\frac{1+| \alpha_1 | ^4+| \beta_1 | ^4+\sqrt{| \alpha_1 | ^8+\left(| \beta_1 | ^4+1\right)^2-2 | \alpha_1 | ^4 \left(| \beta_1 | ^4-1\right)}}{4}\\\times\log_2\left(\frac{1+| \alpha_1 | ^4+| \beta_1 | ^4+\sqrt{| \alpha_1 | ^8+\left(| \beta_1 | ^4+1\right)^2-2 | \alpha_1 | ^4 \left(| \beta_1 | ^4-1\right)}}{4}\right)
    \end{split}
    \label{mirho0k-p}
\end{equation}
\begin{equation}
    \begin{split}
      MI^{0}_{k,p}=-| \beta_1 | ^2 \log _2\frac{| \beta_1 | ^2}{2} -| \alpha_1 | ^2 \log _2\frac{| \alpha_1 | ^2}{2}+\frac{| \beta_1 | ^4}{2}  \log _2\frac{| \beta_1 | ^4}{2} +\frac{| \alpha_1 | ^4}{2}  \log _2\frac{| \alpha_1 | ^4}{2} \\+\frac{1+2 | \alpha | ^2 | \beta | ^2-\sqrt{4 | \alpha | ^2 | \beta | ^2+1}}{4} \log_2\left(\frac{1+2 | \alpha | ^2 | \beta | ^2-\sqrt{4 | \alpha | ^2 | \beta | ^2+1}}{4}\right) \\+\frac{1+2 | \alpha | ^2 | \beta | ^2+\sqrt{4 | \alpha | ^2 | \beta | ^2+1}}{4} \log_2\left(\frac{1+2 | \alpha | ^2 | \beta | ^2+\sqrt{4 | \alpha | ^2 | \beta | ^2+1}}{4}\right)
    \end{split}
    \label{mirho0kp}
    \end{equation}
We have shown the variation of  \ref{mirho0k-p} (right) and \ref{mirho0kp} (left) with respect to the parameter $Q$, for $L^2 \gg M^2$ ($L=100$ and $M=10$) case, in \ref{MIlarge0}. For the particle-particle sector, the mutual information monotonically decreases with the increase in $Q$ and has a small value in contrast to the particle-antiparticle sector, and at a large $Q$ limit, it vanishes. However, for the particle-antiparticle sector \ref{ap'}, unlike the Bell inequality, the mutual information monotonically increases with the increase in $Q$ and converges to unity at large $Q$ values.  For the other case, $M^2 \gg L^2$ $(M=5\; \text{and} \; L=1)$, the mutual information does not show any significant numerical variation with respect to the parameter $Q$ for both sectors. Its value is approximately zero and unity for particle-particle and particle-antiparticle sectors, respectively. This once again corresponds to the fact when the field rest mass is very high, the pair creation is suppressed, and hence nothing very non-trivial happens.

\begin{figure}
    \centering
\includegraphics[scale=0.65]{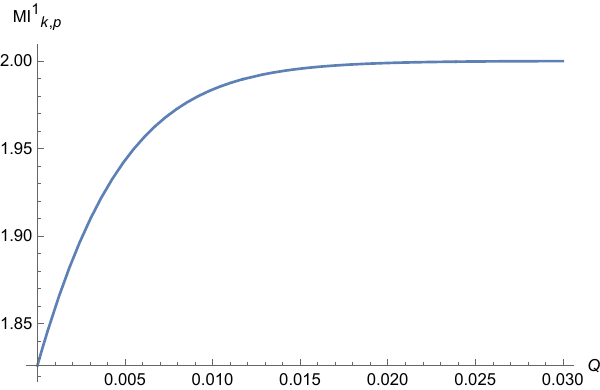}\hspace{1.0cm}
\includegraphics[scale=0.65]{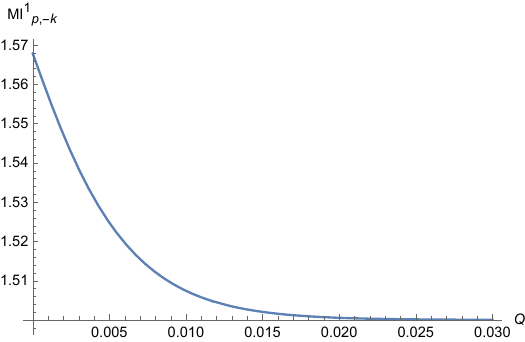}
    \caption{\it{\small Mutual information for the particle-particle sector $\rho_{k,p}^{1}$ (left) and particle-antiparticle sector $\rho_{p,-k}^{1}$ (right)  corresponding to the initial state in \ref{Bell-2}.  We have plotted the mutual information for $\rho_{k,p}^{1}$  ($MI^{1}_{k,p}$) and $\rho_{p,-k}^{1}$  ($MI^{1}_{p,-k}$) concerning to the strength of parameter $Q=2n_LeB/k_z^2$.  This plot correspond to $L^2\gg M^2$  ($L=100$ and $M=10$). For the other case, $M^2 \gg L^2$, the mutual information does not show any significant numerical variation with respect to parameter $Q$.}}
    \label{MIlarge1}
\end{figure}

Now for \ref{Bell-2}, we compute the mutual information for particle-particle and particle-antiparticle sectors represented by density matrices $\rho^{1}_{k,p}$ and $\rho^1_{p,-k}$, are given by
\begin{equation}
MI^1_{k,p}=| \alpha_1 | ^2 \log _2| \alpha_1 | ^2-| \alpha_1 | ^2 \log _2\frac{| \alpha_1 | ^2}{2}-\left(1+| \beta_1 | ^2\right) \log _2\frac{1+| \beta_1 | ^2}{2}+| \beta_1 | ^2 \log _2| \beta_1 | ^2
    \label{mikpbell2}
\end{equation}
and 
\begin{equation}
\begin{split}
MI^{1}_{p,-k}=-\frac{| \alpha_1 | ^2}{2} \log _2\frac{| \alpha_1 | ^2}{2}-\frac{1+| \beta_1 | ^2}{2}  \log _2\frac{1+| \beta_1 | ^2}{2} -\frac{| \beta_1 | ^2 }{2} \log _2\frac{| \beta_1 | ^2}{2}\\-\frac{1+| \alpha_1 | ^2}{2}  \log _2\frac{1+| \alpha_1 | ^2}{2}+\frac{1}{2}
\end{split}
    \label{mikapbell2}
\end{equation}
We have shown the variation of \ref{mikpbell2} with respect to the parameter $Q$, for $L^2 \gg M^2$ case, in \ref{MIlarge1}.  For the particle-particle sector, the mutual information monotonically increases with the increase in $Q$ and at large $Q$ values, it converges to its maximum value, i.e. $2$. Whereas for the particle-antiparticle sector, it decreases with the increase in $Q$. For the other case, $M^2 \gg L^2$ ($M=5$ and $L=1$), the mutual information does not show any significant numerical variation with respect to the parameter $Q$ for both sectors. Its value is approximately two and zero for particle-particle and particle-antiparticle sectors, respectively. For the heavy rest mass case, once again, this analysis corresponds to the fact that the pair creation is suppressed in this case, and in consequence, nothing very non-trivial happens.

Before we conclude, we wish to further extend the above results to the so called fermionic $\alpha$-vacua.

\section{The case of the fermionic \texorpdfstring{$\alpha$}--vacua}
\label{alph}
The fermionic $\alpha$-vacua, like the scalar field~\cite{Allen, alpha-vaccuum1}, corresponds to a Bogoliubov transformation characterised by a parameter $\alpha$ in the `in' mode field quantisation.  Although such vacua may not be very useful to do perturbation theory, for example,~\cite{Mottola:1984ar, Einhorn}, it still attracts attention chiefly from the perspective of the so called trans-Planckian censorship conjecture,~for example,~\cite{Bedroya}. 

In order to construct such vacua, from  \ref{field}, we define a new set of annihilation and creation operators \cite{alpha-vacua},
\begin{eqnarray}
    \label{alphaop1}
    c^{\alpha}({\vec{k}_{\slashed{y}},s,n_L})
    &=&\cos\alpha\, a_{\rm \text{in}}(\vec{k}_{\slashed{y}},s,n_L)-\sin \alpha\, b^{\dagger}_{\rm \text{in}}(\vec{k}_{\slashed{y}},s,n_L) \nonumber\\
   d^{\alpha}({\vec{k}_{\slashed{y}},s,n_L})
&=&\cos\alpha\;b_{\rm \text{in}}(\vec{k}_{\slashed{y}},s,n_L)+\sin\alpha\;a^{\dagger}_{\rm \text{in}}(\vec{k}_{\slashed{y}},s,n_L)  
\end{eqnarray}
where the parameter $\alpha$ is real and $0 \leq \alpha \leq \pi/2$. The above relations indicate that we need to define a new, one parameter family of vacuum state $|0\rangle_{\alpha}$, so that
$$ c^{\alpha}|0\rangle_{\alpha} =0=d^{\alpha}|0\rangle_{\alpha}$$
An $\alpha$-vacuum state is related to the original in-vacuum state via a squeezed state expansion. Note that \ref{alphaop1} does not mix the sectors $s=1$ and $s=2$. Thus as of the previous analysis, we work only with the $s=1$ sector and write for the normalised $\alpha$-vacuum state,
\begin{equation}
\label{alphavacuum}
    |0_k\rangle^{(1)}_{\alpha} = \cos{\alpha}|0_k^{(1)} 0_{-k}^{(1)} \rangle^{\rm \text{in}} + \sin{\alpha}|1_k^{(1)} 1_{-k}^{(1)} \rangle^{\rm \text{in}} 
\end{equation}
Using now \ref{vaccum1} into the above equation, we re-express $|0\rangle^{(1)}_{\alpha} $ in terms of the out states,
\begin{equation}
    \label{alpha_out}
    |0_{k}\rangle^{(1)}_{\alpha} = \alpha' |0_k^{(1)} 0_{-k}^{(1)}\rangle^{\rm \text{out}} + \beta' |1_k^{(1)}1_{-k}^{(1)}\rangle^{\rm \text{out}}
\end{equation}
where 
\begin{eqnarray}
\alpha'=\frac{\alpha_1\cos{\alpha}+\beta_1 \sin{\alpha}}{\sqrt{1+2(\alpha_1 \beta_1^* + \beta_1 \alpha_1^*)\cos{\alpha}\sin{\alpha}}}, \qquad \beta'=\frac{\alpha_1\sin{\alpha}+\beta_1 \cos{\alpha}}{\sqrt{1+2(\alpha_1 \beta_1^* + \beta_1 \alpha_1^*)\cos{\alpha}\sin{\alpha}}}
\label{b'}
\end{eqnarray}
 are the effective Bogoliubov coefficients. Note the formal similarity between \ref{alpha_out} and \ref{vaccum1}. Setting $\alpha=0$ in the first reproduces the second.\\

\noindent
The above mentioned formal similarity thus ensures that the expressions for either the vacuum entanglement entropy or the Bell violation for the $\alpha$-states can be obtained from our earlier results, \ref{entropy}, \ref{bvpm}, \ref{Bell0kp}, \ref{Bell0p-k}, \ref{Bell1p-k}, by simply making the replacements,
$$\alpha_1 \to \alpha'_1, \qquad {\rm and} \qquad \beta_1 \to \beta'_1$$
Some aspects of entanglement for scalar $\alpha$-vacua can be seen in, for example,~\cite{EE for alpha1, EE for alpha, BI for mixed state, Choudhury:2017qyl} (also references therein). We refer our reader also to \cite{SSS} for a discussion on the natural emergence of $\alpha$-like vacua for fermions in the hyperbolic de Sitter spacetime. 

\begin{figure}[h]
	\begin{center}
		\includegraphics[scale=.50]{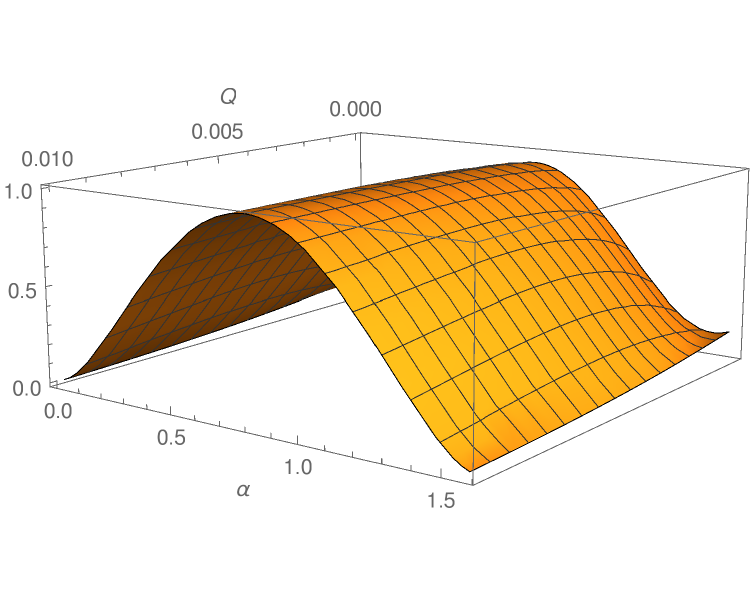}\hspace{1.0 cm}
			\includegraphics[scale=.50]{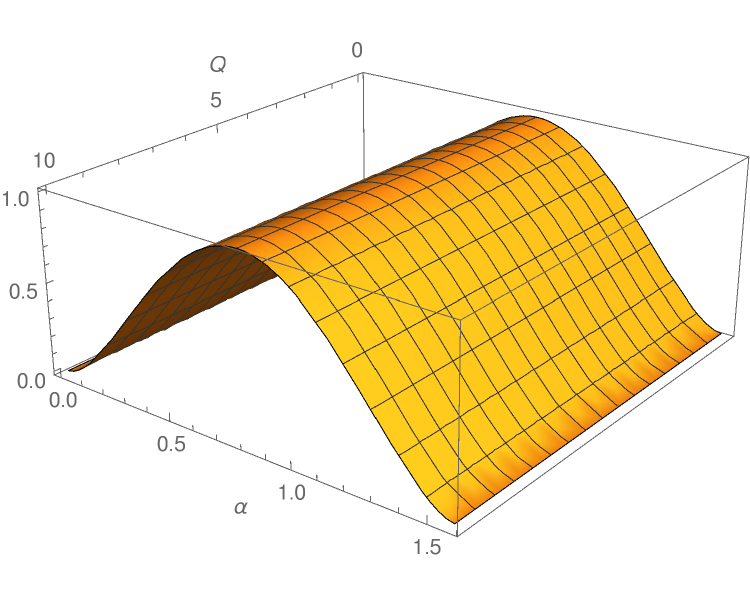}
		\caption{ \it{\small Entanglement entropy for the fermionic  $\alpha$-vacua, \ref{alphavacuum}. We have plotted   \ref{entropy'} vs the parameters $Q=2en_LB/k_z^2$ and $\alpha$. The left one corresponds to $L^2\gg M^2$ ($L=100$ and $M=10$), whereas the right one corresponds to $M^2\gg L^2$ ($M=5$ and $L=1$). See main text for discussions.}
		 	}
		\label{fig:alphaentropy}
		\end{center}
\end{figure}

For the fermionic case the vacuum entanglement entropy, \ref{entropy}, modifies to the $\alpha$-vacua as,
\begin{equation}
    \label{entropy'}
	S_k^{\alpha} =
		-   \left[ \ln	(1-|\beta'_1|^{2})    +|\beta'_1|^2  \ln \frac{|\beta'_1|^{2}}{1-|\beta'_1|^2}\right]
\end{equation}
which is plotted in \ref{fig:alphaentropy} with respect to the parameters $Q=2enB/k_z^2$ and $\alpha$. We see that $S_k^{\alpha}$ first increases with increase in the parameter $\alpha$ and has its maximum at $\alpha = \pi/4$, after which it decreases and becomes vanishing as $\alpha \to \pi/2$.
\begin{figure}[h]
	\begin{center}
		\includegraphics[scale=.50]{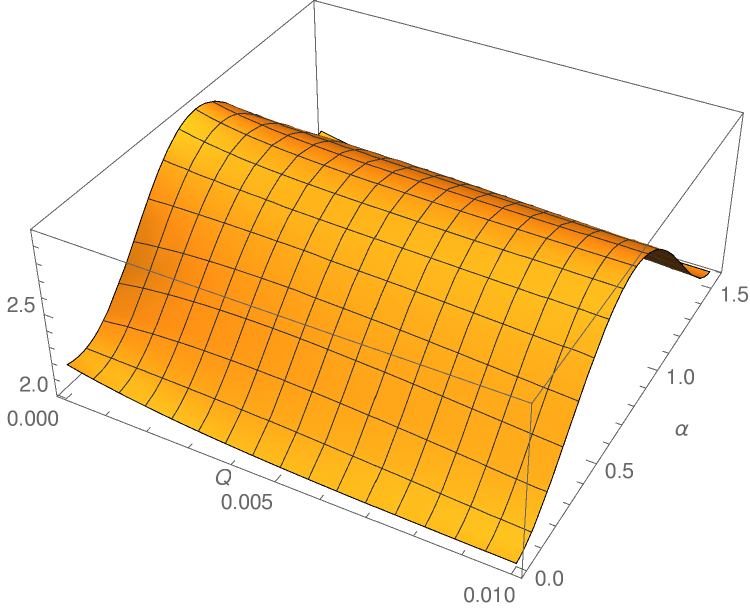}\hspace{1.0cm}
		\includegraphics[scale=.50]{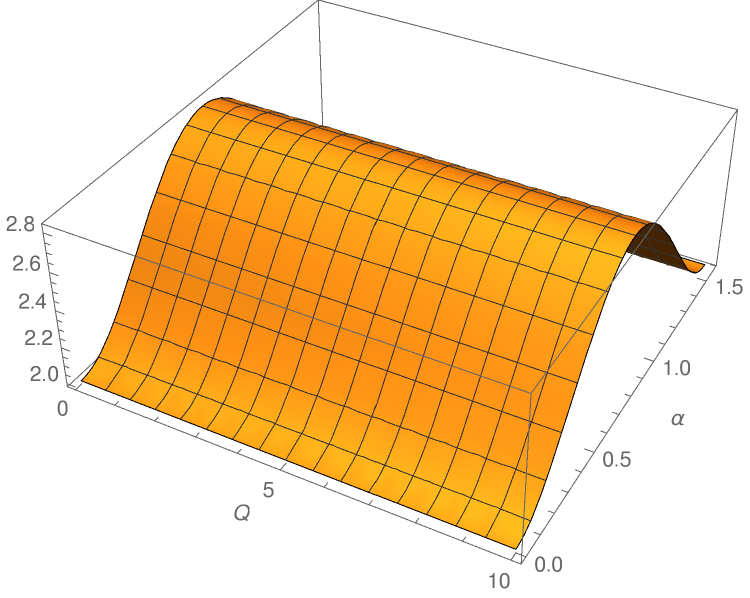}
		\caption{	\it{\small Bell violation for the fermionic $\alpha$-vacua. As we have discussed in the main text,  we have plotted   \ref{bvpm} after replacing $\alpha_1$ and $\beta_1$, respectively by $\alpha'_1$ and $\beta'_1$ given by \ref{b'}. The left one corresponds to $L^2\gg M^2$ ($L=100$ and $M=10$), whereas the right one corresponds to $M^2\gg L^2$ ($M=5$ and $L=1$). $\langle \mathcal{B}\rangle_{\alpha,\,\rm max}>2$ corresponds to the Bell violation.
			}}
		\label{fig:BValphavaccuum}
		\end{center}
\end{figure}
Likewise the Bell violation for the vacuum state, \ref{bvpm}, can be extended to the $\alpha$-vacua and is plotted in \ref{fig:BValphavaccuum}. Like the vacuum entanglement entropy, the vacuum Bell violation also reaches maximum at $\alpha = \pi/4$ and becomes vanishing as $\alpha \to \pi/2$. 

The vanishing of both vacuum entanglement entropy and Bell violation as $\alpha\to\pi/2$ can be understood as follows. In this limit, only the excited state part of \ref{alphavacuum} survives.  \ref{vaccum1} then implies that the corresponding out-basis expansion of this state is not only pure, but also separable. Thus in this limit no entanglement survives, as we discussed in \ref{A short introduction to quantum information}.\\

Let us now come to the case of the maximally entangled states. The states of \ref{Bell-1}, \ref{Bell-2} respectively modify as,
\begin{eqnarray}
\label{Bell-1a}
|\psi\rangle_{\alpha}=\frac{|0_p 0_{-p}0_k0_{-k}\rangle_{\alpha}^{\text{in}}+|1_p0_{-p}0_{k}1_{-k}\rangle_{\alpha}^{\text{in}}}{\sqrt{2}}
\end{eqnarray}
and,
\begin{eqnarray}
\label{Bell-2a}
|\chi\rangle_{\alpha}=\frac{|1_p 0_{-p}0_k0_{-k}\rangle_{\alpha}^{\text{in}}+|0_p0_{-p}1_{k}0_{-k}\rangle_{\alpha}^{\text{in}}}{\sqrt{2}}
\end{eqnarray}

\begin{figure}[h]
	\begin{center}
		\includegraphics[scale=.52]{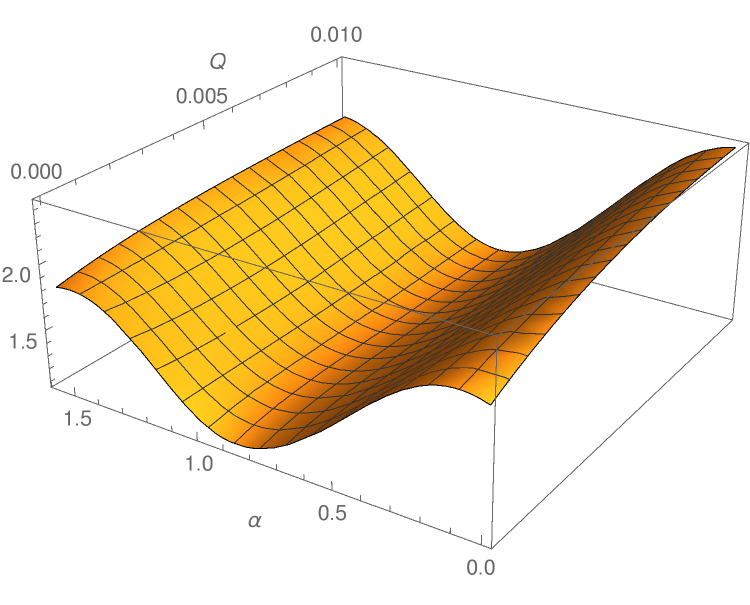}\hspace{1.0cm}
		\includegraphics[scale=.52]{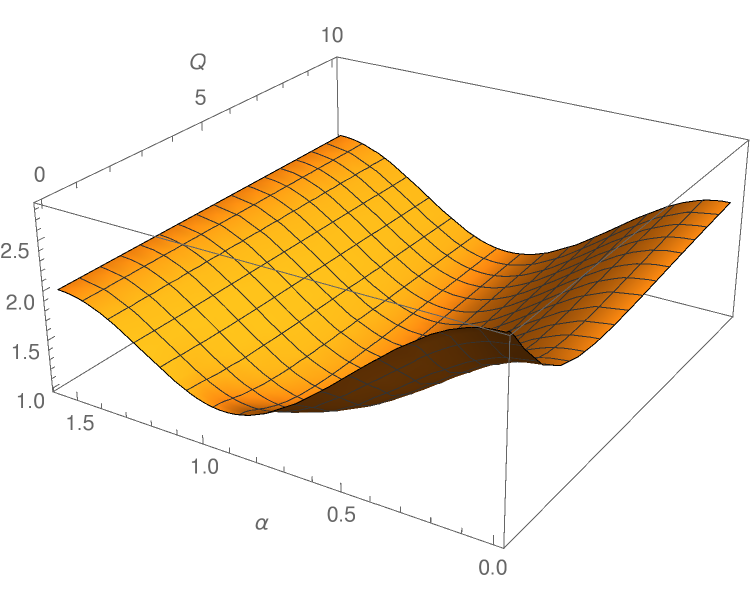}
		\caption{\it{\small Bell violation for the particle-antiparticle sector corresponding to the initial $\alpha$-state in \ref{Bell-1a}. As we have discussed in the main text, we have basically plotted \ref{Bell0p-k} after replacing  $\alpha_1$, $\beta_1$ respectively by $\alpha'_1$ and  $ \beta'_1$ \ref{b'}.  The left plot corresponds to $L^2\gg M^2$  ($L=100$ and $M=10$), whereas the right one corresponds to  $M^2\gg L^2$ ($M=5$ and $L=1$). $\langle \mathcal{B}\rangle_{\alpha,\,{\rm max}}>2$ corresponds to the Bell violation. The particle-particle sector corresponding to this initial state does not show Bell violation, like the  $\alpha=0$ case discussed in \ref{bvme}.}
			}
		\label{fig:BValpharhop-k0}
		\end{center}
\end{figure}

\begin{figure}
	\begin{center}
		\includegraphics[scale=.52]{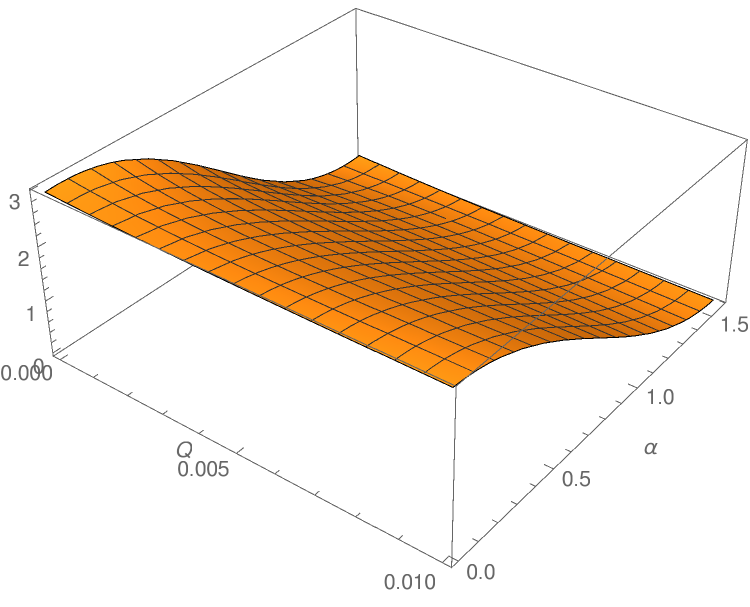}\hspace{1.0cm}
		\includegraphics[scale=.52]{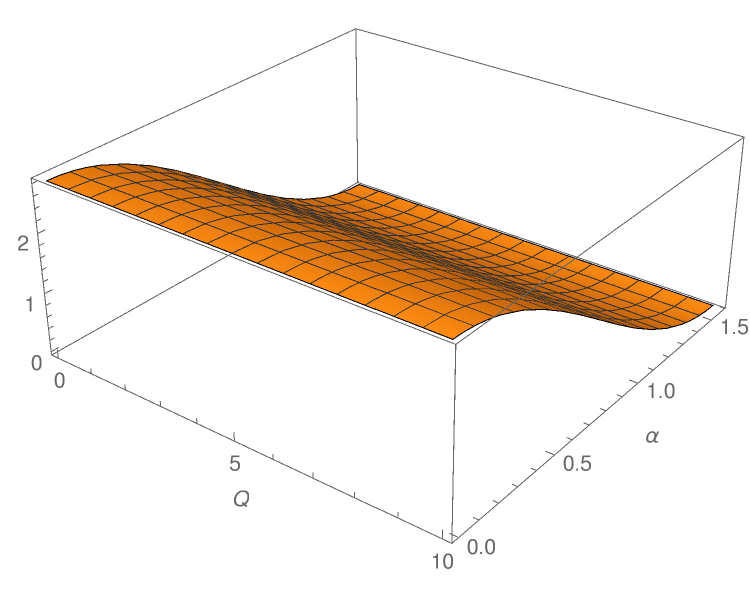}
		\caption{\it{\small Bell violation for the particle-particle sector corresponding to the initial $\alpha$-state in \ref{Bell-2a}. The left plot corresponds to $L^2\gg M^2$  ($L=100$ and $M=10$), whereas the right one corresponds to  $M^2\gg L^2$ ($M=5$ and $L=1$). As earlier, $\langle \mathcal{B}\rangle_{\alpha,{\rm max}}>2$ corresponds to the Bell violation. The particle-antiparticle sector corresponding to this initial state does not show any violation, like the $\alpha=0$ case discussed in \ref{bvme}.
			}}
		\label{fig:BValpharhokp1}
		\end{center}
\end{figure}

\begin{figure}
    \centering
\includegraphics[scale=0.75]{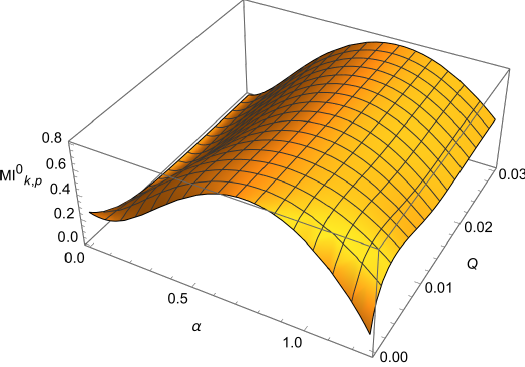}\hspace{1.0cm}
\includegraphics[scale=0.70]{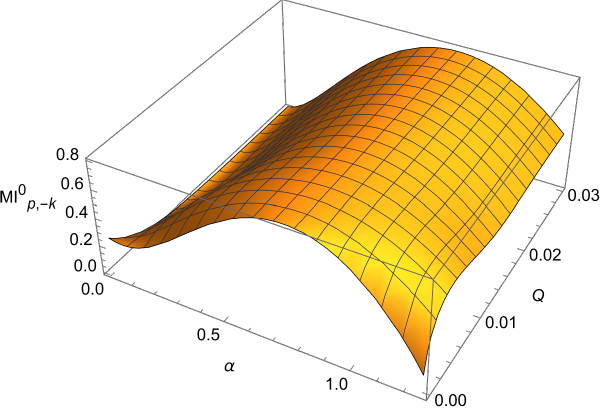 }
    \caption{\it{\small Mutual information for the particle-particle sector $\rho_{k,p}^{0}$ (left) and  particle-anti-particle sector $\rho_{p,-k}^{0}$ (right), corresponding to the initial $\alpha$-state in \ref{Bell-1a}.  These plots correspond to $L^2\gg M^2$  ($L=100$ and $M=10$). For the other case, $M^2 \gg L^2$, the mutual information does not show any significant numerical variation, like the $\alpha=0$ case discussed in \ref{MI}.}}
    \label{MIlarge0alpha}
\end{figure}

\begin{figure}
    \centering
\includegraphics[scale=0.62]{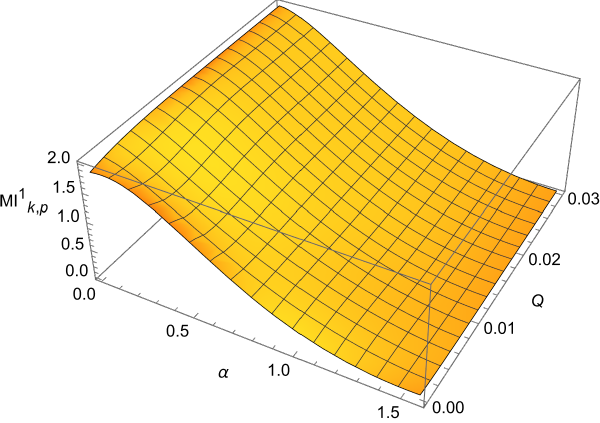}\hspace{1.0cm}
\includegraphics[scale=0.70]{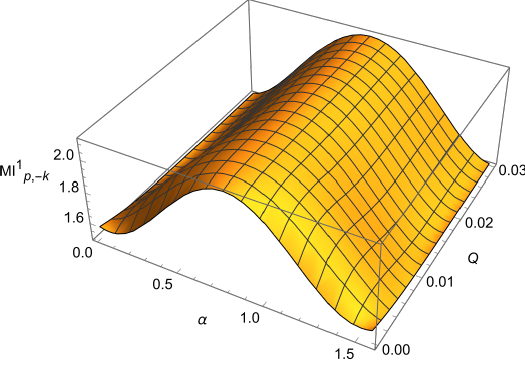}
    \caption{\it{\small Mutual information for particle-particle, $\rho_{k,p}^{1}$ and particle-antiparticle, $\rho_{p,-k}^{1}$ sectors corresponding to the initial $\alpha$-state in \ref{Bell-2a}.  We have plotted the mutual information for $\rho_{k,p}^{1}$  $(MI^{1}_{k,p})$ and $\rho_{p,-k}^{1}$  $(MI^{1}_{p,-k})$ concerning to the strength of parameters $Q=2n_LeB/k_z^2$ and $\alpha$. These plots correspond to the case $L^2\gg M^2$  $(L=100$ and $M=10)$. For the other case, $M^2 \gg L^2$, the mutual information does not show any significant numerical variation, like the $\alpha=0$ case discussed in \ref{MI}.} }
    \label{MIlarge1alpha}
\end{figure}

Using \ref{alphavacuum}, and the method described in~\ref{bvme}, we can easily extend the results of the Bell violation and also the mutual information we found earlier. As we mentioned earlier, this generalisation effectively corresponds to  just replacing  $\alpha_1$, $\beta_1$ respectively by $\alpha'_1$ and  $ \beta'_1$ (\ref{b'}) in appropriate places  (for example, in \ref{Bell0kp}).   We have plotted these Bell violations in \ref{fig:BValpharhop-k0}, \ref{fig:BValpharhokp1} and the mutual information in \ref{MIlarge0alpha}, \ref{MIlarge1alpha}.

\section{Summary and outlook}
\label{sec:SD}
In this Chapter, we have discussed the fermionic entanglement entropy, Bell violation and mutual information in the cosmological de Sitter spacetime, in the presence of primordial electric and magnetic fields of constant strength. We have found relevant in and out orthonormal Dirac mode functions, the Bogoliubov coefficients and the resultant squeezed state relationship between the in and out states in \ref{S2}.  Using these key results, we have computed the vacuum entanglement entropy, the Bell violation and the mutual information (for both vacuum and two maximally entangled initial states), respectively, in \ref{EE} and \ref{BV}. These results are extended further to the so called fermionic $\alpha$-vacua in \ref{alph}. We have focused on two qualitatively distinct cases here -- the `strong' electric field and the `heavy' mass limits (with respect to the Hubble constant), c.f. \ref{case1}, \ref{case2}.  

As we have argued earlier in \ref{Motivation and Overview}, a background magnetic field alone cannot create vacuum instability, but in the presence of spacetime curvature and electric field, it can affect such instability or the rate of the particle pair creation. This is manifest from \ref{number_demsity}, which receives, as we have discussed,  no contribution from the magnetic field if the electric field strength is vanishing. Whereas if the magnetic field strength is very large compared to that of the electric field,  the particle creation rate also becomes independent of the electric and magnetic fields. Our chief aim in this Chapter was to investigate the role of the magnetic field strength on the Bell violation and the mutual information. We have seen that subject to the choices of the initial states, the behaviour of the Bell violation can be qualitatively different, for example,~\ref{fig:bellvofvacuum} and \ref{fig:rhop-k0}. Also, the mutual information depends on the choices of the initial states as well, \ref{MIlarge0} and \ref{MIlarge1}. For the case of the $\alpha$-vacua on the other hand, we have also taken into account the variation of the parameter $\alpha$,~for example,~\ref{fig:alphaentropy}.

The above analysis can be attempted to be extended in a few interesting scenarios. For example, instead of having only constant electric and magnetic fields, can we also have fluctuating ones, like electromagnetic or gravitational radiation? Finally, it also seems interesting to perform a similar analysis in the Rindler spacetime for its relevance to the near horizon geometry of non-extremal black holes. We wish to address this problem for fermions in the Rindler spacetime in the next Chapter. Discussion of the Schwinger pair creation for a complex scalar field coupled to a constant background electric field in the Rindler spacetime can be seen in~\cite{Gabriel:1999yz}. Finally, as we have discussed at the beginning of this Chapter, it will be important to compute the breaking of scale invariance of the cosmological power spectra in the presence of primordial electromagnetic fields and also to compute the Bell violation by the photons (interacting with the entangled fermions) coming from very distant sources, with the hope to constrain the strengths of those background fields. A gauge invariant formulation of an effective action for the second problem seems  to be a non-trivial task. We hope to come back to this issue in future works.

\chapter{Schwinger effect and a uniformly accelerated observer}
\label{Schwinger effect and a uniformly accelerated observer}

In the previous Chapters, we discussed the correlation properties between the particles created by electric and time-dependent gravitational fields. In quantum field theory, particle creation is also possible due to the causal structures of the spacetime endowed with event horizons. These effects lead black holes to create and emit particles, known as the Hawking radiation \cite{Hawking, Hawking1}. Such an effect can also be observed by a uniformly accelerated particle detector moving in the flat spacetime, which perceives the Minkowski vacuum to be thermally populated at temperature $T = a/2\pi$ where $a$ is the acceleration parameter, known as the Unruh effect \cite{dS1, Unruh:1976db, Crispino:2007eb, rindler1, rindler2, leftright, unruhdewitter1, unruhdewitter2, unruh_exp}, as was briefly reviewed in \ref{The Unruh effect}. It is a consequence of the causal structure of the spacetime, as perceived by the accelerated or Rindler observer, who sees the light-cone surface of the Minkowski spacetime as an event horizon.

The effect of a background magnetic field on the Schwinger effect and quantum correlations for a complex scalar field in the flat spacetime was discussed in \ref{Background magnetic field and quantum correlations in the Schwinger effect}. It was shown that a sufficiently high strength of the magnetic field stabilises the vacuum. An interesting question is how a uniformly accelerated observer would see these oppositely moving particle-antiparticle pairs created by an external electric field? And what will be the role of the background magnetic field in this context? 
In other words : how will the electric and magnetic field affect the pair creation due to the Unruh effect for a charged quantum field? Precisely, there will be two sources of particle creation here. One will be the Schwinger effect in a given Rindler wedge. The other source certainly corresponds to the Unruh effect. We wish to incorporate {\it both} of these in our composite scenario.

In \cite{SHG, SHU}, the Schwinger effect is studied in the (anti-)de Sitter spaces and black hole backgounds, representing a unified picture of the same with that of the Hawking radiation. It was also shown in these works that the Schwinger effect from a near-extremal black hole is a  product of the AdS$_2$ Schwinger effect and a correction due to the Hawking radiation from non-extremality. 

In \cite{Gabriel:1999yz}, the quantisation of a complex scalar field interacting with a constant background electric field in the $(1+1)$-dimensional Rindler spacetime was performed
and an expression of the vacuum decay rate was found. As we mentioned above, the main characteristic of this problem is that it involves two acceleration scales : the acceleration of the Rindler observer and the acceleration due to the non-zero electric field. In this work the Schwinger effect and the Rindler vacuum decay rate was studied for either the left ($L$) or the right ($R$) Rindler wedge. 
The mean number density of the particles and antiparticles observed by the Rindler observer was also computed. It was observed that the particles and antiparticles are not equally distributed in a particular Rindler wedge, which leads to a charge polarization. It was also observed that the total charge, i.e., the sum of the charges in $R \cup L$, is always conserved. However, the Unruh effect context of this problem, i.e. the construction of the global modes in $R\cup L$ and the subsiquent Bogoliubov relationship was not explicitly addressed in \cite{Gabriel:1999yz}. 

This Chapter considers a similar scenario for the charged fermionic field with constant background electric and magnetic fields in the $(3+1)$-dimensional Rindler spacetime. 
In addition, we shall also construct the global modes out of the local modes existing in both the causally disconnected Rindler wedges. 
There will be two sets of Bogoliubov transformations relevant to this context. The first would correspond to the Schwinger pair creation in a given Rindler wedge. Clearly, this will involve the local modes only. The other set of the Bogoliubov relationship corresponds to the construction of the global modes out of these local modes. 

In addition to the entanglement between the Schwinger pairs, 
an exciting outcome of the aforementioned global modes is the emergence of further entanglement between the causally disconnected left and right wedges. 
Entanglement in the context of black holes or the Rindler spacetime wedges has been receiving a lot of attention during the past few years \cite{Higuchi:2017gcd, Ueda:2021nln, FuentesSchuller:2004xp, Yi Ling, RL, EE_1, EE_2, MartinMartinez:2010ar, unruh4, unruhgaussian}. For both bosons and fermions, entanglement degradation occurs from the perspective of a uniformly accelerated observer due to the thermal particle creation \cite{MShamirzai, Tel, Dirac2}. Also, as we mentioned earlier, the entanglement for the pairs created in the Schwinger effect was studied in for example, \cite{Ebadi:2014ufa, Li:2016zyv, Dai:2019nzv, Li:2018twv}. We will be interested to understand the effect of the existence of the causally disconnected spacetime wedges, such as the Rindler wedges, into the entanglement structure of the Schwinger pairs.


As an aside, we note that perhaps this scenario can be considered as a crude toy model to investigate the entanglement properties of a charged quantum field near an eternal and non-extremal Riessner-Nordstrom black hole carrying both electric and magnetic charges. Although this problem does not seem to be solvable analytically, we may hope that our current analysis
might give us some tentative insights. We also note that the astrophysical black holes are endowed with background electromagnetic fields due to accreting plasmas. However, such a black hole does not have any past or white hole horizon, so they do not have any left-right Rindler wedge structure \cite{QFTCS2}. Hence, particle creation in only one Rindler wedge will be relevant for the latter case.

 In \ref{section : S2}, the Dirac field is quantised in the right (R) and left (L) Rindler wedges explicitly, with respect to the orthonormal modes found in closed form. The Bogoliubov relations between the local modes due to the background electric and magnetic fields are found explicitly. In \ref{S3}, the global modes existing on both wedges are constructed. The relation between the local and global creation and annihilation operators is obtained, which incorporates the two Bogoliubov transformations, one due to the Schwinger and the other due to the Unruh effect. 
In \ref{S4}, the logarithmic negativity is computed to look at the vacuum instability and entanglement properties between the created particles. In \ref{Logarithmic negativity and mutual information of a zero charge state constructed by two fermionic fields}, we construct a maximally entangled system of two fermionic fields and study the entanglement properties of different sectors corresponding to this system. Finally, we concluded in \ref{S5}. Computational details are provided in the appendices. 

\section{ The Dirac modes}\label{section : S2}
\begin{figure}[h]
\begin{center}
     \includegraphics[scale=.52]{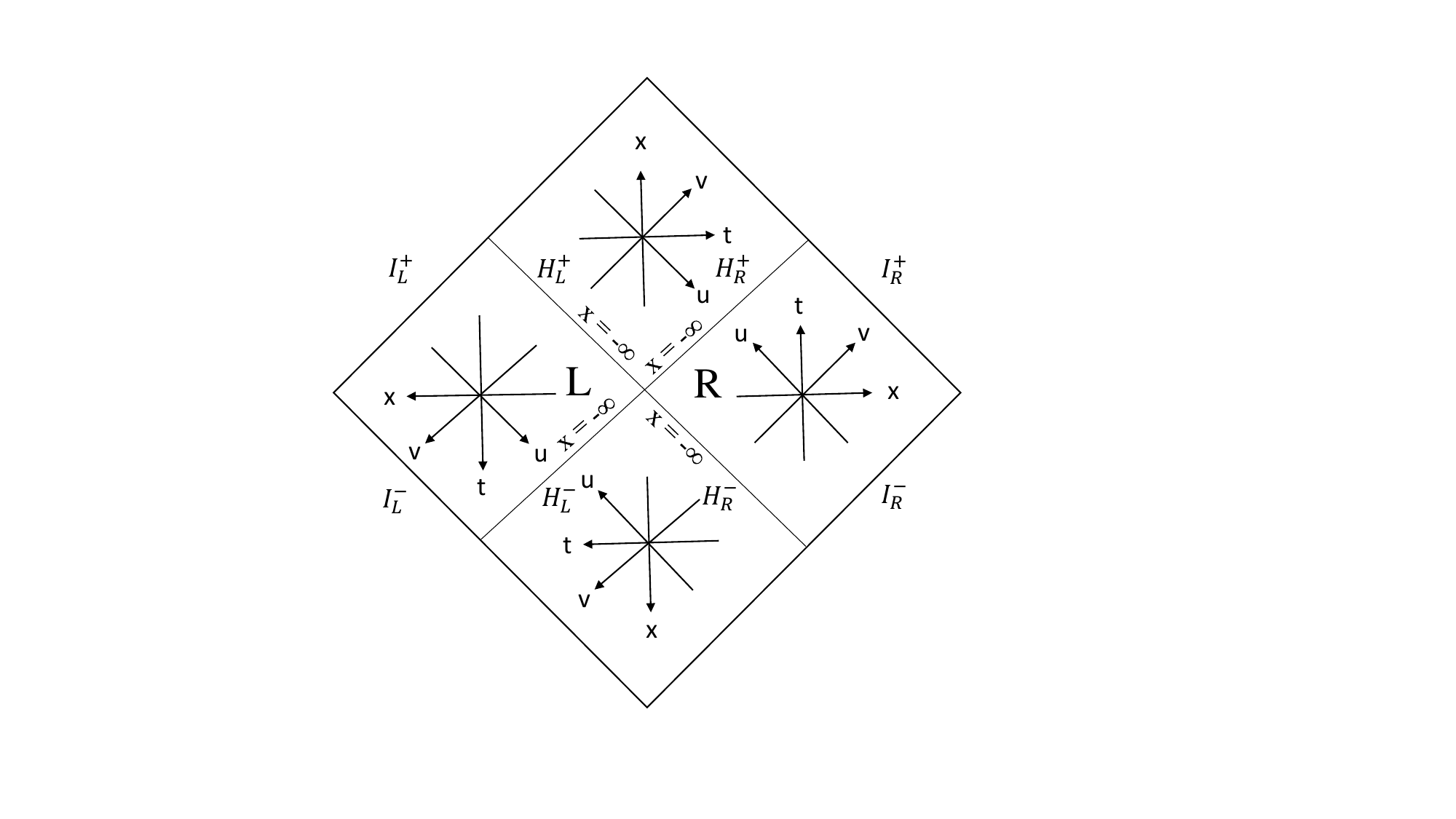}
    \caption{\it{\small The Rindler patches with their coordinates, e.g., \cite{Gabriel:1999yz}. Here $\mathit{I^-_{R,L}}$ and $\mathit{I^+_{R,L}}$ are the past and future null infinities, whereas $\mathit{H^-_{R,L}}$ and $\mathit{H^+_{R,L}}$ are the past and future horizons. The hyperbolic curves represent the trajectories of particles, whereas $u$ and $v$ represent the lightlike coordinates. Our region of interest will be $R \cup L$. }}
    \label{diagram}
    \end{center}
\end{figure}

The Rindler coordinate transformations divide the Minkowski spacetime into four causally disconnected wedges, as was reviewed in \ref{The Rindler spacetime}. Among these four wedges, our region of interest will be the left ($L$) and right ($R$) regions, depicted in \ref{diagram}. As we also discussed earlier in the first Chapter, the coordinate
transformations between the Minkowski $\left(\tau, \rho, y, z\right)$, and the Rindler $\left(t, x, y,z\right)$ left-right wedges are given by
\begin{equation}
\begin{split}
\label{transformation}
    \tau=\frac{e^{ax_R}}{a}\sinh a t_R,\; \;\rho=\frac{e^{ax_R}}{a}\cosh a t_R\; \;{\rm\left(R\right)}\;\;
     \tau=-\frac{e^{ax_L}}{a}\sinh a t_L,\; \;\rho=-\frac{e^{ax_L}}{a}\cosh a t_L\;\; \rm{\left(L\right)}
     \end{split}    
\end{equation}
where for each of these quadrants, the respective Rindler coordinates run from $-\infty$ to $\infty$, shown in \ref{diagram}. On R and L, the vector field $\partial_t$ is timelike. 
The world lines of uniformly accelerated observers in the Minkowski coordinates correspond to hyperbolas to the left and right of the origin, which are
bounded by lightlike asymptotes constituting the Rindler horizon and hence, the two Rindler regions are causally disconnected. An observer undergoing a uniform acceleration remains constrained to either $R$ or $L$ and has no access to the other sector. In \ref{diagram}, $\mathit{I^-_{R,L}}$ and $\mathit{I^+_{R,L}}$ are the past and future null infinities, whereas $\mathit{H^-_{R,L}}$ and $\mathit{H^+_{R,L}}$ are the past and future horizons, respectively. $u$ and $v$ are respectively the retarded and advanced null coordinates (in $t-x$ plane) defined as $u=t-x$ and $v=t+x$. Under the transformation \ref{transformation}, the line element takes the form
\begin{equation}
\label{metric1}
ds^2=g_{\mu\nu}dx^\mu dx^\nu=e^{2ax}(-dt^2+dx^2)+dy^2+dz^2,
\end{equation}
where $a$ is regarded as the acceleration parameter, and the metric represents $(3+1)-$Rindler spacetime.

 Let us now focus on the fermionic field coupled with background electric and magnetic fields of constant strengths in the four-dimensional Rindler spacetime. \\
The Dirac equation in a general spacetime reads \cite{Parker:2009uva},
\begin{equation}
\label{dirac1}
 (i\gamma^{\mu}D_{\mu}-m)\psi(x) = 0
\end{equation}
where the gauge cum spin covariant derivative is defined as
\begin{eqnarray}
\label{spin_conn}
D_\mu=\partial_\mu+i e A_\mu+\Gamma_\mu
\end{eqnarray}
ensuring the local gauge symmetry and the general covariance. The spin connection matrices are given by,
\begin{eqnarray}
\label{conn1'}
\Gamma_\mu=-\frac{1}{8}e^{\mu}_a \left(\partial_{\mu}e_{b\nu}- \Gamma_{\mu\nu}^{\lambda} e_{b\lambda}\right)[\gamma^a, \gamma^b],
\end{eqnarray}
where the Latin indices represent the local inertial frame and  $e^{\mu}_a$ are the tetrads \cite{Peskin:2018}.
The only non-zero component of $\Gamma_\mu$ for \ref{metric1}, is $\Gamma_0$ given by
\begin{equation}
\Gamma_0=\frac{a}{2}\gamma^{1}\gamma^{0},    
\end{equation}
The tetrads relate the Minkowski metric to the curved spacetime metric as
\begin{equation}
\label{tetrad}
g_{\mu\nu}=e^{a}_\mu e^{b}_\nu \eta_{ab}
\end{equation}
Following \ref{tetrad}, we choose the tetrads for \ref{metric1}, as
$e^{\mu}_{a}=\text{diag}(e^{-ax},e^{-ax},1,1)$.\\
On defining a new variable $\psi(x)=e^{e^{-ax}/2} \Tilde{\psi}(x)$ in \ref{dirac1}, it becomes
\begin{eqnarray}
\label{dirac2}
(i e^\mu_a \gamma^{a}\partial_\mu-e e^\mu_a \gamma^{a}A_\mu-m)\Tilde{\psi}(x)=0
\end{eqnarray}
Substituting next
\begin{equation}
\label{de0'}
\widetilde{\psi}(x)= (i e^\mu_a \gamma^{a}\partial_\mu-e A_\mu e^\mu_a \gamma^{a}+m)\chi(x)
\end{equation}
 into \ref{dirac2}, we obtain a squared Dirac equation
\begin{equation}
    \label{dirac3}
    \begin{split}
 \bigg[\frac{1}{e^{2ax}}\Big((\partial_t+i e A_0)^2+a \partial_{x}-\partial_{x}^{2}\Big) -\partial^2_y-(\partial_z+i e A_3)^2+\frac{\gamma^{1} \gamma^{0}}{e^{2 a x}}\big(a \partial_t-i e \partial_x A_0\big)\\-i e \partial_{y}A_3 \gamma^{2}\gamma^{3}-m^2\bigg]\chi(x)=0
  \end{split}
\end{equation}
We choose the gauge to obtain constant strength electric and  magnetic fields along the $x-$axis as
\begin{equation}
    \label{gauge}
    A_\mu \equiv \frac{E  e^{2ax}}{2 a} \delta^t_{\mu} + B y \delta^z_{\mu}
\end{equation}
where $E, \;B $ are constants. Using the coordinate transformations \ref{transformation}, one can easily see that the above gauge reduces to the Minkowski gauge $A_\mu=Ez/2\delta^t_{\mu}+(By-Et/2)\delta^z_\mu$  which leads to electric and magnetic fields of constant strengths $E$ and $B$ respectively \cite{Gabriel:1999yz}. It is also easy to see that the electromagnetic field strength tensor components are given by $F_t{}^x =-E$ and $F_{yz}=B$, and they satisfy the Maxwell equations in \ref{metric1}. Also,
as of the preceding Chapter, we assume that $e$, $E$ and $B$ are positive quantities. 

We now consider the ansatz $\chi(x)=e^{-i\omega t} e^{ik_z z}\zeta_s(x,y) \epsilon_s$ (no sum on $s$) in \ref{dirac3}, to obtain
\begin{equation}
\label{dirac4}
\begin{split}
\bigg[-\frac{1}{e^{2ax}}\big( \omega - \frac{e E e^{2ax}}{2 a}\big)^2+\frac{a}{e^{2 a x}} \partial_{x}-\frac{1}{e^{2 a x}}\partial_{x}^{2}-\partial^2_y+(k_z+ e B y)^2-i\frac{ \gamma^{1} \gamma^{0}}{e^{2ax}}(a \omega+\frac{ e E e^{2ax} }{ 2})\\-i e B \gamma^{2}\gamma^{3}-m^2\bigg]\zeta_s(x,y) \epsilon_s=0 
    \end{split}
\end{equation}
We note using the anticommutation relations for the $\gamma$ matrices that $\gamma^{1}\gamma^{0}$ and $\gamma^{2}\gamma^{3}$ commute. Hence, we take
the $\epsilon_s$ as the simultaneous eigenvectors of these two matrices. We thus have four eigenvalue equations
\begin{center}
    $\gamma^{1}\gamma^{0}\epsilon_1=-\epsilon_1$, $\gamma^{1}\gamma^{0}\epsilon_2=-\epsilon_2$, $\gamma^{1}\gamma^{0}\epsilon_3=\epsilon_3 $, $\gamma^{1}\gamma^{0}\epsilon_4=\epsilon_4$, $\gamma^{2}\gamma^{3}\epsilon_1=-i \epsilon_1$, $\gamma^{2}\gamma^{3}\epsilon_2=i \epsilon_2$, $\gamma^{2}\gamma^{3}\epsilon_3=-i \epsilon_3$ and $\gamma^{2}\gamma^{3}\epsilon_4=i \epsilon_4$
    \end{center}
The explicit form of the  $\epsilon_s$ are given by
\begin{center}$\epsilon_1= \frac{1}{\sqrt{2}}\left( {\begin{array}{cccc}
  0\\
 0\\
1\\
   1\\
  \end{array} } \right)$,  
  $\epsilon_2= \frac{1}{\sqrt{2}}\left( {\begin{array}{cccc}
  -1\\
 1\\
0\\
   0\\
  \end{array} } \right)$,
$\epsilon_3=\frac{1}{\sqrt{2}} \left( {\begin{array}{cccc}
  0\\
 0\\
-1\\
   1\\
  \end{array} } \right) $and $\epsilon_4= \frac{1}{\sqrt{2}}\left( {\begin{array}{cccc}
  1\\
 1\\
0\\
   0\\
  \end{array} } \right)$.
 \end{center}
Using the eigenvalue equations and separation of variables as done in \ref{Fermionic Bell violation in de Sitter spacetime with background electric and mangnetic fields}, we obtain the variable decoupled equations 
 \begin{equation}
 \label{rho1d}
     \bigg[\partial^2_{x}-a \partial_x+\omega^2-\frac{e E \omega e^{2 a x }}{a}-i \omega a-\bigg(\frac{e E e^{2 a x}}{2 a}\bigg)^2 -\frac{i e E e^{2 a x}}{2}+(m^2+S_s) e^{2 a x}\bigg]\zeta_s(x)=0
 \end{equation} 
\begin{equation}
    \label{y1}
    \bigg[\partial^2_y-(k_z+qBy)^2+e B-S_s\bigg]H_s(y)=0
\end{equation}
with the separation constants,
\begin{equation}
\label{seapration}
  S_1=-2 n_L e B \;\text{and}\; S_2=-(2 n_L+1) e B  
\end{equation} 
where $n_L=0,1,2,3,.....$ denotes the Landau levels. 
The general solution of \ref{rho1d}, is 
\begin{equation}
    \label{zetasol}
    \zeta_s(x)=C_1e^{a x}e^{-\frac{i e E e^{2 a x}}{4 a^2}} e^{i \omega x} U(\lambda_s, \nu, \xi ) +C_2 e^{a x} e^{-\frac{i e E e^{2 a x} }{4 a^2}}  e^{i \omega x} L(-\lambda_s, \nu-1, \xi) 
\end{equation}
where $U$ and $L$ are the confluent hypergeometric and the generalized Laguerre functions, respectively, whose explicit forms are given in \cite{AS}. 
The solution of \ref{y1} is given by the Hermite polynomials as earlier 
\begin{equation}
    \label{soly1}
    H_s(y)=\left(\frac{\sqrt{eB}}{2^{n_L+1}\sqrt{\pi}(n_L+1)!}\right)^{1/2}e^{-\Tilde{y}^2/2}\mathcal{H}_{n_L}(\Tilde{y})=h_{n_L}(\Tilde{y})~({\rm say})
\end{equation}
In \ref{zetasol} $\lambda_s$'s and $\nu$ are parameters defined as
\begin{equation}
\label{coeff'}
\lambda_1=\lambda_3=\frac{i(m^2 + S_1)+2 e E}{2 e E}, \;
\lambda_2=\lambda_4=\frac{i(m^2 + S_2)+2 e E}{2 e E}, \;
\nu=\frac{3}{2}+\frac{i \omega}{a}
\end{equation}
The variables $\Tilde{y}$ and $\xi$ in \ref{zetasol} and \ref{soly1}, respectively, are defined as
\begin{equation}
\label{ell}
\xi=-\frac{i e E e^{2a x}}{2 a^2},\;
    \Tilde{y}=\left(\sqrt{eB}y+\frac{k_z}{\sqrt{eB}}\right) 
\end{equation}
Let us now find out the `in' modes for $R$ wedge, at $x \to \infty$ and  $x \to -\infty$, that corresponds to $\mathit{I^-_R}$ and $\mathit{H^-_R}$, respectively in \ref{diagram}. For a mode emerging from $\mathit{H}^-_R$ and moving towards $\mathit{I^+_R}$, the relevant part of the same is proportional to $e^{- i \omega u}$ 
\begin{center}
   $ \zeta_s(x) \sim e^{a x} e^{-\frac{i e E e^{2 a x}}{4 a^2}} e^{-i \omega (t-x)} \xi^{-\lambda_s},\;s=1,2 $ 
\end{center}
where we have used the limiting values of the special functions provided in \cite{AS}. Similarly, for a mode emerging from $\mathit{I^-_R}$ and moving towards $\mathit{H^+_R}$, the relevant part of the same is proportional to $e^{- i \omega v}$
\begin{center}
      $ \zeta_s(x) \sim e^{ax} e^{\frac{i e E e^{2 a x} }{4 a^2}} e^{-i \omega (t+x)} \frac{\Gamma(-\lambda_s^*+ \nu^*)}{\Gamma(\nu^*) \Gamma(-\lambda_s^* + 1)},\;s=1,2 $  
\end{center}
Putting these together, we have four `in' modes from which two correspond to $\mathit{I^-_R}$ and two to $\mathit{H^-_R}$ written as
\begin{eqnarray}
\label{chi'}
\chi(x)_{\mathit{H^{-}_{R,s}}}&=&e^{-i\omega( t-x)} e^{i k_z z} e^{a x} e^{-\frac{i e E e^{2 a x}}{4 a^2}} U(\lambda_s, \nu, \xi) H_s(y) \epsilon_s\\
\label{chi1}
 \chi(x)_{\mathit{I^{-}_{R,s}}}&=&e^{-i\omega ( t+x)} e^{i k_z z} e^{a x} e^{\frac{i e E e^{2 a x}}{4 a^2}} (L(-\lambda_s, \nu-1, \xi))^*  H_s(y) \epsilon_s
\end{eqnarray}
where $s=1,2$ in \ref{chi'} and \ref{chi1}. These modes are assumed to have zero support other than the place of their origin (i.e., $H^-_R$ or $I^-_R$), initially. 
\subsection{ Field quantization in the right Rindler wedge}
For computing the full modes, we need to substitute $\chi(x)$ into \ref{dirac3} and then using our earlier  definition $\psi(x)=e^{e^{-ax}/2} \Tilde{\psi}(x)$, the final particle `in' modes are given by
\begin{equation}
    \label{mode1}
    U_{s,n_L}(x)_{\mathit{H^-_R}}=\frac{ e^{e^{-ax}/2}  }{N_s} (ie^\mu_a \gamma^{a}\partial_\mu-e A_\mu e^\mu_a \gamma^{a}+m) e^{-i\omega( t-x)} e^{i k_z z} e^{a x} e^{-\frac{i e E e^{2 a x}}{4 a^2}} U(\lambda_s, \nu, \xi) H_s(y) \epsilon_s,
\end{equation}
\begin{equation}
    \label{mode2}
    U_{s,n_L}(x)_{\mathit{I^-_R}}=\frac{e^{e^{-ax}/2} }{M_s} (i e^\mu_a \gamma^{a}\partial_\mu-e A_\mu e^\mu_a \gamma^{a}+m) e^{-i\omega ( t+x)} e^{i k_z z} e^{a x} e^{\frac{i e E e^{2 a x}}{4 a^2}} (L(-\lambda_s, \nu-1, \xi))^*  H_s(y) \epsilon_s,
\end{equation}
where $s=1, 2$ for \ref{mode1} and \ref{mode2} and the parameter $\xi$ is defined in \ref{ell}. They are the positive frequency modes with respect to a future-directed timelike Killing vector field, $\partial_t$. 

The antiparticle `in' modes are found by simply taking the charge conjugation (i.e., $U \to i \gamma^2 U^*$), of \ref{mode1} and \ref{mode2}, given respectively by
\begin{eqnarray}
   \label{mode1a}
   V_{s,n_L}(x)_{\mathit{H^-_R}}&=&\frac{ e^{e^{-ax}/2} }{P_s} (ie^\mu_a \gamma^{a}\partial_\mu-e A_\mu e^\mu_a \gamma^{a}+m) e^{i\omega(t-x)}e^{ax}e^{i k_z z}e^{-\frac{i e Ee^{2ax}}{4a^2}}e^{\xi}\\&& \times U(\nu-\lambda_s,\nu,\xi)H_s(y)\epsilon_s \nonumber \\
     V_{s,n_L}(x)_{\mathit{I^-_R}}&=&\frac{e^{e^{-ax}/2} }{R_s} (i e^\mu_a \gamma^{a}\partial_\mu-e A_\mu e^\mu_a \gamma^{a}+m) e^{i\omega(t+x)}e^{ax}e^{i k_z z}e^{\frac{i e Ee^{2ax}}{4a^2}}\xi^{1-\nu^*}\nonumber \\&&\times (L(\nu-\lambda_s-1,1-\nu,\xi))^*H_s(y)\epsilon_s 
\end{eqnarray}
where $s=3, 4$ and $N_s$, $M_s$, $P_s$ and $R_s$ are the normalization constants obtained at constant $u$ and $v$ surfaces, as shown explicitly in \ref{A}. Accordingly, these modes are orthonormal,
\begin{equation}
(U_{s,n_L}(x)_{\mathit{H^-_R}, \mathit{I^-_R}},U_{s^{\prime},n_L^{\prime}}(x)_{\mathit{H^-_R, \mathit{I^-_R}}})= (V_{s,n_L}(x)_{\mathit{H^-_R}, \mathit{I^-_R}},V_{s^,n_L^{\prime}}(x)_{\mathit{H^-_R}, \mathit{I^-_R}} )= \delta(k_z-k'_z)\delta(\omega-\omega ')\delta_{n_L n_L^{\prime}}\delta_{ss^{\prime}}
\end{equation}
Next, we have computed another set of orthonormal `out' modes corresponding to the regions $\mathit{I^+_R}$ and $\mathit{H^+_R}$ of \ref{diagram}.  Following the analysis of \cite{out_def, out_def1} done for a complex scalar field, we define the time-reversed versions modes as\begin{eqnarray}
   U_{s,n_L}(t,\vec{x})_{\mathit{I^+_R},\mathit{H^+_R}}=(U_{s,n_L}(-t,\vec{x})_{\mathit{H^-_R},\mathit{I^-_R}})^{*},\quad 
   V_{s,n_L}(t,\vec{x})_{\mathit{I^+_R},\mathit{H^+_R}}=(V_{s,n_L}(-t,\vec{x})_{\mathit{H^-_R},\mathit{I^-_R}})^{*}
   \label{outdefn}
\end{eqnarray}
and simply take them to be our `out' modes. The explicit form of these modes and the calculation of the normalization constants are shown in \ref{A1}. These modes also turn out to be orthonormal. As we shall see below, the above `out' modes are simply given by the linear combinations of the `in' modes via some Bogoliubov relationship.

In terms of these `in' and `out' modes, we now make the field quantisation on $R$ wedge, \begin{equation}
\label{FQR1}
    \begin{split}
       \psi_R(x) = \sum_{n_L; s}\int\frac{d\omega d k_z}{2\pi } \Bigg[a_{\rm }(\omega,k_z,s,n_L)_{\mathit{H^-_R}}U_{s,n_L}(x;\omega,k_z)_{\mathit{H^-_R}}+b^{\dagger}_{\rm}(\omega,k_z,s,n_L)_{\mathit{H^-_R}}V_{s,n_L}^*(x;\omega,k_z)_{\mathit{H^-_R}}\\ + a_{\rm}(\omega,k_z,s,n_L)_{\mathit{I^-_R}}U_{s,n_L}(x;\omega,k_z)_{\mathit{I^-_R}}+b^{\dagger}_{\rm }(\omega,k_z,s,n_L)_{\mathit{I^-_R}}V_{s,n_L}^*(x;\omega,k_z)_{\mathit{I^-_R}}\Bigg] \\
=\sum_{n_L; s}\int\frac{d\omega d k_z}{2\pi } \Bigg[a_{\rm}(\omega,k_z,s,n_L)_{\mathit{H^+_R}}U_{s,n_L}(x;\omega,k_z)_{\mathit{H^+_R}}+b^{\dagger}_{\rm }(\omega,k_z,s,n_L)_{\mathit{H^+_R}}V_{s,n_L}^*(x;\omega,k_z)_{\mathit{H^+_R}}\\ +a_{\rm }(\omega,k_z,s,n_L)_{\mathit{I^+_R}}U_{s,n_L}(x;\omega,k_z)_{\mathit{I^+_R}}+b^{\dagger}_{\rm}(\omega,k_z,s,n_L)_{\mathit{I^+_R}}V_{s,n_L}^*(x;\omega,k_z)_{\mathit{I^+_R}}\Bigg] 
    \end{split}
\end{equation}
where creation and annihilation operators are assumed to satisfy the usual canonical anti-commutation relations.
Using the properties of confluent hypergeometric functions \cite{AS}, we can write the Bogoliubov relation between `in' and `out' modes as \begin{equation}
     \label{BTM}
     U_{s,n_L}(x)_{\mathit{H^-_R}}=\alpha_{s}^{*}  U_{s,n_L}(x)_{\mathit{I^+_R}}+\beta_s^{*}  ( V_{s,n_L}(x)_{\mathit{I^+_R}})^*
\end{equation}
    \begin{equation}
         V_{s,n_L}(x)_{\mathit{H^-_R}}=\alpha_{s}^{*}  V_{s,n_L}(x)_{\mathit{I^+_R}}+\beta_s^{*}  ( U_{s,n_L}(x)_{\mathit{I^+_R}})^*
         \label{BTM1}
    \end{equation}
where $s=1,2$ in \ref{BTM} and \ref{BTM1}, and, $\alpha_s$ and $\beta_s$ are the Bogoliubov coefficients given by
\begin{align}
\label{Bcoeff}
\alpha_s = \frac{N_s \Gamma(1-\lambda_s)\sin\pi(\lambda_s-\nu)}{M_s\sin \pi  \nu },
\beta_s = \frac{N_s \sin \pi  \lambda_s \Gamma (\nu -\lambda_s )}{R_s\sin \pi  \nu  }
\end{align}
It is easy to check using the expressions given in \ref{A1} that $|\alpha_s|^2+|\beta_s|^2=1$. We note from \ref{FQR1}, that there exists another independent set of the Bogoliubov transformations connecting the `in' modes on $I^-_R$ to the `out' modes on $H^+_R$, analogous to \ref{BTM} and \ref{BTM1}. However, since these two sets are independent of each other, we shall only focus below on the first for the sake of simplicity. \\
From \ref{FQR1}, the Bogoliubov transformations between the creation and annihilation operators are given by
\begin{equation}
\label{Bogoin_out}
a_{\rm }(\omega,k_z,s,n_L)_{\mathit{H^-_R}}\;=\;\alpha_s a_{\rm }(\omega,k_z,s,n_L)_{\mathit{I^+_R}}-\beta_{s}^{*} b_{\rm }^{\dagger}(-\omega,-k_z,s,n_L)_{\mathit{I^+_R}}
\end{equation}
\begin{equation}
\label{Bogoin_out1}
b_{\rm}(\omega,k_z,s,n_L)_{\mathit{H^-_R}}\;=\;\alpha_s b_{\rm }(\omega,k_z,s,n_L)_{\mathit{I^+_R}}+\beta_{s}^{*} a^{\dagger}_{\rm }(-\omega,-k_z,s,n_L)_{\mathit{I^+_R}}
\end{equation}
 The coefficient $\beta_s$ is responsible for pair creation, and the quantity $\lvert \beta_s \rvert^2$ is the mean number density of particles for the $R$ wedge 
\begin{equation}
    \label{beta2}
\lvert\beta_s\rvert^2= \frac{\sinh^3{\pi \Delta}}{e^{\pi \Delta } \cosh^3{\pi(\Delta-\frac{\omega}{a})}+\sinh^3{\pi \Delta}} 
\end{equation}
where the parameter $\Delta$ is defined as,
\begin{equation}
\label{delta}
   \Delta = \text{Im}{(\lambda_s)}=\frac{m^2+S_s}{2eE} 
\end{equation}
where $\lambda_s$ is defined in \ref{coeff'}. We note that $\lvert\beta_s\rvert^2$ is independent of the spatial momentum. For $\Delta \to \infty$, which corresponds to $E \to 0$ or $e \to 0$ or $B\to \infty$, the number density for local vacuum vanishes, i.e., $\lvert\beta_s\rvert^2 \to 0$  and this behaviour is similar to the earlier scenario of the Minkowski vacuum in \ref{Background magnetic field and quantum correlations in the Schwinger effect}. In particular, we also note that $\lvert\beta_s\rvert^2$ depends on  the acceleration of the Rindler observer. Thus \ref{beta2}, quantifies the Schwinger effect in a uniformly accelerated frame, in a given wedge. However, it does not contain any contribution due to the Unruh effect, as we have not considered yet the other Rindler wedge or the global vacuum. We wish to perform this task below.  

\subsection{Field quantization in the left Rindler wedge}
The field equations and their solutions are the same on the left and right wedges. The only difference between $R$ and $L$ wedges is that in $L$, the timelike Killing vector field $\partial_t$ is past directed compared to that of in $R$ \cite{leftright}. This implies that the sign of the charges and the `in' and `out' labels have to be interchanged in $L$ compared to that of $R$. Hence, from \ref{mode1}, \ref{mode2} and \ref{mode1a} the complete set of `in' modes for the $L$ wedge is given by
\begin{equation}
\label{fullmode1L}
\begin{split}
    U_{s,n_L}(x_L)_{\mathit{H^-_L}}=\frac{e^{e^{-ax_L}/2} }{N_s} (ie^\mu_a \gamma^{a}\partial_\mu-e A_\mu e^\mu_a \gamma^{a}+m)e^{-i \omega (t_L+x_L)} e^{i k_z z} e^{a x_L} e^{-\frac{i e E e^{2 a x_L}}{4 a^2}}\\ \times(U(\lambda_s, \nu, \xi_L))^*H_s(y) \epsilon_s,
    \end{split}
\end{equation}
\begin{equation}
\label{fullmode2L}
\begin{split}
    U_{s,n_L}(x_L)_{\mathit{I^-_L}}=\frac{e^{e^{-ax_L}/2} }{M_s} (ie^\mu_a \gamma^{a}\partial_\mu-e A_\mu e^\mu_a \gamma^{a}+m) e^{-i \omega (t_L-x_L)} e^{i k_z z} e^{a x_L} e^{-\frac{i e E e^{2 a x_L}}{4 a^2}} \\ \times L(-\lambda_s, \nu-1, \xi_L) H_s(y) \epsilon_s,
    \end{split}
\end{equation}
\begin{equation}
\begin{split}
\label{fullmode1aL}
    V_{s,n_L}(x_L)_{\mathit{H^-_L}}=\frac{e^{e^{-ax_L}/2} }{P_s} (ie^\mu_a \gamma^{a}\partial_\mu-e A_\mu e^\mu_a \gamma^{a}+m)e^{i \omega (t_L+x_L)} e^{-i k_z z} e^{a x_L} e^{-\frac{i e E e^{2 a x_L}}{4 a^2}} \\ \times (e^{\xi_L}U(\nu-\lambda_s,\nu,\xi_L))^* H_s(y) \epsilon_s,
    \end{split}
\end{equation}
\begin{equation}
\label{fullmode2aL}
\begin{split}
    V_{s,n_L}(x_L)_{\mathit{I^-_L}}=\frac{e^{e^{-ax_L}/2} }{R_s} (ie^\mu_a \gamma^{a}\partial_\mu-e A_\mu e^\mu_a \gamma^{a}+m) e^{i \omega (t_L-x_L)} e^{-i k_z z} e^{a x_L} e^{-\frac{i e E e^{2 a x_L}}{4 a^2}} \\ \times   \xi_L^{1-\nu} L(\nu-\lambda_s-1,1-\nu,\xi_L)  H_s(y) \epsilon_s,
    \end{split}
    \end{equation}
where $s=1,2$ for \ref{fullmode1L}, \ref{fullmode2L} and $s=3, 4$ for \ref{fullmode1aL},  \ref{fullmode2aL} and the parameter $\xi$ in defined by \ref{ell}. The above two equations, respectively, denote the positive and negative frequency `in' modes in the left Rindler wedge. Likewise, using the definition for the `out' modes given in the preceding section, we find the `out' modes for $L$ wedge, shown explicitly in \ref{A1}. The Bogoliubov coefficients and the transformations between modes at $\mathit{I^-_L}$ and $\mathit{H^+_L}$ will remain similar to that of the $R$ wedge \ref{Bcoeff} and \ref{Bogoin_out}.

We note that as of the preceding Chapter, in this case, as well, there is no spin mixing. Thus, for simplicity, we will only work with a single spin, say with $s=1$. 
\noindent

\section{The global modes and further Bogoliubov relationship}
\label{S3}
Neither the region $R$ nor the region $L$ can cover the whole of the Minkowski spacetime, \ref{diagram}. Thus, in order to construct a sensible quantum field theory in the whole of the Minkowski spacetime, we need to construct global modes which are analytic in $R\cup L$. As was reviewed in \ref{The Unruh effect}, following \cite{Unruh:1976db, Global_1}, we shall achieve this via the analytic continuation and hybridisation of the $R$ and $L$ local modes derived above.

We construct the global `in' modes using the asymptotic behaviour of the local modes shown explicitly in \ref{B}. We also note from \ref{BTM} and \ref{BTM1} the Bogoliubov relationship between the modes at infinity and on the horizon. By taking into account similar relationships for the $L$ region, we note that the `in' mode at $\mathit{I^-_L}$ will eventually reach $\mathit{H^+_L}$. Considering this, for example, we make the linear combination of the local `in' modes at $\mathit{H^-_R}$ and $\mathit{I^-_L}$ to construct one of the global `in' modes. It is clear that even though this addition is non local initially, they will become local eventually at $\mathit{H_R^-} \cap \mathit{H_L^+} $, when the $\mathit{I_L^-}$ modes reaches $\mathit{H_L^+}$ as dictated by the aforementioned Bogoliubov relationships. Certainly, this combination, apart from the usual Unruh effect, will carry information about the Schwinger effect as well. That's why we have regarded them as the global {\it in modes}. According to this scheme, the normalised hybrid global `in' modes in terms of the local `in' modes are given by
\begin{equation}
    \label{Gmode1}
    \phi^{G}_{1}(x)=\frac{1}{\sqrt{2\; \cosh\frac{\omega \pi}{a}}}\big(e^{\frac{\pi \omega}{2a}} U_{1,n_L}(x)_{\mathit{H^-_R}}+e^{-\frac{\pi \omega}{2a}}V_{1,n_L}(x)_{\mathit{I^-_L}}\big)
\end{equation}
\begin{equation}
    \label{Gmode11}
    \phi^{G}_{2}(x)=\frac{1}{\sqrt{2\; \cosh\frac{\omega \pi}{a}}}\big(e^{\frac{\pi\omega}{2a}} U_{1,n_L}(x)_{\mathit{I^-_L}}+e^{-\frac{\pi\omega}{2a}}V_{1,n_L}(x)_{\mathit{H^-_R}}\big)
\end{equation}
\begin{eqnarray}
    \label{Gmode111}
     \phi^{G}_{3}(x)=\frac{1}{\sqrt{2\; \cosh\frac{\omega \pi}{a}}}\big(e^{\frac{\pi\omega}{2a}} V_{1,n_L}(x)_{\mathit{H^-_R}}-e^{-\frac{\pi \omega}{2a}}U_{1,n_L}(x)_{\mathit{I^-_L}}\big)\\
     \label{Gmode1111}
    \phi^{G}_{4}(x)=\frac{1}{\sqrt{2\; \cosh\frac{\omega \pi}{a}}}\big(e^{\frac{\pi \omega}{2a}}V_{1,n_L}(x)_{\mathit{I^-_L}}-e^{-\frac{\pi \omega}{2a}}U_{1,n_L}(x)_{\mathit{H^-_R}}\big)
\end{eqnarray} 
It is also easy to see that the above global modes are orthonormal. These global modes can further be expressed in terms of the local  `out' modes by using the Bogoliubov transformations from \ref{BTM} in \ref{Gmode1}, \ref{Gmode11}, \ref{Gmode111} and \ref{Gmode1111}. 

Further, we write the field quantization of the Dirac field $\psi$ in $R\cup L$ in terms of the local modes in $R$ and $L$ as well as in terms of the global modes (we have suppressed the subscript for $s$ from now onwards), given as follows
\begin{align}
\label{localmode}
    \psi(x)&=\sum_{n_L}\int\frac{d\omega d k_z}{2\pi } \Bigg[a_{\rm }(\omega,k_z,n_L)_{\mathit{H^-_R}}U_{n_L}(x;\omega,k_z)_{\mathit{H^-_R}}+b^{\dagger}_{\rm }(\omega,k_z,n_L)_{\mathit{H^-_R}}V_{n_L}(x;\omega,k_z)_{\mathit{H^-_R}}\nonumber\\&+a_{\rm}(\omega,k_z,n_L)_{\mathit{I^-_L}}U_{n_L}(x;\omega,k_z)_{\mathit{I^-_L}}+b^{\dagger}_{\rm }(\omega,k_z,n_L)_{\mathit{I^-_L}}V_{n_L}(x;\omega,k_z)_{\mathit{I^-_L}}\Bigg] \nonumber\\
&=\sum_{n_L}\int\frac{d\omega d k_z}{2\pi } \Bigg[a_{\rm }(\omega,k_z,n_L)_{\mathit{I^+_R}}U_{n_L}(x;\omega,k_z)_{\mathit{I^+_R}}+b^{\dagger}_{\rm }(\omega,k_z,n_L)_{\mathit{I^+_R}}V_{n_L}(x;\omega,k_z)_{\mathit{I^+_R}} \nonumber\\ &+ a_{\rm}(\omega,k_z,n_L)_{\mathit{H^+_L}}U_{n_L}(x;\omega,k_z)_{\mathit{H^+_L}}+b^{\dagger}_{\rm}(\omega,k_z,n_L)_{\mathit{H^+_L}}V_{n_L}(x;\omega,k_z)_{\mathit{H^+_L}}\Bigg]
   \end{align}
   \begin{equation}
\label{glabalfield}
\begin{split}
    \psi(x)=\sum_{n_L}\int\frac{d\omega d k_z}{2\pi } \Bigg[c_{\rm 1}(\omega,k_z,n_L)\phi_1^G(x)+d^{\dagger}_{\rm 1}(\omega,k_z,n_L)\phi_3^G(x)+c_{\rm 2}(\omega,k_z,n_L)\phi_4^{G}(x)\\+d^{\dagger}_{\rm 2}(\omega,k_z,n_L)\phi_2^{G}(x)\Bigg]
    \end{split}
\end{equation}
Comparing \ref{localmode} and \ref{glabalfield}, we obtain the Bogoliubov relations given by
\begin{equation}
\label{Bogoglobal}
    \begin{split}
    c_1=\frac{1}{\sqrt{2 \cosh \frac{\omega \pi}{a}}}\big(e^{\frac{\pi \omega}{2a}}a_{\mathit{H^-_{R}}}\left(\omega,k_z,n_L\right)-e^{-\frac{\pi \omega}{2a}}b^{\dagger}_{\mathit{I^-_{L}}}\left(-\omega,-k_z,n_L\right)\big),\\ d_1^{ \dagger}=\frac{1}{\sqrt{2 \cosh \frac{\omega \pi}{a}}}\big(e^{\frac{\pi \omega}{2a}}a_{\mathit{I^-_{L}}}(\omega,k_z,n_L)-e^{-\frac{\pi \omega}{2a}} b^{\dagger}_{\mathit{H^-_{R}}}(-\omega,-k_z,n_L)\big)
    \end{split}
\end{equation}
Substituting further \ref{Bogoin_out} and \ref{Bogoin_out1} into \ref{Bogoglobal},
we obtain the relationship between the global and the local `out' operators 
\begin{equation}
    \label{Global_outBC}
    c_1=\frac{1}{\sqrt{2 \cosh\frac{\omega \pi}{a}}}\Bigg(e^{\frac{\pi \omega}{2a}}\alpha_1a_{\mathit{I^+_{R}}}-e^{\frac{\pi \omega}{2a}}\beta_1^*b^{\dagger}_{\mathit{I^+_{R}}}-e^{-\frac{\pi \omega}{2a}}\alpha_1^*b^{\dagger}_{\mathit{H^+_L}}+e^{-\frac{\pi \omega}{2a}}\beta_1a_{\mathit{H^+_L}}\Bigg)
\end{equation}
Similarly, there will be another set of creation and annihilation operators $\left(c_2, d_2^{\dagger}\right)$ corresponding to another set of the global modes. The global vacuum can therefore be defined as $|0\rangle=|0\rangle^{1} \otimes |0\rangle^{2} $, where $|0\rangle^{1}$ is annihilated by $\left(c_1, d_1\right)$ and $|0\rangle^{2}$ is annihilated by $\left(c_2, d_2\right)$. We will work with only $|0\rangle^{1}$ as the other will have similar characteristics. Using the Bogoliubov relationship, we can write $|0\rangle^{1}$ in terms of the local `out' basis. We are now ready to compute the number density and correlations.
\noindent
\section{The logarithmic negativity for the vacuum}\label{S4}
The local `in' vacuum is defined as
\begin{equation}
    \label{in vaccum}
    a_{\mathit{H^-_{R}}}\lvert 0\rangle^{\mathit{H^-_R}}=b_{\mathit{H^-_{R}}}\lvert0\rangle^{\mathit{H^-_R}}=0 \;,\;\;a_{\mathit{I^-_{L}}}\lvert0\rangle^{\mathit{I^-_L}}=b_{\mathit{I^-_{L}}}\lvert0\rangle^{\mathit{I^-_L}}=0
\end{equation}
and the local `out' vacuum is defined as,
\begin{equation}
    \label{out vaccum}
a_{\mathit{I^+_{R}}}\lvert0\rangle^{\mathit{I^+_R}}=b_{\mathit{I^+_{R}}}\lvert 0\rangle^{\mathit{I^+_R}}=0,\;\;a_{\mathit{H^+_{L}}}\lvert 0\rangle^{\mathit{H^+_L}}=b_{\mathit{H^+_{L}}}\lvert0\rangle^{\mathit{H^+_L}}=0
\end{equation}
Using \ref{Bogoin_out} and \ref{Bogoin_out1}, $\lvert 0\rangle^{\mathit{H^-_R}}$ can be expressed in terms of the local `out' basis as
\begin{equation}
   \label{in_out_R}
    \lvert 0 \rangle^{\mathit{H^-_R}} = \alpha_1 \lvert 0_k 0_{-k} \rangle^{\mathit{I^+_R}}+\beta_1 \lvert 1_k 1_{-k} \rangle^{\mathit{I^+_R}} 
\end{equation}
Next, the global vacuum is defined as
\begin{equation}
    \label{globalvac1}
    c_1|0\rangle^1 = d_1|0\rangle^1=0
\end{equation}
Following the Bogoliubov relationship for $R$ wedge \ref{Global_outBC} and by taking into account the similar relationships for $L$ region, we express the global `in' vacuum in terms of the local `out' basis as follows
\begin{equation}
\label{global_vacua}
\begin{split}
    |0\rangle^{1} \equiv|0_k0_{-k}\rangle^1 =\frac{1}{(1+e^{-\frac{2\pi \omega}{a}})^{\frac{1}{2}}} \bigg(\alpha_1^{2}|0_k 0_{-k};0_k 0_{-k}\rangle^{\mathit{I^+_R};\mathit{H^+_L}}+\beta_1^{* 2}|1_k 1_{-k};1_k 1_{-k}\rangle^{\mathit{I^+_R};\mathit{H^+_L}}\\+\alpha_1 \beta_1^{*}\big(|1_k 1_{-k};0_k 0_{-k}\rangle^{\mathit{
    I^+_R};\mathit{H^+_L}}+|0_k 0_{-k};1_k 1_{-k}\rangle^{\mathit{I^+_R};\mathit{H^+_L}}\big)+e^{-\frac{\pi \omega}{a}}|1_k 0_{-k};0_k 1_{-k}\rangle^{\mathit{I^+_R};\mathit{H^+_L}}\bigg)
    \end{split}
\end{equation}
where the first two and last two entries correspond to $R$ and $L$ wedges, respectively, and $^1\langle 0|0\rangle^{1}=1$. The state space $\mathcal{H}$ is constructed by the tensor product, $\mathcal{H} =  \mathcal{H}_k^\mathit{R} \otimes \mathcal{H}_{-k}^\mathit{R}\otimes \mathcal{H}_k^\mathit{L} \otimes \mathcal{H}_{-k}^\mathit{L}$, where $\mathcal{H}_{k}^\mathit{R}$ ($\mathcal{H}_k^\mathit{L}$) and $\mathcal{H}_{-k}^\mathit{R}$ ($\mathcal{H}_{-k}^\mathit{L}$) are the state spaces of the modes of the particle and the antiparticle in $R(L)$ wedge, respectively. We find the spectra of particles and antiparticles in the Minkowski vacuum from the perspective of a Rindler observer near $\mathit{I}^+_R$ and $\mathit{H}^+_L$, respectively, given by 
 \begin{equation}
     \label{vaccum_ND}
  N=\;^1\langle0|\;a_{I^+_{R}}^\dagger a_{I^+_{R}}|0\rangle^1=\frac{ \lvert\beta_1\rvert^2 e^{\frac{2 \pi \omega}{a}} }{1+e^{\frac{2\pi \omega}{a}}}+\frac{1}{1+e^{\frac{2\pi \omega}{a}}}=\;^1\langle0|\;b_{H^+_{L}}^\dagger b_{H^+_{L}}|0\rangle^1
 \end{equation}
 where $\lvert\beta_1\rvert^2$ is given by \ref{beta2}, in terms of variable $\Delta$, defined in \ref{ell}. In \ref{vaccum_ND}, the first term on the right-hand side depends on the parameter $\Delta$ and $a$, whereas the second term is independent of parameter $\Delta$. 
In the limit $\Delta \to \infty$ (i.e., $E \to 0$ or $e \to 0$ or $B \to \infty$), \ref{vaccum_ND} reduces to
 \begin{equation}
     \label{Planck_spectrum}
 N = \frac{1}{1+e^{\frac{2 \pi \omega}{a}}}
\end{equation}
The result \ref{Planck_spectrum} is the fermionic Planck spectrum with temperature $T=a/2\pi$, which is the usual Unruh temperature observed by an observer moving with uniform acceleration in the Minkowski spacetime \cite{ SHN:2020, roy}. Similar fermionic spectra are obtained in \ref{Fermionic Bell violation in de Sitter spacetime with background electric and mangnetic fields} for zero electric field in the de Sitter spacetime; there, the non-zero number density was due to the gravitational field. \ref{Planck_spectrum}, thus serves as a consistency check of our main result.

At the limit of vanishing acceleration, i.e., $a\to0$, the number density given by \ref{vaccum_ND} vanishes. 
This apparent ambiguity is due to the fact that the quantization of a charged field in the Rindler coordinates differs from the Minkowski coordinates. It leads to an unequal distribution of particles and antiparticles in different Rindler wedges in the presence of background electromagnetic fields, known as the charge polarization \cite{Gabriel:1999yz}. Of course, the total charge in the Minkowski spacetime is always conserved. 
Thus in order to get the full scenario, we also need to compute the particle and antiparticle spectra at all the possible places in $R \cup L$. \ref{vaccum_ND}, represents the number density of particles and antiparticles near $\mathit{I^{+}_R}$ and $\mathit{H^{+}_L}$, respectively. A similar analysis has been done for different regions of the particular Rindler wedge for the charged scalar field; for details, we refer our reader to\cite{Gabriel:1999yz}.
A complete analysis of particle and antiparticle spectra pertaining to this problem seems interesting, and we shall address it in a future publication.
\noindent 
\subsection{The logarithmic negativity}
In this thesis, our focus is on computing bipartite entanglement, but the global vacuum given by \ref{global_vacua} is a quadripartite state. To gain insights into the entanglement structure of the global vacuum, our objective is to compute the logarithmic negativity for the density matrix of particles and antiparticles in $R$ and $L$ wedges, respectively.

In order to understand the entanglement structure of the global vacuum, we wish to compute the logarithmic negativity for the particles and antiparticles in $R$ and $L$ wedges, respectively. 

\begin{figure}[h]
\begin{center}
		\includegraphics[scale=.55]{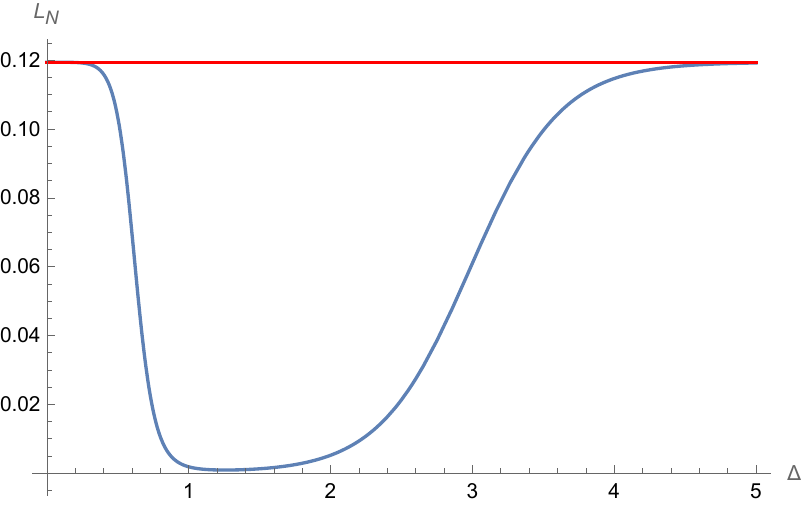}\hspace{1.0cm}
		\caption{\it{\small Logarithmic negativity between the particles and antiparticles in $R$ and $L$ wedges, respectively. As we have discussed in the main text, we have plotted \ref{LN} (blue curve) and \ref{LNE} (red curve) variation with respect to the parameter $\Delta= \frac{m^2 + (2 n_L + 1) e B}{e E}$, where we have taken $\frac{\omega}{a}=1$. Logarithmic negativity first monotonically decreases with an increase in $\Delta$ then reaches a plateau after that increases monotonically with increasing $\Delta$.}
		}
		\label{fig:LNforVacuum}
		\end{center}
\end{figure}

The density operator for the global vacuum is defined as $\rho_{\text{global}}=|0\rangle^{1} ~^{1}\langle0|$. We obtain the reduced density operator for particles and antiparticles in $R$ and $L$ wedges, respectively, by tracing out antiparticles and particles of $R$ and $L$ wedges, respectively, given by
\begin{equation}
 \label{densityRL}
\begin{split}
    \rho^{p;a}_{R;L}= \frac{1}{1+e^{-\frac{2 \pi \omega}{a}}}\Big[\lvert\alpha_1\rvert^4 \lvert00\rangle \langle00\rvert +(\lvert\beta_1\rvert^4+e^{-\frac{2 \pi \omega}{a}})\lvert 11\rangle \langle11\rvert+\lvert\alpha_1\rvert^2 \lvert\beta_1\rvert^2(\lvert10\rangle \langle10\rvert+\lvert01\rangle \langle01\rvert) \\
    +e^{-\frac{\pi \omega}{a}}(\alpha_1^2\lvert00\rangle \langle11\rvert+\alpha_1^{*2}\lvert11\rangle \langle00\rvert)\Big]
\end{split}
\end{equation}
Following the definition of logarithmic negativity given in \ref{L-N}. We computed $L_N$ for $\rho^{p;a}_{R;L}$, given by
\begin{equation}
    \label{LN}
    L_N= \log_2\Big[1+\frac{e^{-\frac{\pi \omega}{a}}(\alpha_1^2+\alpha_1^{*2})}{1+e^{-\frac{2\pi \omega}{a}}}\Big]
\end{equation}
where $L_N$ is a function of $\Delta$ and $a$. This clarifies that the entanglement depends on the motion of the observer as well as on the strength of background electric and magnetic fields. For $\Delta \to \infty$,
\begin{equation}
\label{LNE}
L_N=\log_2\Big[1+\frac{2}{e^{\frac{\pi \omega}{a}}+e^{-\frac{\pi \omega}{a}}}\Big]
\end{equation}
It represents the usual $R-L$ entanglement.We have plotted \ref{LN} (blue curve)
and \ref{LNE} (red curve) vs. $\Delta$ keeping the parameter $\omega/a$ fixed, in \ref{fig:LNforVacuum}. In the presence of the background electric and magnetic fields, initially, $L_N$ decreases monotonically with an increase in $\Delta$, near $\Delta \approx 1$, it reaches the minima and further increases monotonically with the increase in $\Delta$. Whereas the red curve represents \ref{LNE}.  
This non-monotonic behaviour of logarithmic negativity is probably due to the mixed-state structure of the density matrix. 
We next wish to investigate the entanglement properties of some Bell states constructed from two fermionic fields, in this context.
\noindent
\section{A zero charge state}
\label{Logarithmic negativity and mutual information of a zero charge state constructed by two fermionic fields}
Based on the argument of the gauge transformation properties given in \ref{Background magnetic field and quantum correlations in the Schwinger effect}, we consider a maximally entangled zero-charge state, $
	\rho^{(0)}
=
	\ketN{\psi}
	\braN{\psi}
$, which is a pure state, with
\begin{equation}
    \label{zerochargeRinler}
|\psi\rangle=\frac{|0_k 1_{-k} 1_p 0_{-p}\rangle^{\text{in}}+|0_k0_{-k}0_p0_{-p}\rangle^{\text{in}}}{\sqrt{2}}
\end{equation}
In our notation, the first (second) pair of entries appearing in the kets stands for the first (second) fermion. For a specific pair, the first (second) entry represents particle (antiparticle) degrees of freedom. Using \ref{global_vacua}, we re-express \ref{zerochargeRinler} in the `out' basis as
\begin{equation}
\begin{split}
    \label{zero_out_R}
|\psi\rangle=\frac{1}{P}\Big[(\alpha_k^2|0_k1_{-k}0_k0_{-k}\rangle^{\mathit{I^+_R};\mathit{H^+_L}}+\alpha_k \beta_k^*|0_k1_{-k}1_{k}1_{-k}\rangle^{\mathit{I^+_R};\mathit{H^+_L}}+e^{-\frac{\pi \omega}{a}}|1_k1_{-k}0_k1_{-k}\rangle^{\mathit{I^+_R};\mathit{H^+_L}})\\ \otimes (\alpha_p^2|0_p1_{-p}0_p0_{-p}\rangle^{\mathit{I^+_R};\mathit{H^+_L}}+\alpha_p \beta_p^*|0_p1_{-p}1_{p}1_{-p}\rangle^{\mathit{I^+_R};\mathit{H^+_L}}+e^{-\frac{\pi \omega}{a}}|1_p1_{-p}0_p1_{-p}\rangle^{\mathit{I^+_R};\mathit{H^+_L}})\Big]\\+\frac{1}{P^\prime}\Big[\Big(\alpha_k^{2}|0_k 0_{-k};0_k 0_{-k}\rangle^{\mathit{I^+_R};\mathit{H^+_L}}+\beta_k^{* 2}|1_k 1_{-k};1_k 1_{-k}\rangle^{\mathit{I^+_R};\mathit{H^+_L}}+\alpha_k \beta_k^{*}\big(|1_k 1_{-k};0_k 0_{-k}\rangle^{\mathit{
    I^+_R};\mathit{H^+_L}}\\+|0_k 0_{-k};1_k 1_{-k}\rangle^{\mathit{I^+_R};\mathit{H^+_L}}\big)+e^{-\frac{\pi \omega}{a}}|1_k 0_{-k};0_k 1_{-k}\rangle^{\mathit{I^+_R};\mathit{H^+_L}}\Big)\otimes\Big(\alpha_p^{2}|0_p 0_{-p};0_p 0_{-p}\rangle^{\mathit{I^+_R};\mathit{H^+_L}}\\+\beta_p^{* 2}|1_p 1_{-p};1_p 1_{-p}\rangle^{\mathit{I^+_R};\mathit{H^+_L}}+\alpha_p \beta_p^{*}\big(|1_p 1_{-p};0_p 0_{-p}\rangle^{\mathit{
    I^+_R};\mathit{H^+_L}}+|0_p 0_{-p};1_p 1_{-p}\rangle^{\mathit{I^+_R};\mathit{H^+_L}}\big)\\+e^{-\frac{\pi \omega}{a}}|1_p 0_{-p};0_p 1_{-p}\rangle^{\mathit{I^+_R};\mathit{H^+_L}}\Big)\Big]
\end{split}
\end{equation}
where we have suppressed the subscript $s=1$ from $\alpha_1$ and $\beta_1$ to shorten the notation. The coefficients $P$ and $P^\prime$ are given by
$$P=\Big(2(|\alpha_k|^2+e^{-\frac{2\pi\omega}{a}})(|\alpha_p|^2+e^{-\frac{2\pi\omega}{a}})\Big)^{1/2} \quad\text{and}\quad P^\prime=\sqrt{2}(1+e^{-\frac{2\pi\omega}{a}})$$
We shall focus below only on the correlations between the particle-particle and particle-antiparticle sectors corresponding to the density matrix of the above state. 
Accordingly, tracing out first the antiparticle-antiparticle degrees  of freedom from $\rho^{(0)}$, we have the reduced density matrix for the particle-particle (particles near $I^+_R$ and $H^+_L$) sector,
\begin{equation}
    \label{rho11}
    \rho^{R;L}_{k;p} =\left(
\begin{array}{cccc}
 A_1 & E_1^* & 0 & 0 \\
E_1 & B_1 & 0 & 0 \\
 0 & 0 & C_1 & 0 \\
 0 & 0 & 0 & D_1 \\
\end{array}
\right)
\end{equation}
These matrix elements are given by
$$A_1=\frac{| \alpha | ^2}{2}  \left[\frac{1}{| \alpha | ^2+e^{-\frac{2 \pi  \omega }{a}}}+\frac{| \beta | ^2}{1+e^{-\frac{2 \pi  \omega }{a}}}\right],\;B_1=\frac{2 +| \alpha | ^2  | \beta | ^2 \left(| \beta | ^4+1\right)}{2 \left(1+e^{\frac{2 \pi  \omega }{a}}\right)}$$
$$
C_1=\frac{1}{2} \left[\frac{1}{| \alpha | ^2+e^{\frac{2 \pi  \omega }{a}}}+\frac{| \beta | ^2 \left(1+| \beta | ^2e^{\frac{2 \pi  \omega }{a}}\right)}{1+e^{\frac{2 \pi  \omega }{a}}}\right],\;D_1=\frac{| \alpha | ^2 \left(1+| \alpha | ^2 e^{\frac{2 \pi  \omega }{a}}\right)}{2 \left(1+e^{\frac{2 \pi  \omega }{a}}\right)}$$ $$ E_1=\frac{\alpha \beta}{2}\Big(\frac{1+| \alpha | ^2 e^{\frac{2 \pi  \omega }{a}}}{1+e^{\frac{2 \pi  \omega }{a}}}\Big)^{1/2}$$
For computational simplicity, we assume that both fields have
the same rest mass, and we consider modes in which their momenta along the $z$-direction
and the Landau levels are the same.

\begin{figure}[h]
    \centering
\includegraphics{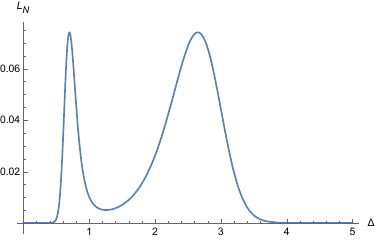}\hspace{1.0cm}
\includegraphics{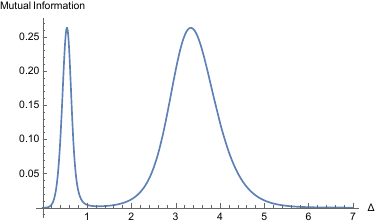}
    \caption{\it{\small Logarithmic negativity and mutual information of $\rho^{R;L}_{k;p}$ with respect to the parameter $\Delta=\frac{m^2+(2n_L+1)eB}{eE}$, where we have taken $\frac{\omega}{a}=2$. Both are non-monotonic and, at a very small or large value of $\Delta$, it vanishes.}}
    \label{rho1R}
\end{figure}

 Using the definition of mutual information and logarithmic negativity given in \ref{A short introduction to quantum information}. The mutual information and the logarithmic negativity for $\rho^{R;L}_{k;p}$ are given by 
\begin{eqnarray}
    \label{mi1}
    MI(\rho^{R;L}_{k;p})&=&\frac{1}{2} \Bigg[-(A_1+D_1)\log(A_1+D_1)-(B_1+C_1)\log(B_1+C_1)-(A_1+C_1)\nonumber\\ \nonumber&&\times \log(A_1+C_1)-(B_1+D_1)\log(B_1+D_1)+C_1\log C_1+D_1 \log D_1
   \nonumber \\\nonumber&& +\frac{A_1 + B_1- \sqrt{A_1^2 - 2 A_1 B_1 + B_1^2 + 4 |E_1|^2}}{2} \\ \nonumber&&\times \log\Big(\frac{A_1 + B_1 - \sqrt{A_1^2 - 2 A_1 B_1 + B_1^2 + 4 |E_1|^2}}{2}\Big)\\ \nonumber&&+\frac{A_1 + B_1+ \sqrt{A_1^2 - 2 A_1 B_1 + B_1^2 + 4 |E_1|^2}}{2} \\ &&\times \log\Big(\frac{A_1 + B_1 + \sqrt{A_1^2 - 2 A_1 B_1 + B_1^2 + 4 |E_1|^2}}{2}\Big)\Bigg]
\end{eqnarray}
\begin{equation}
\begin{split}
\label{ln1}
    L_N(\rho^{R;L}_{k;p})=\log \Big[A_1+B_1+\frac{1}{2} (C_1+D_1-\sqrt{C_1^2-2 C_1 D_1+4 |E_1|^2+D_1^2})\\+\frac{1}{2} (C_1+D_1+\sqrt{C_1^2-2 C_1 D_1+4 |E_1|^2+D_1^2}) \Big]
    \end{split}
\end{equation}
The variation of \ref{mi1} and \ref{ln1} with respect to the parameter $\Delta$ is shown in \ref{rho1R}. It is non-monotonic and vanishes at large and small values of $\Delta$.

\begin{figure}
    \centering
\includegraphics{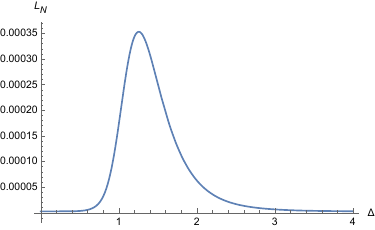}\hspace{1.0cm}
\includegraphics{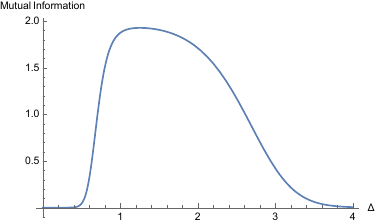}
    \caption{\it{\small Logarithmic negativity and mutual information of $\rho^{R;L}_{k;p}$ with respect to the parameter $\Delta=\frac{m^2+(2n_L+1)eB}{eE}$, where we have taken $\omega/a=2$. Both are non-monotonic and, at a very small or large value of $\Delta$, they vanish.}}
    \label{rho2R}
\end{figure}

Further, the reduced density matrix for the particle-antiparticle (particle near $I^+_R$ and antiparticle near $H^+_L$) sector is given by
\begin{equation}
    \label{rho22}
    \rho^{R;L}_{k;-p} =\left(
\begin{array}{cccc}
 A_2 & E_2^* & 0 & 0 \\
E_2 & B_2 & 0 & 0 \\
 0 & 0 & C_2 & 0 \\
 0 & 0 & 0 & D_2 \\
\end{array}
\right)
\end{equation}
The matrix elements are given by
\begin{center}
$A_2=\frac{| \alpha | ^4}{2}  \left[\frac{1}{\left(| \alpha | ^2+e^{-\frac{2 \pi  \omega }{a}}\right)^2}+\frac{1}{1+e^{-\frac{2 \pi  \omega }{a}}}\right], B_2=| \alpha | ^2 \left[\frac{e^{-\frac{2 \pi  \omega }{a}}}{\left(| \alpha | ^2+e^{-\frac{2 \pi  \omega }{a}}\right)^2}+\frac{| \beta | ^2}{2 \left(1+e^{-\frac{2 \pi  \omega }{a}}\right)}\right]$
$C_2=\frac{e^{-\frac{4 \pi  \omega }{a}} }{2} \left[\frac{1}{\left(| \alpha | ^2+e^{-\frac{2 \pi  \omega }{a}}\right)^2}+\frac{1}{1+e^{-\frac{2 \pi  \omega }{a}}}\right]+\frac{|\beta|^2(|\beta|^2+2e^{-\frac{2 \pi  \omega }{a}})}{2(1+e^{-\frac{2 \pi  \omega }{a}})}, $
$D_2=\frac{| \alpha | ^2 | \beta | ^2}{2 \left(e^{-\frac{2 \pi  \omega }{a}}+1\right)}+\frac{e^{-\frac{2 \pi  \omega }{a}} | \alpha | ^2}{2} \left[\frac{1}{\left(| \alpha | ^2+e^{-\frac{2 \pi  \omega }{a}}\right)^2}+\frac{1}{1+e^{-\frac{2 \pi  \omega }{a}}}\right],\; E_2=\frac{\alpha ^2 \beta ^2}{2 (1+e^{-\frac{2 \pi  \omega }{a}})^{1/2}\left(| \alpha | ^2+e^{-\frac{2 \pi  \omega }{a}}\right)}$
\end{center}
The expressions of the mutual information and the logarithmic negative for this sector can be achieved by replacing the coefficients $A_1, B_1, C_1, D_1, E_1$ by $A_2, B_2, C_2, D_2, E_2$, respectively, in \ref{mi1} and \ref{ln1}, since the matrix structure is the same for both sectors.
The variation of these measures with respect to parameter $\Delta$ is shown in \ref{rho2R}. It is non-monotonic and vanishes at large and small values of $\Delta$. We note that its behaviour differs from the preceding case, which corresponds to the fact that the correlations depend upon the choice of the sector.  
\noindent

\section{Summary and outlook}
\label{S5}
In this Chapter, we have investigated the effect of constant background electric and magnetic fields on particle creation in the Rindler spacetime. Also, some aspects of quantum entanglement between the created particles are discussed. We have computed the  `in' and `out' local modes for both $R$ and $L$ wedges, and the Bogoliubov relationship between them is derived in \ref{section : S2}. We construct global `in' and `out' modes from local modes, and the Bogoliubov transformations between relevant creation and annihilation operators are obtained in \ref{S3}. These Bogoliubov transformations not only carries information about the Unruh effect, but also the Schwinger effect taking place in each of the Rindler wedges.  Using these results, we have written the squeezed state expansion of the global vacuum in terms of local `out' state basis. Further, we have computed the logarithmic negativity for the created particles in the global vacuum in \ref{S4}. Next, in \ref{Logarithmic negativity and mutual information of a zero charge state constructed by two fermionic fields}, we have computed the correlations for different sectors of a maximally entangled system of two fermionic fields in terms of the logarithmic negativity and the mutual information. 

The main characteristic of this problem is that it involves two acceleration parameters: the acceleration of the Rindler observer $a$ and the acceleration of the charged quanta $(eE/m)$ due to the background electric field. There are two sources of particle creation in this case: the Schwinger and Unruh effects. As we have discussed in \ref{Motivation and Overview}, the magnetic field alone cannot be expected to create vacuum instability, but it may affect the  impact of the electric field on the same. Indeed, from \ref{beta2}, which represents the number density for local vacuum, it is clear that the magnetic field holds no role in pair creation without an electric field. Also, in the presence of the electric field, the magnetic field opposes the effect of the electric field. Likewise, in the previous two cases \ref{Background magnetic field and quantum correlations in the Schwinger effect} and \ref{Fermionic Bell violation in de Sitter spacetime with background electric and mangnetic fields}, in this case also, magnetic field plays no role in the absence of an electric field. Moreover, from \ref{beta2}, it is clear that the particle creation due to the Schwinger effect depends upon the observer's motion characterized by the acceleration parameter, i.e. $a$. 

The behaviour of the logarithmic negativity is non-monotonic for the global vacuum, as shown in \ref{fig:LNforVacuum}. Next, we construct a maximally entangled zero charge system of two fermionic fields and compute the correlations of different sectors corresponding to this state in terms of the logarithmic negativity and the mutual information. The variations of these measures with respect to the parameter $\Delta$ are shown in \ref{rho1R} and \ref{rho2R}. As we have mentioned earlier a complete analysis of the particle creation at different places for the $R$ and $L$ wedges is important in this context. One can further extend this analysis wit time-dependent electromagnetic fields.

\chapter{Summary and Outlook}
\label{Summary and Outlook}

As we have discussed in \ref{Motivation and Overview}, the magnetic field alone cannot lead to pair creation in the quantum field theoretical system, but it may affect particle creation rate in the presence of a background electric field. The main goal of the thesis is to describe how the background magnetic field affects the correlations or entanglement between the created Schwinger pairs. We have computed these correlations for complex scalar and fermionic fields in the presence of background electric and magnetic fields of constant strength for various spacetime backgrounds. Throughout the thesis, we have used entanglement entropy, logarithmic negativity, mutual information, and Bell's inequality violation to quantify correlations or entanglement for the pure and mixed states. This investigation would be relevant for getting an insight into the relativistic entanglement in the early inflationary universe scenario, where such background fields might exist due to primordial fluctuations, and in the near-horizon of non-extremal black holes, which are often endowed with background electromagnetic fields due to the accretion of plasma onto them.

In \ref{Background magnetic field and quantum correlations in the Schwinger effect}, we consider a complex scalar field in the presence of background electric and magnetic fields of constant strength. We computed the number density of the created particles and the entanglement entropy of the vacuum state. The number density vanishes, and the entanglement entropy degrades with increased magnetic field strength. Next, we have constructed some maximally entangled states of different charge content from two complex scalar fields. We compute the correlations between different sectors of these states regarding logarithmic negativity and mutual information. We find that the correlations between different sectors depend on the choice of the initial state and the sector. 


In \ref{Fermionic Bell violation in de Sitter spacetime with background electric and mangnetic fields}, we take a fermionic field in the presence of background electric and magnetic fields of constant strengths in the cosmological de Sitter spacetime. There were two sources of particle creation: the background electric and gravitational fields. We have computed the correlations between the pairs created in the vacuum state in terms of entanglement entropy and Bell's inequality violation. We find that the magnetic field does not affect the correlations between the particles created by the gravitational field. In contrast, it degrades the correlations between the particles created by the electric field and does not play any role in the absence of an electric field. Next, we construct some maximally entangled states from two fermionic fields and compute the correlations between different sectors of these states in terms of Bell's inequality violation and mutual information. We find that these correlations depend on the choice of states and the sectors. 

In \ref{Schwinger effect and a uniformly accelerated observer}, we have discussed the Schwinger effect in the presence of a background magnetic field of constant strength from the perspective of a non-inertial or Rindler observer. We consider a fermionic field in the presence of background electric and magnetic fields of constant strengths in the $(1+3)$-dimensional Rindler spacetime. We have quantized our field in the right ($R$) and left ($L$) Rindler regions and computed the local `in' and `out' modes corresponding to these regions. Next, we obtain the Bogoliubov transformation between them due to the presence of background electric and magnetic fields. Using the analytical continuation technique, we construct the global modes from the local ones. We have obtained the number density of particles and antiparticles in the global vacuum from the perspective of the Rindler observer at different regions of a particular Rindler wedge. We find that particles and antiparticles are not equally distributed in different regions of a particular Rindler wedge, and this ambiguity is due to the charge polarization. Next, we compute the logarithmic negativity to study the correlations between the particles and antiparticles in $R$ and $L$ wedges in the global vacuum state. We observe that its behaviour is non-monotonic. We find that the Minkowski vacuum's entanglement structure from the Rindler observer's perspective is far different from the inertial observer's. Next, we construct a maximally entangled state from two fermionic fields and compute the correlations of different sectors of this state in terms of logarithmic negativity and mutual information. The behaviour of these correlations is non-monotonic concerning the strength of the electric and magnetic fields. 

We wish to extend this study to AdS spacetime, and it would be interesting to see how these correlations will affect by the negative curvature of spacetime. It will also be interesting to extend this work in the AdS-Rindler spacetime, and this would be a crude toy model to get some insight into the characteristics of the near horizon of the near-extremal non-rotating black holes. 
We also wish to extend these studies with quantum electromagnetic fields instead of background or classical electric and magnetic fields. 
There exist some non-equilibrium systems for which the vacuum is unstable, and by treating them as an open quantum system, they can interact with their surroundings, for which it can be very interesting to see how background electromagnetic fields affect such systems in the cosmological scenario in terms of change in cosmic decoherence rate.
\appendix
\chapter{}
\section{Explicit form of the Dirac mode functions and normalisations in the de Sitter spacetime}
\label{A}
The four orthonormal simultaneous eigenvectors $\omega_s$ of the operators $(M\gamma^0+L \gamma^0 \gamma^3)$ and $\gamma^1 \gamma^2$ appearing in \ref{dirac41} are given by,
\begin{eqnarray}
\label{eigenkets}
\omega_1&=&\frac{1}{P_1}\left( {\begin{array}{cccc}
   \frac{\sqrt{M^2+L^2}-L}{M}\\
0\\
1\\
   0\\
  \end{array} } \right), \qquad 
  \;\omega_2=\frac{1}{P_2}\left( {\begin{array}{cccc}
  0\\
 \frac{\sqrt{M^2+L^2}+L}{M}\\
0\\
   1\\
  \end{array} } \right), \nonumber\\
  \omega_3&=&\frac{1}{P_1}\left( {\begin{array}{cccc}
  0\\
 \frac{\sqrt{M^2+L^2}-L}{M}\\
0\\
   -1\\
  \end{array} } \right), \qquad
  \;\omega_4=\frac{1}{P_2}\left( {\begin{array}{cccc}
 \frac{\sqrt{M^2+L^2}+L}{M}\\
 0\\
-1\\
   0\\
  \end{array} } \right),
\end{eqnarray}
where $P_1$ and $P_2$ are normalisation constants. $\omega_3$ and $\omega_4$ are respectively related to $\omega_1$ and $\omega_2$ via the charge conjugation, $\omega_{3,4}= {\cal C}\, \omega^{*}_{1,2}$, where ${\cal C}=i\gamma^2$. The explicit representation of the gamma matrices we are using is given by,
\begin{eqnarray}
\gamma^0\;=\;\left(\begin{array}{ccc}
0&I\\
I&0\\
\end{array}\right), \qquad 
\gamma^i\;=\;\left(\begin{array}{ccc}
0&\sigma^i\\
-\sigma^i&0\\ 
\end{array}\right)\qquad (i=1,2,3)
\end{eqnarray}
The explicit forms of the positive and negative frequency in and out modes appearing in \ref{mainmodes}, as follows,
\begin{equation}
\label{positive1in}
\begin{split}
U_{1,n_L}^{\rm in}=\frac{ \gamma^0}{N_1 a^{3/2}}\left\lbrace\left(i\partial_\eta-k_z\gamma^0 \gamma^3+aH\left(M \gamma^0 + L \gamma^0 \gamma^3 \right)\right)+\left(i\gamma^0 \gamma^2 \partial_y-\left(k_x+eBy\right)\gamma^0 \gamma^1\right)\right\rbrace \\ \times e^{-iHLz}e^{i\vec{k}\slashed{y}\cdot\vec{x}} W_{\kappa_1,i|\mu|}(z_1)h_{n_L}(\overline{y})\, \omega_1
  \end{split}
\end{equation}
\begin{equation}
\begin{split}
U_{2,n_L}^{\rm in}=\frac{\gamma^0}{N_2a^{3/2}} \left\lbrace\left(i\partial_\eta-k_z\gamma^0 \gamma^3+aH\left(M \gamma^0 + L \gamma^0 \gamma^3 \right)\right)+\left(i\gamma^0 \gamma^2 \partial_y-\left(k_x+eBy\right)\gamma^0 \gamma^1\right)\right\rbrace \\  \times e^{-iHLz}e^{i\vec{k}\slashed{y}\cdot\vec{x}} W_{\kappa_2,i|\mu|}(z_2)h_{n_L}(\overline{y})\, \omega_2
\end{split}
\end{equation}
\begin{equation}
\label{positive1out}
\begin{split}
U_{1,n_L}^{\rm out}=\frac{\gamma^0}{M_1a^{3/2}} \left\lbrace\left(i\partial_\eta-k_z\gamma^0 \gamma^3+aH\left(M \gamma^0 + L \gamma^0 \gamma^3 \right)\right)+\left(i\gamma^0 \gamma^2 \partial_y-\left(k_x+eBy\right)\gamma^0 \gamma^1\right)\right\rbrace \\ \times e^{-iHLz}e^{i\vec{k}\slashed{y}\cdot\vec{x}} M_{\kappa_1,i|\mu|}(z_1)h_{n_L}(\overline{y})\,\omega_1
%
  \end{split}
\end{equation}
\begin{equation}
\begin{split}
U_{2,n_L}^{\rm out}=\frac{\gamma^0}{M_2 a^{3/2}} \left\lbrace\left(i\partial_\eta-k_z\gamma^0 \gamma^3+aH\left(M \gamma^0 + L \gamma^0 \gamma^3 \right)\right)+\left(i\gamma^0 \gamma^2 \partial_y-\left(k_x+eBy\right)\gamma^0 \gamma^1\right)\right\rbrace \\ \times  e^{-iHLz}e^{i\vec{k}\slashed{y}\cdot\vec{x}} M_{\kappa_2,i|\mu|}(z_2)h_{n_L}(\overline{y})\,\omega_2
  \end{split}
\end{equation}

\begin{equation}
\begin{split}
V_{1,n_L}^{\rm in}=\frac{\gamma^0}{N_1a^{3/2}} \left\lbrace\left(i\partial_\eta-k_z\gamma^0 \gamma^3+aH\left(M \gamma^0 + L \gamma^0 \gamma^3 \right)\right)+\left(i\gamma^0 \gamma^2 \partial_y-\left(k_x-eBy\right)\gamma^0 \gamma^1\right)\right\rbrace \\  \times e^{iHLz}e^{-i\vec{k}\slashed{y}\cdot\vec{x}} W_{-\kappa_1,-i|\mu|}(-z_1)h_{n_L}(y_-) \,\omega_3
  \end{split}
\end{equation}
\begin{equation}
\begin{split}
V_{2,n_L}^{\rm in}=\frac{\gamma^0}{N_2a^{3/2}} \left\lbrace\left(i\partial_\eta-k_z\gamma^0 \gamma^3+aH \left(M \gamma^0 + L \gamma^0 \gamma^3 \right)\right)+\left(i\gamma^0 \gamma^2 \partial_y-\left(k_x-eBy\right)\gamma^0 \gamma^1\right)\right\rbrace \\  \times  e^{iHLz}e^{-i\vec{k}\slashed{y}\cdot\vec{x}} W_{-\kappa_2,-i|\mu|}(-z_2)h_{n_L}(y_-)\omega_4
%
   \end{split}
\end{equation}
\begin{equation}
\label{negative1out}
\begin{split}
V_{1,n_L}^{\rm out}=\frac{\gamma^0}{M_1a^{3/2}} \left\lbrace\left(i\partial_\eta-k_z\gamma^0 \gamma^3+aH\left(M \gamma^0 + L \gamma^0 \gamma^3 \right)\right)+\left(i\gamma^0 \gamma^2 \partial_y-\left(k_x-eBy\right)\gamma^0 \gamma^1\right)\right\rbrace \\  \times  e^{iHLz}e^{-i\vec{k}\slashed{y}\cdot\vec{x}} M_{-\kappa_1,-i|\mu|}(-z_1)h_{n_L}(y_-)\,\omega_3
%
 \end{split}
\end{equation}
\begin{equation}
\begin{split}
V_{2,n_L}^{\rm out}=\frac{\gamma^0}{M_2a^{3/2}} \left\lbrace\left(i\partial_\eta-k_z\gamma^0 \gamma^3+aH\left(M \gamma^0 + L \gamma^0 \gamma^3 \right)\right)+\left(i\gamma^0 \gamma^2 \partial_y-\left(k_x-eBy\right)\gamma^0 \gamma^1\right)\right\rbrace \\  \times  e^{iHLz}e^{-i\vec{k}\slashed{y}\cdot\vec{x}} M_{-\kappa_2,-i|\mu|}(-z_2)h_{n_L}(y_-)\,\omega_4
%
 \end{split}
\end{equation}
The normalisation constants, $N_1,\,N_2,\, M_1,\, M_2$ are given by \ref{nc}. We shall explicitly evaluate $N_1$ below. The rest can be derived in a similar manner. Using \ref{eigenkets} into \ref{positive1in}, we find after some algebra
\begin{equation}
\label{N11}
    \begin{split}
     \int d^3x a^3\,U_{1}^{\dagger}{^\text{in}}U_{1}^\text{in}{^\prime}=\frac{1}{N_1^2}  \int d^3x\,e^{-i(\vec{k}_\slashed{y}-\vec{k}_\slashed{y}^\prime)\cdot\vec{x}}\times \Big[\Big(-i\partial_\eta{W_{\kappa_1,i|\mu|}(z_1)}^*-(\frac{k_zL}{\sqrt{M^2+L^2}}\\-aH\sqrt{M^2+L^2})W_{\kappa_1,i|\mu|}^*(z_1)\Big)h_{n_L}(\overline{y})\omega_1^{\dagger}+\frac{k_zM}{\sqrt{M^2+L^2}}X_1^{\dagger} W_{\kappa_1,i|\mu|}^*(z_1)h_{n_L}(\overline{y})\\-\left(\partial_y{h_{n_L}(\overline{y})}+\left(k_x+eBy\right) h_{n_L}(\overline{y})\right)W_{\kappa_1,i|\mu|}^*(z_1)\omega_3^{\dagger}\Big] \times \Big[\Big(-i\partial_\eta{W_{\kappa_1,i|\mu|}(z_1)}\\-(\frac{k_zL}{\sqrt{M^2+L^2}}-aH\sqrt{M^2+L^2})W_{\kappa_1,i|\mu|}(z_1)h_{n_L}(\overline{y})\omega_1
+\frac{k_zM}{\sqrt{M^2+L^2}}X_1 W_{\kappa_1,i|\mu|}(z_1)h_{n_L}(\overline{y})\Big)\\-\left(\partial_y{h_{n_L}(\overline{y})}+\left(k_x+eBy\right) h_{n_L}(\overline{y})\right)W_{\kappa_1,i|\mu|}(z_1)\omega_3\Big]
    \end{split}
\end{equation}
where $X_1= i \gamma^2 \omega_1$. The $x$ and $z$ integrals trivially give, $\delta(k_x-k'_x)\delta (k_z-k'_z)\equiv \delta^2(\vec{k}_\slashed{y}-\vec{k}_\slashed{y}^\prime)$. Using the orthonormality of $\omega_1$ and $\omega_3$, and the definition of the variable $\overline{y}$ appearing below \ref{dirac9'}, the $y$ integral is extracted to be, 
\begin{equation}
 \int d \overline{y} \left[\partial_{y}{h_{n_L}(\overline{y})}\partial_{y} h_{n_L^\prime}(\overline{y})+\overline{y} h_{n_L}(\overline{y})\partial_{y} h_{n_L^\prime}(\overline{y})+\overline{y} h_{n_L^\prime}(\overline{y})\partial_{y} h_{n_L}(\overline{y})+\overline{y}^2h_{n_L^\prime}(\overline{y}) h_{n_L}(\overline{y})\right]   
 \label{hint}
\end{equation}
Using some properties of the Hermite polynomials given as~\cite{AS},
$$
 \int_{-\infty}^{\infty}H_n(y)H_m(y)=\delta_{nm},\quad \int_{-\infty}^{\infty}H_n(y)\frac{dH_m(y)}{dy}=\begin{cases}
    \frac{1}{a}\sqrt{\frac{n+1}{2}},\quad m=n+1\\
  - \frac{1}{a}\sqrt{\frac{n}{2}},\quad m=n-1\\
    0, \quad \text{otherwise}
      \end{cases}
$$
$$
\frac{d H_n(y)}{dy}=2n H_{n-1}(y),\quad\int_{-\infty}^{\infty}u_{n}(y) y u_m (y)dy=\begin{cases}
    \frac{1}{a}\sqrt{\frac{n+1}{2}},\quad m=n+1\\
   \frac{1}{a}\sqrt{\frac{n}{2}},\quad m=n-1\\
    0, \quad \text{otherwise}
    \end{cases}
$$
where $u_n(y)=\sqrt{\frac{a}{\pi^{1/2}n!2^n}}H_n(ay)e^{-\frac{a^2y^2}{2}}$.
The first integral of \ref{hint} equals
%
$$\int d\overline{y}\, \partial_{y}{h_{n_L}(\overline{y})}\partial_{y} h_{n_L^\prime}(\overline{y})=3eB\left(n_L+\frac16\right)\delta_{n_L n_L^\prime}$$
%
The second and third integrals vanish,
%
$$\int d\overline{y}\, \overline{y} h_{n_L}(\overline{y})\partial_{y} h_{n_L^\prime}(\overline{y}) =0=\int d\overline{y}\, \overline{y} h_{n_L^\prime}(\overline{y})\partial_{y} h_{n_L}(\overline{y})$$
%
whereas the fourth integral equals,
%
$$\int d\overline{y}\, \partial_{y}{h_{n_L}(\overline{y})}\partial_{y} h_{n_L^\prime}(\overline{y}) y^2_ {+}   = eB \left(n_L+\frac12\right) \delta_{n_Ln_L^\prime}$$
%
Collecting all the pieces, the $\overline{y}$ integral becomes
\begin{eqnarray}
\label{inthermite}
 4eB\left(n_L+\frac14\right) \delta_{n_Ln_L^\prime}
\end{eqnarray}
Since normalisation is time-independent, we may choose the arguments of the Whittaker functions in \ref{N11} as per our convenience.  Accordingly, we choose $\eta \to -\infty $, for which $W_{\kappa_1,i|\mu|}\approx e^{-z_1/2}z_{1}^{\kappa_1}$. We have 
\begin{eqnarray}
\label{Whit1}
&&W_{\kappa_1,i|\mu|}(z_1)(W_{\kappa_1,i|\mu|}(z_1))^*=
e^{\pi|\kappa_1| {\rm sgn}(k_z)}, \partial_\eta W_{\kappa_1,i|\mu|}(z_1)\partial_\eta (W_{\kappa_1,i|\mu|}(z_1))^*=e^{\pi|\kappa_1| {\rm sgn}(k_z)}(k_z^2+S_1)\nonumber\\
&&\partial_\eta (W_{\kappa_1,i|\mu|}(z_1))^* W_{\kappa_1,i|\mu|}(z_1)-\partial_\eta W_{\kappa_1,i|\mu|}(z_1) (W_{\kappa_1,i|\mu|}(z_1))^*=2i e^{\pi|\kappa_1| {\rm sgn}(k_z)}\sqrt{k_z^2+S_1}
\end{eqnarray}
The normalisation factor is $N_1=e^{ \pi|\kappa_1|{\rm sgn}(k_z)/2} $. Putting it in \ref{N11}, makes the normalisation integral $\delta^2(\vec{k}_\slashed{y}-\vec{k}_\slashed{y}^\prime)\delta_{n_Ln_L^\prime}$. 
The normalisation for the other `in' modes can be found similarly. 

For the normalisation of the out modes, we choose the integration hypersurface to be in the asymptotic future, $\eta \to 0^-$, for our convenience and use in this limit, $M_{\kappa_1,i|\mu|}\approx\left(2i\eta\sqrt{k^2_z+S_1}\right)^{1/2+i|\mu|}$.
The rest of the calculations remain the same.
\noindent
\section{Explicit form of the Dirac mode functions and normalizations in the Rindler spacetime}
\label{A1}
\begin{equation}
\label{fullmode1}
\begin{split}
    U_{1,n_L}(x)_{\mathit{H^-_R}}=\frac{e^{e^{-ax}/2} }{N_1}\bigg(\frac{i \epsilon_4}{e^{a x}}\partial_t -\frac{i \epsilon_4 }{e^{a x}}\partial_x-\epsilon_2\partial_2-i\epsilon_2\partial_3-\frac{e E e^{a x}}{2 a} \epsilon_4+ e B y \epsilon_2 + m \epsilon_1\bigg) \\ \times e^{-i \omega (t-x)} e^{-i k_z z} e^{a x} e^{-\frac{i e E e^{2 a x}}{4 a^2}} U(\lambda_1, \nu, \xi)H_1(y) 
    \end{split}
\end{equation}
 \begin{equation}
\begin{split}
\label{fullmode2}
    U_{1,n_L}(x)_{\mathit{I^-_R}}=\frac{e^{e^{-ax}/2} }{M_1}\bigg(\frac{i \epsilon_4}{e^{a x}}\partial_t -\frac{i \epsilon_4 }{e^{a x}}\partial_x-\epsilon_2\partial_2-i\epsilon_2\partial_3-\frac{e E e^{a x}}{2 a} \epsilon_4+ e B y \epsilon_2 + m \epsilon_1\bigg)\\ \times e^{-i \omega (t+x)} e^{-i k_z z} e^{a x} e^{\frac{i e E e^{2 a x}}{4 a^2}} (L(-\lambda_1, \nu-1, \xi))^* H_1(y) 
    \end{split}
\end{equation}
\begin{equation}
\begin{split}
\label{fullmode3}
    U_{2,n_L}(x)_{\mathit{H^-_R}}=\frac{e^{e^{-ax}/2} }{N_2}\bigg(\frac{i \epsilon_3}{e^{a x}}\partial_t -\frac{i \epsilon_3 }{e^{a x}}\partial_x-\epsilon_1\partial_2-i\epsilon_1\partial_3-\frac{e E e^{a x}}{2 a} \epsilon_3+ e B y \epsilon_1 + m \epsilon_2\bigg)\\ \times e^{-i \omega (t-x)} e^{-i k_z z} e^{a x} e^{-\frac{i e E e^{2 a x}}{4 a^2}} U(\lambda_2, \nu, \xi) H_2(y)
    \end{split}
\end{equation}
\begin{equation}
\begin{split}
\label{fullmode4}
     U_{2,n_L}(x)_{\mathit{I^-_R}}=\frac{e^{e^{-ax}/2} }{M_2}\bigg(\frac{i \epsilon_3}{e^{a x}}\partial_t -\frac{i \epsilon_3 }{e^{a x}}\partial_x-\epsilon_1\partial_2-i\epsilon_1\partial_3-\frac{e E e^{a x}}{2 a} \epsilon_3+ e B y \epsilon_1 + m \epsilon_2\bigg)\\ \times e^{-i \omega (t+x)} e^{-i k_z z} e^{a x} e^{\frac{i e E e^{2 a x}}{4 a^2}} (L(-\lambda_2, \nu-1, \xi))^* H_2(y) 
    \end{split}
\end{equation}
\begin{equation}
\begin{split}
\label{fullmode1a}
   V_{1,n_L}(x)_{\mathit{H^-_R}}=\frac{e^{e^{-ax}/2} }{P_1}\bigg(-\frac{i \epsilon_2}{e^{a x}}\partial_t +\frac{i \epsilon_2 }{e^{a x}}\partial_x-\epsilon_4\partial_2-i\epsilon_4\partial_3+\frac{e E e^{a x}}{2 a} \epsilon_2+ e B y \epsilon_4 + m \epsilon_3\bigg)\\ \times e^{i \omega (t-x)} e^{i k_z z} e^{a x} e^{\frac{i e E e^{2 a x}}{4 a^2}} e^{\xi}U(\nu-\lambda_1,\nu,\xi)H_1(y_-)
    \end{split}
\end{equation}
\begin{equation}
\begin{split}
\label{fullmode2a}
   V_{1,n_L}(x)_{\mathit{I^-_R}}=\frac{e^{e^{-ax}/2}} {R_1}\bigg(-\frac{i \epsilon_2}{e^{a x}}\partial_t +\frac{i \epsilon_2 }{e^{a x}}\partial_x-\epsilon_4\partial_2-i\epsilon_4\partial_3+\frac{e E e^{a x}}{2 a} \epsilon_2+ e B y \epsilon_4 + m \epsilon_3\bigg)\\ \times e^{i \omega (t+x)} e^{i k_z z} e^{a x} e^{-\frac{i e E e^{2 a x}}{4 a^2}}(\xi^{1-\nu} L(\nu-\lambda_s-1,1-\nu,\xi))^* H_1(y_-)
    \end{split}
\end{equation}
\begin{equation}
\begin{split}
\label{fullmode3a}
   V_{2,n_L}(x)_{\mathit{H^-_R}}=\frac{e^{e^{-ax}/2} }{P_2}\bigg(-\frac{i \epsilon_1}{e^{a x}}\partial_t +\frac{i \epsilon_1 }{e^{a x}}\partial_x-\epsilon_3\partial_2-i\epsilon_3\partial_3+\frac{e E e^{a x}}{2 a} \epsilon_1+ e B y \epsilon_3 + m \epsilon_4\bigg)\\ \times e^{i \omega (t-x)} e^{i k_z z} e^{a x} e^{\frac{i e E e^{2 a x}}{4 a^2}}e^{\xi}U(\nu-\lambda_2,\nu,\xi)H_2(y_-)
    \end{split}
\end{equation}
\begin{equation}
\begin{split}
\label{fullmode4a}
    V_{2,n_L}(x)_{\mathit{I^-_R}}=\frac{e^{e^{-ax}/2} }{R_2}\bigg(-\frac{i \epsilon_1}{e^{a x}}\partial_t +\frac{i \epsilon_1 }{e^{a x}}\partial_x-\epsilon_3\partial_2-i\epsilon_3\partial_3+\frac{e E e^{a x}}{2 a} \epsilon_1+ e B y \epsilon_3 + m \epsilon_4\bigg)\\ \times e^{i \omega (t+x)} e^{i k_z z} e^{a x} e^{-\frac{i e E e^{2 a x}}{4 a^2}} (\xi^{1-\nu}L(\nu-\lambda_2-1,1-\nu,\xi))^* H_2(y_-)
    \end{split}
\end{equation} 
where in \ref{fullmode1a}, \ref{fullmode2a}, \ref{fullmode3a} and \ref{fullmode4a} $y_-=\left(\sqrt{e B}y-\frac{k_z}{\sqrt{e B}}\right)$ and $H_1(y_-)= H_2(y_-)= \left(\frac{\sqrt{e B}}{2^{n+1}\sqrt{\pi}(n+1)!}\right)^{1/2}
e^{-y_-^2/2}\mathit{H}_{n_L}(y_-).$ The normalisation constants, $N_1,\,N_2,\,M_1$ and $M_2$ are given in the previous section. We shall explicitly evaluate $N_1$ below, for which we choose constant time hypersurface 
\begin{equation}
    \begin{split}
 \int_{\vec{x}}  n_\mu \sqrt{|g|}  (U_{s,n_L}(x)_{\mathit{H^-_R}})^{\dagger}\gamma^{0} \gamma^\mu U_{s,n_L^\prime}(x)_{\mathit{H^-_R}}=\frac{1}{|N_s|^2}  \int \Bigg[\Big(\frac{\omega}{e^{a x}}-\frac{e E}{2 a}-\frac{i}{e^{a x}}\big(a-\frac{i e e^{2 a x}}{2 a}+i \omega-2 a \lambda_1\big)\Big)\\\times\bigg(\frac{\omega^{\prime}}{e^{a x}}-\frac{e E}{2 a}+\frac{i}{e^{a x}}\big(a+\frac{i e e^{2 a x}}{2 a}-i \omega^{\prime}-2 a \lambda_1^*\big)\bigg) H_1^{\prime}(y) H_1(y) + (\partial_y + y_{+}\sqrt{e B})H_1^{\prime}(y)\\(\partial_y + y_{+}\sqrt{e B})H_1(y)+m^2\Bigg] e^{i(\omega-\omega^{\prime})(t-x)} e^{i(k_z-k_z^{\prime})z}\bigg(\frac{2 a^2}{i e E}\bigg)^{\lambda}\bigg(\frac{2 a^2}{-i e E}\bigg)^{\lambda^*}e^{-a x}e^{2e^{-ax/2}}  \\
 =\frac{1}{|N_s|^2} \bigg(\frac{a^2}{ E^2}e^{-\pi\frac{m^2+S_s}{4 q E}}\bigg) \delta_{n_L n_L'} \delta(\omega-\omega') \delta(k_z-k_z')
    \end{split}
\end{equation}
where we have used the asymptotic limit of the $U(\lambda, \nu, \xi)$  function at $x \to \infty\; (|\xi| \to \infty)$, i.e. $U(\lambda, \nu, \xi) \approx \xi^{-\lambda}$, the variable $\xi$ is defined in \ref{ell}. Note that the normalization of $U_{s,n_L}(x)_{\mathit{I^-_R}}$ can be done in the same way as of $U_{s,n_L}(x)_{\mathit{H^-_R}}$ for which we have used the asymptotic form of $L(\lambda, \nu, \xi)$ at $x \to -\infty\; (|\xi| \to 0) $, i.e. $L(\lambda, \nu, \xi) = \frac{\Gamma(\nu+\lambda+1)}{\Gamma(\lambda+1) \Gamma (\nu+1)}$ which gives $M_s$, similarly we normalize all other in modes. The set of modes for $R$ and $L$ regions are given as follows
\begin{equation}
\begin{split}
\label{fullmode1out}
    U_{1,n_L}(x)_{\mathit{H^+_R}}=\frac{e^{e^{-ax}/2}}{N_1}\bigg(-\frac{i \epsilon_4}{e^{a x}}\partial_t +\frac{i \epsilon_4 }{e^{a x}}\partial_x-\epsilon_2\partial_2+i\epsilon_2\partial_3-\frac{e E e^{a x}}{2 a} \epsilon_4+ e B y \epsilon_2 + m \epsilon_1\bigg)\\ \times e^{-i \omega (t+x)} e^{-i k_z z} e^{a x} e^{\frac{i e E e^{2 a x}}{4 a^2}} (U(\lambda_1, \nu, \xi))^*H_1(y) 
    \end{split}
\end{equation}
 \begin{equation}
\begin{split}
\label{fullmode2out}
    U_{1,n_L}(x)_{\mathit{I^+_R}}=\frac{e^{e^{-ax}/2}}{M_1}\bigg(-\frac{i \epsilon_4}{e^{a x}}\partial_t +\frac{i \epsilon_4 }{e^{a x}}\partial_x-\epsilon_2\partial_2+i\epsilon_2\partial_3-\frac{e E e^{a x}}{2 a} \epsilon_4+ e B y \epsilon_2 + m \epsilon_1\bigg)\\ \times e^{-i \omega (t-x)} e^{-i k_z z} e^{a x} e^{-\frac{i e E e^{2 a x}}{4 a^2}}L(-\lambda_1, \nu-1, \xi) H_1(y)
    \end{split}
\end{equation}
\begin{equation}
\begin{split}
\label{fullmode3out}
    U_{2,n_L}(x)_{\mathit{H^+_R}}=\frac{e^{e^{-ax}/2}}{N_2}\bigg(-\frac{i \epsilon_3}{e^{a x}}\partial_t +\frac{i \epsilon_3 }{e^{a x}}\partial_x-\epsilon_1\partial_2+i\epsilon_1\partial_3-\frac{e E e^{a x}}{2 a} \epsilon_3+ e B y \epsilon_1 + m \epsilon_2\bigg)\\ \times e^{-i \omega (t+x)} e^{-i k_z z} e^{a x} e^{\frac{i e E e^{2 a x}}{4 a^2}} (U(\lambda_2, \nu, \xi))^* H_2(y)
    \end{split}
\end{equation}
\begin{equation}
\begin{split}
\label{fullmode4out}
     U_{2,n_L}(x)_{\mathit{I^+_R}}=\frac{e^{e^{-ax}/2}}{M_2}\bigg(-\frac{i \epsilon_3}{e^{a x}}\partial_t+\frac{i \epsilon_3 }{e^{a x}}\partial_x-\epsilon_1\partial_2+i\epsilon_1\partial_3-\frac{e E e^{a x}}{2 a} \epsilon_3+ e B y \epsilon_1 + m \epsilon_2\bigg)\\ \times e^{-i \omega (t-x)} e^{-i k_z z} e^{a x} e^{-\frac{i e E e^{2 a x}}{4 a^2}}L(-\lambda_2,\nu-1,\xi)  H_2(y) 
    \end{split}
\end{equation}
\begin{equation}
\begin{split}
\label{fullmode1aout}
   V_{1,n_L}(x)_{\mathit{H^+_R}}=\frac{e^{e^{-ax}/2}}{P_1}\bigg(\frac{i \epsilon_2}{e^{a x}}\partial_t -\frac{i \epsilon_2 }{e^{a x}}\partial_x-\epsilon_4\partial_2+i\epsilon_4\partial_3+\frac{e E e^{a x}}{2 a} \epsilon_2+ e B y \epsilon_4 + m \epsilon_3\bigg)\\ \times e^{i \omega (t+x)} e^{i k_z z} e^{a x} e^{\frac{-i e E e^{2 a x}}{4 a^2}} (e^{\xi}U(\nu-\lambda_1,\nu,\xi))^*H_1(y_-)
    \end{split}
\end{equation}
\begin{equation}
\begin{split}
\label{fullmode2aout}
   V_{1,n_L}(x)_{\mathit{I^+_R}}=\frac{e^{e^{-ax}/2}}{R_1}\bigg(\frac{i \epsilon_2}{e^{a x}}\partial_t -\frac{i \epsilon_2 }{e^{a x}}\partial_x-\epsilon_4\partial_2+i\epsilon_4\partial_3+\frac{e E e^{a x}}{2 a} \epsilon_2+ e B y \epsilon_4 + m \epsilon_3\bigg)\\ \times e^{i \omega (t-x)} e^{i k_z z} e^{a x} e^{\frac{i e E e^{2 a x}}{4 a^2}}\xi^{1-\nu}L(\nu-\lambda_1-1,1-\nu,\xi) H_1(y_-)
    \end{split}
\end{equation}
\begin{equation}
\begin{split}
\label{fullmode3aout}
   V_{2,n_L}(x)_{\mathit{H^+_R}}=\frac{e^{e^{-ax}/2}}{P_2}\bigg(\frac{i \epsilon_1}{e^{a x}}\partial_t -\frac{i \epsilon_1 }{e^{a x}}\partial_x-\epsilon_3\partial_2+i\epsilon_3\partial_3+\frac{e E e^{a x}}{2 a} \epsilon_1+ e B y \epsilon_3 + m \epsilon_4\bigg)\\ \times e^{i \omega (t+x)} e^{i k_z z} e^{a x} e^{\frac{-i e E e^{2 a x}}{4 a^2}} (e^{\xi} U(\nu-\lambda_2,\nu,\xi))^*H_2(y_-)
    \end{split}
\end{equation}
\begin{equation}
\begin{split}
\label{fullmode4aout}
    V_{2,n_L}(x)_{\mathit{I^+_R}}=\frac{e^{e^{-ax}/2}}{R_2}\bigg(\frac{i \epsilon_1}{e^{a x}}\partial_t -\frac{i \epsilon_1 }{e^{a x}}\partial_x-\epsilon_3\partial_2+i\epsilon_3\partial_3+\frac{e E e^{a x}}{2 a} \epsilon_1+ e B y \epsilon_3 + m \epsilon_4\bigg)\\ \times e^{i \omega (t-x)} e^{i k_z z} e^{a x} e^{\frac{i e E e^{2 a x}}{4 a^2}} \xi^{1-\nu} L(\nu-\lambda_2-1,1-\nu,\xi) H_2(y_-)
    \end{split}
\end{equation}
\begin{equation}
\begin{split}
\label{fullmode1LL}
    U_{1,n_L}(x_L)_{\mathit{H^+_L}}=\frac{e^{e^{-ax_L}/2} }{N_1}\bigg(\frac{i \epsilon_4}{e^{a x_L}}\partial_{t_{L}} -\frac{i \epsilon_4 }{e^{a x_L}}\partial_{x_{L}}-\epsilon_2\partial_2-i\epsilon_2\partial_3-\frac{e E e^{a x_L}}{2 a} \epsilon_4+ e B y \epsilon_2 + m \epsilon_1\bigg)\\ \times e^{-i \omega (t_L-x_L)} e^{i k_z z} e^{a x_L} e^{-\frac{i e E e^{2 a x_L}}{4 a^2}} U(\lambda_1, \nu, \xi_L)H_1(y) 
    \end{split}
\end{equation}
 \begin{equation}
\begin{split}
\label{fullmode2LL}
    U_{1,n_L}(x_L)_{\mathit{I^+_L}}=\frac{e^{e^{-ax_L}/2} }{M_1}\bigg(\frac{i \epsilon_4}{e^{a x_L}}\partial_{t_{L}} -\frac{i \epsilon_4 }{e^{a x_L}}\partial_{x_{L}}-\epsilon_2\partial_2-i\epsilon_2\partial_3-\frac{e E e^{a x_L}}{2 a} \epsilon_4+ e B y \epsilon_2 + m \epsilon_1\bigg)\\ \times e^{-i \omega (t_L+x_L)} e^{i k_z z} e^{a x_L} e^{\frac{i e E e^{2 a x_L}}{4 a^2}}  (L(-\lambda_1, \nu-1, \xi_L))^* H_1(y) 
    \end{split}
\end{equation}
\begin{equation}
\begin{split}
\label{fullmode3L}
    U_{2,n_L}(x_L)_{\mathit{H^+_L}}=\frac{e^{e^{-ax_L}/2} }{N_2}\bigg(\frac{i \epsilon_3}{e^{a x_L}}\partial_{t_{L}} -\frac{i \epsilon_3 }{e^{a x_L}}\partial_{x_{L}}-\epsilon_1\partial_2-i\epsilon_1\partial_3-\frac{e E e^{a x_L}}{2 a} \epsilon_3+ e B y \epsilon_1 + m \epsilon_2\bigg)\\ \times e^{-i \omega (t_L-x_L)} e^{i k_z z} e^{a x_L} e^{-\frac{i e E e^{2 a x_L}}{4 a^2}} U(\lambda_2, \nu, \xi_L) H_2(y)
    \end{split}
\end{equation}
\begin{equation}
\begin{split}
\label{fullmode4L}
     U_{2,n_L}(x_L)_{\mathit{I^+_L}}=\frac{e^{e^{-ax_L}/2} }{M_1}\bigg(\frac{i \epsilon_3}{e^{a x_L}}\partial_{t_{L}} -\frac{i \epsilon_3 }{e^{a x_L}}\partial_{x_{L}}-\epsilon_1\partial_2-i\epsilon_1\partial_3-\frac{e E e^{a x_L}}{2 a} \epsilon_3+ e B y \epsilon_1 + m \epsilon_2\bigg)\\ \times e^{-i \omega (t_L+x_L)} e^{i k_z z} e^{a x_L} e^{\frac{i e E e^{2 a x_L}}{4 a^2}}  (L(-\lambda_1, \nu-1, \xi))^* H_2(y) 
    \end{split}
\end{equation}
\begin{equation}
\begin{split}
\label{fullmode1aLL}
   V_{1,n_L}(x_L)_{\mathit{H^+_L}}=\frac{e^{e^{-ax_L}/2} }{P_1}\bigg(-\frac{i \epsilon_2}{e^{a x_L}}\partial_{t_L} +\frac{i \epsilon_2 }{e^{a x}}\partial_x-\epsilon_4\partial_2-i\epsilon_4\partial_3+\frac{e E e^{a x_L}}{2 a} \epsilon_2+ e B y \epsilon_4 + m \epsilon_3\bigg)\\ \times e^{i \omega (t_L-x_L)} e^{-i k_z z} e^{a x_L} e^{\frac{i e E e^{2 a x_L}}{4 a^2}} e^{\xi}U(\nu-\lambda_1,\nu,\xi)H_1(y_-)
    \end{split}
\end{equation}
\begin{equation}
\begin{split}
\label{fullmode2aLL}
   V_{1,n_L}(x_{L})_{\mathit{I^+_L}}=\frac{e^{e^{-ax_L}/2} }{R_1}\bigg(-\frac{i \epsilon_2}{e^{a x_L}}\partial_{t_L} +\frac{i \epsilon_2 }{e^{a x_L}}\partial_{x_{L}}-\epsilon_4\partial_2-i\epsilon_4\partial_3+\frac{e E e^{a x_L}}{2 a} \epsilon_2+ e B y \epsilon_4 + m \epsilon_3\bigg)\\
   \times e^{i \omega (t_{L}+x_{L})} e^{-i k_{z} z} e^{a x_{L}} e^{-\frac{i e E e^{2 a x_{L}}}{4 a^2}}(\xi_{L}^{1-\nu} L(\nu-\lambda_{1}-1,1-\nu,\xi))^{*} H_{1} (y_{-})
    \end{split}
\end{equation}
\begin{equation}
\begin{split}
\label{fullmode3aL}
   V_{2,n_L}(x_L)_{\mathit{H^+_L}}=\frac{e^{e^{-ax_L}/2} }{P_2}\bigg(-\frac{i \epsilon_1}{e^{a x_L}}\partial_{t_L} +\frac{i \epsilon_1 }{e^{a x_L}}\partial_{x_L}-\epsilon_3\partial_2-i\epsilon_3\partial_3+\frac{e E e^{a x_L}}{2 a} \epsilon_1+ e B y \epsilon_3 + m \epsilon_4\bigg)\\ \times e^{i \omega (t_L-x_L)} e^{-i k_z z} e^{a x} e^{\frac{i e E e^{2 a x_L}}{4 a^2}} e^{\xi}U(\nu-\lambda_2,\nu,\xi)H_2(y_-)
    \end{split}
\end{equation}
\begin{equation}
\begin{split}
\label{fullmode4aL}
    V_{2,n_L}(x_L)_{\mathit{I^+_L}}=\frac{e^{e^{-ax_L}/2} }{R_2}\bigg(-\frac{i \epsilon_1}{e^{a x_L}}\partial_{t_L} +\frac{i \epsilon_1 }{e^{a x_L}}\partial_{x_L}-\epsilon_3\partial_2-i\epsilon_3\partial_3+\frac{e E e^{a x_L}}{2 a} \epsilon_1+ e B y \epsilon_3 + m \epsilon_4\bigg)\\ \times  e^{i \omega (t_L+x_L)} e^{-i k_z z} e^{a x_L} e^{-\frac{i e E e^{2 a x_L}}{4 a^2}} (\xi_L^{1-\nu} L(\nu-\lambda_2-1,1-\nu,\xi_L))^*  H_2(y_-)
    \end{split}
\end{equation} 
where the normalization constants are 
\begin{center}
    $N_s=e^{-\frac{\pi \Delta}{2}} \cosh{\frac{\pi \omega}{a}} \Big(\frac{\sinh{\pi \Delta} \cosh{\pi (\Delta - \frac{\omega}{a})}}{\cosh^3{\pi(\Delta-\frac{\omega}{a})}+e^{-\pi \Delta} \sinh^3{\pi \Delta}}\Big)^{\frac{1}{2}} $, $ M_s=e^{-\frac{\pi \Delta}{2}}  \sqrt{\frac{\pi}{\Delta}}$, $P_s= \Big(\frac{6n_L+1}{eB}+m^2\Big)^{\frac{1}{2}}$  and $R_s =\sqrt{\pi}$ \quad \quad $(s=1,2)$
\end{center}
\noindent
\section{Limits of modes near null infinities}\label{B}
\begin{equation}
\begin{split}
\label{fullmode1IRlim}
    U_{s,n_L}(x)_{\mathit{H^-_R}}=\frac{e^{e^{-ax}/2} }{N_s}\bigg(ie^\mu_a \gamma^{a}\partial_\mu-e A_\mu e^\mu_a \gamma^{a}+m\bigg)e^{-i \omega (t-x)} e^{-i k_z z} e^{-a x} e^{-\frac{i e E e^{2 a x}}{4 a^2}} \Big(-\frac{i e E}{2a^2}\Big)^{-\lambda_s}\\ \times e^{-\frac{ia (m^2+S_s)x}{qE}}H_s(y)\epsilon_s
   \end{split}
\end{equation}
\begin{equation}
\begin{split}
\label{fullmode1aIRlim1}
    V_{s,n_L}(x)_{\mathit{H^-_R}}=\frac{e^{e^{-ax}/2} }{P_s} \bigg(ie^\mu_a \gamma^{a}\partial_\mu-e A_\mu e^\mu_a \gamma^{a}+m\bigg)e^{i \omega (t+x)} e^{i k_z z} e^{-a x} e^{-\frac{i e E e^{2 a x}}{4 a^2}} \Big(-\frac{i e E}{2a^2}\Big)^{-\nu^*+\lambda_s}\\
    \times e^{\frac{i a (m^2+S_s) x}{qE}} H_s(y) \epsilon_s
    \end{split}
\end{equation}
\begin{equation}
\begin{split}
\label{fullmode1ILlim}
    U_{s,n_L}(x)_{\mathit{I^-_L}}=\frac{e^{e^{-ax_L}/2} }{M_s}\bigg(ie^\mu_a \gamma^{a}\partial_\mu-e A_\mu e^\mu_a \gamma^{a}+m\bigg)e^{-i \omega (t_L+x_L)} e^{i k_z z} e^{a x_L} e^{-\frac{i e E e^{2 a x_L}}{4 a^2}}\\ \times \frac{\Gamma(-\lambda_s^* + \nu^*)}{\Gamma\left(-\lambda_s^*+1\right) \Gamma (\nu^*)} H_s(y)\epsilon_s
    \end{split}
\end{equation}
\begin{equation}
\begin{split}
\label{fullmode11ILlim}
    V_{s,n_L}(x)_{\mathit{I^-_L}}=\frac{e^{e^{-ax_L}/2}}{R_s}\bigg(ie^\mu_a \gamma^{a}\partial_\mu-e A_\mu e^\mu_a \gamma^{a}+m\bigg)e^{i \omega (t_L+x_L)} e^{i k_z z} e^{-a x_L} e^{-\frac{i e E e^{2 a x_L}}{4 a^2}}\\ \times \frac{\Gamma(-\lambda_s^* + 1)}{\Gamma(-\lambda_s^*+\nu^*) \Gamma (2-\nu^*)} H_s(y)\epsilon_s
    \end{split}
\end{equation}

\bibliographystyle{cas-model2-names}



\end{document}